%% file: main.tex
\documentclass[runningheads]{llncs}
\usepackage{amsmath}
\usepackage{amsfonts}
\usepackage{amssymb}
\usepackage{graphicx}
\usepackage{tikz}
\usetikzlibrary{positioning,calc,shapes,decorations.pathreplacing,fit}
\makeatletter
\newcommand{\gettikzxy}[3]{%
	\tikz@scan@one@point\pgfutil@firstofone#1\relax
	\edef#2{\the\pgf@x}%
	\edef#3{\the\pgf@y}%
}
\makeatother

\usepackage[normalem]{ulem}
\newcommand{\inserted}[1]{#1}

\usepackage{xcolor}
\usepackage{rwthcolors}
\usepackage{xspace}
\usepackage{multirow}
\usepackage{array}
\usepackage{booktabs}
\usepackage{longtable}
\usepackage{pdflscape} 
\usepackage{scrextend}
\usepackage{ifthen}
\usepackage{wrapfig}
\usepackage{subcaption}
\usepackage{listings}
\usepackage{url}
\usepackage{hyperref}

\usepackage{todonotes}

\usepackage{tcolorbox}
\tcbuselibrary{breakable} 

\usepackage{simulink}

\usepackage{stmaryrd}

\newcommand{\sldv}{{\small\texttt{SLDV}}\xspace}
\newcommand{\btc}{{\small\texttt{BTC}}\xspace}
\newcommand{\btcembedded}{{\small\texttt{BTC EmbeddedPlatform}}\xspace}

\newcommand{\initially}{\texttt{initially}\xspace}
\newcommand{\globally}{\texttt{globally}\xspace}
\newcommand{\beforeR}{\texttt{before R}\xspace}
\newcommand{\afterQ}{\texttt{after Q}\xspace}
\newcommand{\betweenQandR}{\texttt{between Q and R}\xspace}
\newcommand{\afterQuntilR}{\texttt{after Q until R}\xspace}

\newcommand{\absence}{\texttt{absence}\xspace}
\newcommand{\universality}{\texttt{universality}\xspace}

\newcommand{\response}{\texttt{response}\xspace}

\newcommand{\invariant}{\texttt{invariant}\xspace}
\newcommand{\responseTbEb}{\texttt{responseTbEb}\xspace}

\newcommand{\customHochkommata}[1]{``#1''}

\input{simulinkTikzStyle.tex}
\usepackage{simulink}

\newcommand{\simulinkBlock}[7]{	
	\begin{tcolorbox}[colback=white,colframe=black75,title=#1 (Category: #2),
		subtitle style={boxrule=0.4pt,colback=black50}] 
		#3
		\tcbsubtitle{Interface}
		\begin{center} #4\end{center}
		Parameters:
		\begin{itemize} #5\end{itemize}
		\tcbsubtitle{Implementation}\begin{center}#6\end{center}
		\tcbsubtitle{Comments}\begin{itemize}#7\end{itemize}
	\end{tcolorbox}
}

\newcommand{\simulinkFunction}[3]{	
	\begin{tcolorbox}[colback=white,colframe=black75,title=\texttt{#1},
		subtitle style={boxrule=0.4pt,colback=black50}] 
		Parameters:
		\begin{itemize} #2\end{itemize}
		\tcbsubtitle{Description}#3
	\end{tcolorbox}
}

\begin{document}
\title{Multiple Analyses, Requirements Once:}
\subtitle{simplifying testing \& verification\\in automotive model-based development}
\titlerunning{Multiple Analyses, Requirements Once}
%
\author{
  Philipp Berger\inst{1} \and
  Johanna Nellen\inst{1} \and
  Joost-Pieter Katoen\inst{1} \and 
  Erika \'Abrah\'am\inst{1}\\
  Md Tawhid Bin Waez\inst{2} \and
  Thomas Rambow\inst{2}
}

\institute{
RWTH Aachen University, Germany\\\email{ \{berger, johanna.nellen, katoen, abraham\}@cs.rwth-aachen.de}\vspace*{1em}
\and
Ford Motor Company, USA/Germany\\\email{ \{mwaez, trambow\}@ford.com}
}

\authorrunning{P. Berger et al.}
%
%
\maketitle              
\begin{abstract}

  In industrial model-based development (MBD) frameworks, requirements are
  typically specified informally using textual descriptions. To enable
  the application of formal methods, these specifications need to be
  formalized in the input languages of all formal tools that should be
  applied to analyse the models at different development levels. 
  In this paper we propose a unified approach for the
  computer-assisted formal specification of requirements and their
  fully automated translation into the specification languages of
  different verification tools.  We consider a two-stage MBD scenario
  where first Simulink models are developed from which executable code
  is generated automatically. We (i) propose a specification language
  and a prototypical tool for the formal but still textual
  specification of requirements, (ii) show how these requirements can
  be translated automatically into the input languages of Simulink
  Design Verifier for verification of Simulink models and BTC EmbeddedValidator
  for source code verification, and (iii) show how our unified
  framework enables besides automated formal verification also the
  automated generation of test cases.

\end{abstract}

\section{Introduction}
\label{sec:introduction}
\input{chapters/introduction.tex}

\section{Related Work}
\label{sec:relatedWork}	
\input{chapters/relatedWork.tex}


\section{Pattern-Based Requirement Specification Language}
\label{sec:specificationLanguage}
\input{chapters/specificationLanguage.tex}

\section{Pattern-Based Requirement Specification Tool}
\label{sec:specificationTool}
\input{chapters/specificationTool.tex}
\section{Requirement Specification Export to Verification Tools}
\label{sec:exportSldv}
\input{chapters/exportToSldv.tex}

\section{Conclusion and Future Work}
\label{sec:conclusion}
\input{chapters/conclusion.tex}


%

%
%
%

\bibliographystyle{splncs04}
\bibliography{references}
%


\newpage
\appendix
\input{chapters/appendix/eventGrammar.tex}
\newpage
\input{chapters/appendix/specificationLibrary.tex}

\end{document}

%% file: simulinkTikzStyle.tex
\definecolor{slGrey}{RGB}{217,217,217}
\definecolor{slTestObj}{RGB}{153,153,153}

\tikzset{
	link/.style={
		->,arrows={-Triangle}
	},
	testObj/.style={
		draw=none,circle,fill=slTestObj,text=white
	},
	testObjLabel/.style={
		draw=none,text=slTestObj
	},
	toTestObj/.style={
		-, shorten >=2pt,
	},
	fromTestObj/.style={
		link, shorten <=2pt,
	},
	verificationObj/.style={
		draw=none,circle,fill=slTestObj,text=white
	},
	toVerificationObj/.style={
		-, shorten >=2pt,
	}
}

%% file: chapters/introduction.tex
In the automotive industry, software units for controllers are often
im\-ple\-men\-ted using \textcolor{blue}{\emph{model-based development}} (\emph{MBD}). 
The industry standard ISO26262 recommends \emph{formal verification} to
ensure that such safety-critical software is implemented in accordance
with the functional requirements.  
The work \inserted{of} our previous two papers~\cite{fm_paper_johanna,fm_paper_philipp} and this paper not only applies to safety critical automotive software but also to quality management (QM) or non-safety critical automotive software. 
In fact, we worked only on Ford QM software features in our papers.
To optimally exploit recent academic developments as well as the capabilities of state-of-the-art
verification tools, Ford Motor Company and RWTH Aachen University
initiated an alliance research project to analyze how formal
verification techniques for discrete-time systems can be embedded into
Ford's model-based controller development framework, and to
experimentally test their feasibility for industrial-scale C code
controllers for mass production.

In our previous works~\cite{fm_paper_johanna,fm_paper_philipp}, we considered an MBD process starting with the development of Simulink controller models and using Simulink's code generation functionality to derive C code for software units. 
For formal verification, we analyzed the feasibility of both \emph{Simulink Design Verifier} (\emph{\sldv}) for Simulink models as well as \emph{\btc EmbeddedPlatform} verification tool for the generated C code. 
Our papers~\cite{fm_paper_johanna,fm_paper_philipp} present our observations and give recommendations for requirement engineers, model developers and tool vendors how they can contribute to a formal verification process that can be smoothly integrated into MBD.


The most serious pragmatic obstacles that we identified for the integration of formal methods are related to the \textcolor{blue}{\emph{requirement specifications}}.
The requirement specifications were given informally in natural language. 
All the considered natural language requirements described time-bounded linear temporal logic (LTL) properties, which we manually formalized for both the \sldv and the \btc verification tools. 
During the formalization we detected ambiguity, incompleteness or inconsistency for roughly half of the textual requirements.

The manual formalizations needed discussions with requirement engineers to clarify and correct these flaws. 
However, a high degree of automation is a prerequisite for mass production and the integration of formal methods into the established MBD process at Ford.
Automation allows the usage of formal verification within a development team of engineers with little knowledge of formal verification.  
Ideally, verification is automatically triggered whenever changes have been made to either the requirements, the Simulink model, or the used verification tools.  
Verification results can then be stored and compared with previous runs, making deviations from previous results easily detectable. All deviations can then be reported to a person with a strong background in formal methods for thorough investigation.


We also encountered problems rooted in the fact that the formalizations for the two different formal tools were done independently due to syntactic differences: in Simulink, requirements are themselves Simulink models that need to be embedded into the models that should satisfy them, whereas in \btc the requirements can be
specified either using a graphical interface for pattern-based specification or directly in an XML-based file input format. 

The independence of multiple requirement formalizations has several disadvantages. 
First and foremost, basically the same work is done multiple times, using different input languages. 
In addition, the formalizations have the risk to be slightly different. 
This may result in potentially incompatible analysis results requiring a deep and
time-consuming analysis. 
\textcolor{blue}{\emph{When the formalizations are done independently, they cost additional resources in time and expert knowledge, raising development cost.}}


In addition, typically several programming and modeling languages are used within a company such as Ford. 
The preference of these languages changes over time and each language has its own analyses tools. 
Different teams within a company like Ford may use different tools for the same purpose. 
The fact that almost every formal verification tool has its unique input language is a big obstacle to introduce formal methods into versatile companies like Ford. 
\textcolor{blue}{\emph{A common requirement language for all formal verification tools may help to take advantage of the strengths of different tools}}.

To diminish these problems, this paper presents a \textcolor{blue}{\emph{common formal requirement specification framework}}. 
We focus on Simulink and C code verification in the automotive domain, but our framework
is naturally extensible to further languages and tools.
Concretely, the paper makes the following main contributions:
\begin{enumerate}
\item We identify \textcolor{blue}{\emph{a small fragment of LTL}} as a
  formal specification language that is expressive enough \textcolor{blue}{\emph{for the formalization of typical requirements in the context of MBD in the automotive sector}}.
\item We describe our \textcolor{blue}{\emph{tool}} that was designed as a prototype for use inside this research project as a proof of concept. Similar to BTC EmbeddedSpecifier it assists users who are not experts in formal methods to specify unambiguous and complete formal requirements using textual descriptions according to a pattern-based syntax.
\item We propose an approach for the \textcolor{blue}{\emph{fully automated translation of the above-specified formal requirements into Simulink models}} that can be embedded in \sldv verification processes.
\item We describe how to \textcolor{blue}{\emph{automatically generate \btc models}} from those formal requirements for source code verification.
\item We show how to \textcolor{blue}{\emph{automatically generate test objectives}} from formal requirements that can be used for automated test-case generation.
\end{enumerate}

\begin{wrapfigure}[13]{r}[0pt]{7.2cm}
	\centering
\vspace*{-6ex}
\hspace*{-3ex}
	\scalebox{0.89}{\input{images/pattern-basedSpecification.tex}}
	\caption{The structure of our unified specification and analysis framework.}
	\label{fig:patternBasedSpec}
\end{wrapfigure}
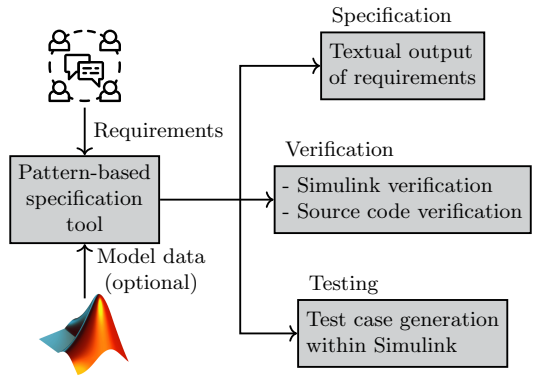
Our framework is illustrated in Fig.
\ref{fig:patternBasedSpec}. While compu\-ter-assisted approaches for
formal requirement specification have been proposed (see
Sec. \ref{sec:relatedWork}), we believe that our approach supporting
direct analysis using multiple tools at different development levels
is novel, especially the automated specification export to Simulink
and the generation of test-cases.

%% file: images/pattern-basedSpecification.tex
%
%
%
%
%
%

\tikzstyle{toolOutput}=[rectangle, draw=black100, thick, fill=black25, inner sep = 2pt, minimum width = 2.0cm, minimum height = 1.0cm, text=black100]
\tikzstyle{toolOutputDescription}=[draw=none, text=black100]
\tikzstyle{tool}=[rectangle, draw=black100, thick, fill=black25, inner sep = 2pt, minimum width = 2.0cm, minimum height = 1.0cm, text=black100]
\tikzstyle{line}=[->, thick]

\begin{tikzpicture}[scale=1.0, font=\footnotesize, text=black100,node distance=2cm]

\node[toolOutput] (textualReqs) at (0,0) {\begin{tabular}{l}Textual output\\of requirements\end{tabular}};
\node[toolOutputDescription] (textualReqsTitle) at (textualReqs.north west) [yshift=-0.1cm,anchor=south west] {\begin{tabular}{c}Specification\end{tabular}};

\node[toolOutput] (verificationOutput) [below of=textualReqs] {\begin{tabular}{l}- Simulink verification\\- Source code verification\end{tabular}};
\node[toolOutputDescription] (verificationOutputTitle) at (verificationOutput.north west) [yshift=-0.1cm,anchor=south west] {\begin{tabular}{c}Verification\end{tabular}};

\node[toolOutput] (testingOutput) [below of=verificationOutput] {\begin{tabular}{l}Test case generation\\within Simulink\end{tabular}};
\node[toolOutputDescription] (testingOutputTitle) at (testingOutput.north west) [yshift=-0.1cm,anchor=south west] {\begin{tabular}{c}Testing\end{tabular}};

\coordinate (MiddleOutput) at ($(textualReqs.north)!0.5!(testingOutput.south)$);

\node[tool] (tool) [left of=MiddleOutput,xshift=-2.75cm] {\begin{tabular}{c}Pattern-based\\specification\\tool\end{tabular}};

\gettikzxy{(textualReqs.west)}{\textx}{\texty}
\gettikzxy{(verificationOutput.west)}{\verx}{\very}
\gettikzxy{(testingOutput.west)}{\testx}{\testy}

\draw[thick] ($(tool.east)+(0.0,-0.0cm)$) -| ($(verificationOutput.west)+(-0.5,0)$);
\draw[line] ($(verificationOutput.west)+(-0.5,0)$) -| ($(\verx,\texty)+(-0.5,0)$) -- (textualReqs.west);
\draw[line] ($(verificationOutput.west)+(-0.5,0)$) -- (verificationOutput.west);
\draw[line] ($(verificationOutput.west)+(-0.5,0)$) -| ($(\verx,\testy)+(-0.5,0)$) -- (testingOutput.west);

\node[draw=none] (requirements) [above of=tool] {\pgftext{\includegraphics[scale=0.08]{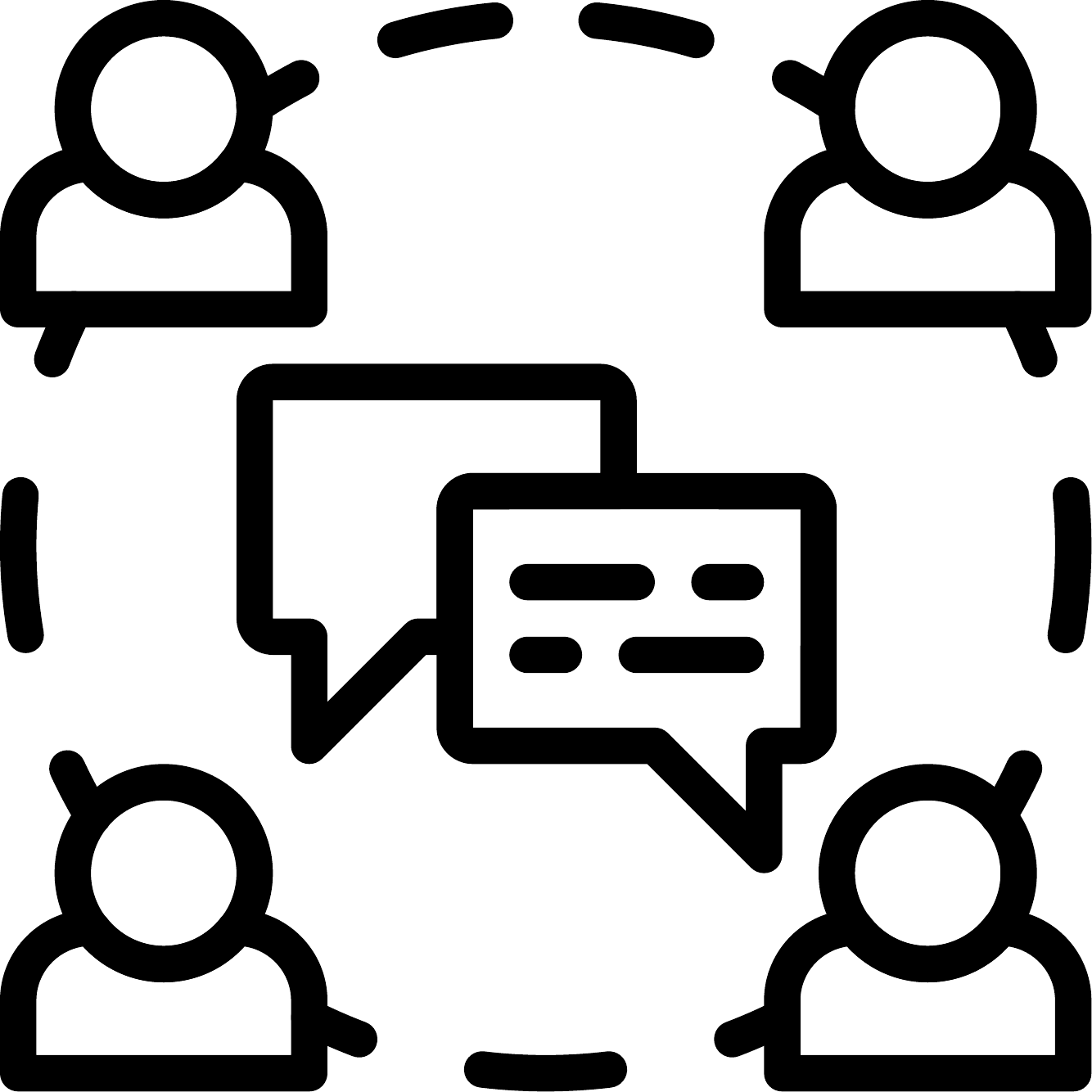}}};
\draw[line] ($(requirements.south)+(0.0,-0.5cm)$) -- ($(tool.north)$) node [midway, right] (TextNode1) {Requirements};

\node[draw=none] (model) [below of=tool] {\pgftext{\includegraphics[scale=0.3]{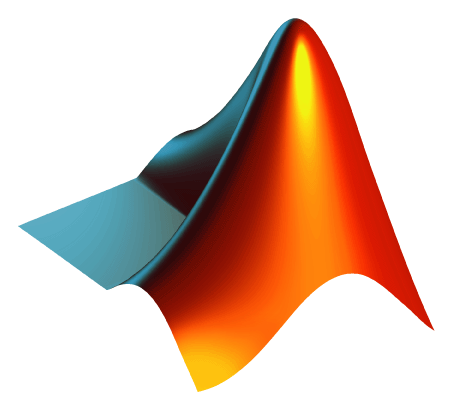}}};
\draw[line] ($(model.north)+(0.0,+0.4cm)$) -- ($(tool.south)$)node [midway, right] (TextNode2) {\begin{tabular}{c}Model data\\(optional)\end{tabular}};

\end{tikzpicture}%

%% file: chapters/relatedWork.tex
Patterns for specifying properties in formal verification were introduced by Dwyer et al. \cite{dwyer}. 
Cheng et al. has extended this work to real-time properties \cite{cheng} and Grunske introduced a pattern system for probabilistic properties \cite{grunske}. 
Autili et al. \cite{autili} recently presented a unified pattern catalog that combines known patterns for qualitative, real-time and probabilistic properties. Then the work has been extended by some new patterns.
Our works relies on the pattern catalogs from \cite{dwyer,autili}. 
Inspired by our experience with Ford \cite{fm_paper_johanna,fm_paper_philipp}, we selected a set of patterns that covers more than $90\%$ of our investigated automotive requirements.


Several tools are available for pattern-based specifications. 
The PSPWizard \cite{psp_wizard} and the SESAMM Specifier \cite{sesamm_specifier} provide for a given pattern library export functionalities to a formal logic or to a textual representation. 
The SESAMM specifier has been integrated into an industrial toolchain in the automotive domain. 
The tool PASS (\textbf{P}roperty \textbf{ASS}istant) \cite{pass_tool} guides the user by a set of questions towards a suitable pattern from which a $\mu$-calculus formula and a UML sequence diagram can be generated. 
The tool PROPEL \cite{propel} represents patterns in natural language and by finite-state automata. 
The COMPASS toolset \cite{DBLP:journals/cj/BozzanoCKNNR11} supports the original patterns by Dwyer, real-time- and probabilistic patterns.
While the previous mentioned tools use the pattern catalog from \cite{autili,dwyer,DBLP:journals/cj/BozzanoCKNNR11}, the work \cite{tool} presents different patterns and a tool for the requirement specification and automated translation to a formal language.
The tool DDPSL \cite{ddpsl_tool} goes a step further by allowing the user to fill the templates in a pattern with assumptions on the variables using a defined set of logical and temporal operators.
The ReSA Tool allows an automated consistency check of requirements on multiple abstraction levels using SAT checking.
The commercial tool \btc EmbeddedPlatform\footnote{\url{https://www.btc-es.de/en/products/btc-embeddedplatform/}} also offers the possibility to formalize textual requirements in a pattern-based language. 
Former versions of the tool support a pattern catalog but the latest release uses the \emph{universal pattern} \cite{universal_pattern} that offers a graphical specification for trigger-action based requirements.
Our tool focuses on the key patterns but allows for automated generation of test cases, as well as properties for Simulink model and source code verification.

Besides the tools for pattern-based specifications, several experience reports on using specification patterns have been published. 
In \cite{patterns_to_industry}, a case study in the field of service-based applications is presented. 
\cite{property_specification} reports on an approach using pattern-based specifications in the area of work flow modeling. 
Bosch company investigated the suitability of the pattern catalog from \cite{cheng} for $289$ informal behavioral requirements from their automotive projects.
A report on the integration of a pattern-based specification tool in an industrial (automotive) tool chain is given in \cite{sesamm_specifier,automotive_case_study}.
A restricted set of patterns was used for the formal specifications within the PICASSOS project \cite{picassos}.
\inserted{A system for modeling and testing flight critical software was presented in \cite{DBLP:conf/re/MoitraSCCDLYMM18}, but their focus lies on test-case generation and modeling structural aspects of the software system, whereas our focus is on the automated translation of requirements.}

%% file: chapters/specificationLanguage.tex
Requirement documents are commonplace in the automotive industry and
are usually written in natural language by a large number of
stakeholders.  These can include engineers and other people without a
strong background in formal methods, which may lead to ambiguous
requirements.  
Specification patterns may assist engineers in writing complete and unambiguous textual requirements.
A pattern defines the behavior that is described by a requirement and uses templates for additional information like the specification of events and durations.
In contrast to most existing approaches, events are specified by a constrained grammar, and higher-order operators, e.g. hysteresis\inserted{\footnote{\inserted{Hysteresis is a functionality often used to prevent rapid toggling when
observing an input signal against some threshold by introducing an upper and a lower delta.}}}, are supported to enable specifying new operations on events.

\smallskip\noindent\textbf{Goals.}\quad The pattern-based
specification language should produce human readable specifications.
A formal semantics avoids ambiguities and allows
the automated generation of tool-specific requirement specifications.
Our aim is to keep the pattern language simple such that no expert
knowledge is needed and the learning curve for requirement engineers
is low.  
We believe that a limited number of simple patterns reduces
incorrect choices of patterns or scopes when writing requirements
while still covering a high percentage of requirements.

\smallskip\noindent\textbf{Why yet another specification
  language?}\quad Tools like \btcembedded come with their own,
existing, pattern-based specification language and there are existing
tools for pattern-based specification.  Nonetheless we decided that
creating our own language and tool was the better choice.  A key
difference from many established pattern-based specification languages
is that we also require the events to be specified using a constrained
grammar, enforcing the events to be formalizable properties.  This, in
turn, allows us to immediately export the entire property to a
supported format without the need for any further user interaction.

Adding new features or constructs like higher-order operators (e.g.
hysteresis) is easy to achieve, requiring only very modular changes to
the grammar and the back-end exporter classes.  We want to be able to
create our own pool of higher-order operators for event
specification that can be used to ease the burden of formalization for
the engineers.
Our own language allowed us to do rapid prototyping while coming up with new ideas, without the burden of getting all stakeholders of an established language on board beforehand.

\begin{figure}[t]
\begin{center}
\tt
\begin{tabular}{ll}
specification: & scope pattern;\\
scope:         &initially | globally;\\
pattern:       &invariant | response;\\
initially:     &'\emph{At system start,}';\\
globally:      &'\emph{At each time step,}';\\
invariant:     &'\emph{[}' event '\emph{] holds.}';\\
response:      &'\emph{if [}' event '\emph{] has been valid for [}' duration '\emph{],}'\\
& '\emph{then in response, after a delay of [}' duration '\emph{],}'\\
               & '\emph{[}' event '\emph{] is valid for [}' duration '\emph{].}';\\
event:         & identifier | event AND event | \ldots | term $ \leq $ term | \ldots \\
term:          & identifier | term + term |\ldots\\
duration:      &uint unit;\\
uint:          &[1..9] [0..9]*;\\
unit:          &'\emph{simulation steps}' | '\emph{milliseconds}' | '\emph{seconds}' \\
               & | '\emph{minutes}' | '\emph{hours}';
\end{tabular}
\end{center}
\caption{Syntax of our pattern-based requirement specification language.}
\label{def:syntaxSpecification}
\end{figure}

\smallskip\noindent\textbf{Syntax.}\quad We used \cite{autili,dwyer,cheng} as a starting
point to design our \emph{pattern-based requirement specification language} $\mathcal{L}$, whose grammar 
is shown in Fig.
\ref{def:syntaxSpecification}; for more details see also Appendix
\ref{sec:eventGrammar}.

\begin{table}[t]
	\centering
	\caption{Pattern distributions for three different controller models.}
	\scalebox{0.9}{
		\begin{tabular}{l @{\hskip 0.4cm} c c @{\hskip 0.8cm} c c @{\hskip 0.8cm} c c}
			\toprule
			Pattern					& \multicolumn{2}{c@{\hskip 0.8cm}}{LSC} 
			& \multicolumn{2}{c@{\hskip 0.8cm}}{DSR}
			& \multicolumn{2}{c@{\hskip 0.8cm}}{ECC}\\\midrule
			Invariant				& $35$	& $85.4\%$ 		& $50$	& $92.6\%$		& $80$	& $97.6\%$\\
			Time-bounded response (exact bound)	& $5$	& $12.2\%$ 		& $4$	& $7.4\%$		& $2$	& $2.4\%$\\
			Event-bounded response		& $1$	& $2.4\%$ 		& $0$	& $0.0\%$		& $0$	& $0.0\%$\\
		\end{tabular}
	}
	\label{tab:patternDistribution}
\end{table}

Requirement \emph{specifications} consist of a scope followed by a
pattern. We start with a limited set of scopes and patterns
that can be extended later to cover further specification types.
However, these limited sets were sufficient to formalize more than
$90\%$ of the requirements in all three case studies (Low Speed Control for Parking (LSC), Driveline State Request (DSR) and E-Clutch Control (ECC)) we
considered in \cite{fm_paper_philipp,fm_paper_johanna} (see Table
\ref{tab:patternDistribution}). Other internal case studies from Ford show similar results.

Currently two \emph{scopes} are supported: the
\initially scope is used to express that a property should hold at
system start, i.e. at time step $0$ of a simulation before any operations have been performed, while the
\globally scope expresses that a property should hold at each time
step of an execution, but starting after the first execution.  In \cite{autili,dwyer,cheng} there are further
scopes like \beforeR, \afterQ, \betweenQandR and \afterQuntilR that
can be considered for future inclusion.

We support two \emph{patterns} for defining which property is required
to hold for a given scope.  The \invariant pattern allows to state that a certain event
holds (at each time step within the specified
scope), and covers both the
\absence and the \universality patterns from \cite{dwyer} if the negation
of events is supported.  
%
The \response pattern specifies causalities between two events:
the continuous validity of a trigger event for a given trigger
duration implies that after a fixed separative duration the response
event holds continuously for a given response duration. 
The \emph{events} in the above patterns are built from identifiers (\texttt{signals}, \texttt{con\-stants} and
(calibration) \texttt{parameters}) using a set of \texttt{functions} and \texttt{operators}.
%
We support those operators and functions that were used in our case studies,
including the Boolean operators AND, OR, NOT and IMPLIES, the relational operators $<$, $\leq$, $>$, $\geq$ and $=$, the arithmetic operators $+$, $-$, $\cdot$ and $/$, absolute value, minimum, maximum, functions for bit extraction (bit $x$ of variable $y$)  and time delays (value of $x$ $n$ steps ago).
The complete ANTLR grammar for events is presented in Appendix \ref{sec:eventGrammar}.
We plan in future work to incorporate more advanced operators like \emph{state change} (``the value of [param] transitions from [const1] to [const2]''), different variants of \emph{hysteresis} functions, \emph{saturation}, \emph{rate limiter} and \emph{ramping up} functions and \emph{lookup tables}. 
Note that though custom operators and functions allow users a more efficient specification, special operators (e.g. lookup tables) might not be realizable in all specification languages for which export is provided.


%


\begin{figure}[t]
\begin{center}
\begin{tabular}{llll}
scopes: & $\llbracket \initially\ \texttt{pattern} \rrbracket$ & 
$=$ &
$\llbracket\texttt{pattern}\rrbracket$\\
& $\llbracket \globally\ \texttt{pattern} \rrbracket$ & 
$=$ & 
$\bigcirc\ \Box\ \llbracket\texttt{pattern}\rrbracket$\\[1ex]
patterns: & $\llbracket [e]\ \mathit{holds}.\rrbracket$ & 
$=$ & 
$\llbracket e\rrbracket$\\
& \multicolumn{3}{l}{$\llbracket \mathit{if}\ [e_P]\ \textit{has been valid for } [d_P],\ \textit{then in response, after a delay of}\ [d],$}\\
& \multicolumn{3}{l}{$\phantom{\llbracket}[e_Q]\ \textit{is valid for}\ [d_Q].\rrbracket$}\\
& & $=$ & $\left(\Box^{[\leq \llbracket d_P \rrbracket ]}\ \llbracket e_P  \rrbracket\right) \rightarrow \left( \lozenge^{[=\llbracket d_P \rrbracket + \llbracket d \rrbracket ]}\ \Box^{[\leq \llbracket d_Q \rrbracket]}\ \llbracket e_Q \rrbracket \right)$\\[1ex]
events: & $\llbracket\texttt{identifier}\rrbracket$ & $=$ & $\texttt{identifier}$\quad\ldots\\
& $\llbracket e_1\ \texttt{AND}\ e_2\rrbracket$ & $=$ & $\llbracket e_1\rrbracket \wedge \llbracket e_2\rrbracket$ \quad \ldots\\
& $\llbracket t_1 \leq t_2\rrbracket$ & $=$ & $\llbracket t_1\rrbracket \leq \llbracket t_2\rrbracket$ \quad \ldots\\
durations: & $\llbracket n\ \textit{seconds}\rrbracket$ & $=$ & $\frac{1000 \cdot n}{D_{\text{Step}}}$ \quad \ldots\\
\end{tabular}
\end{center}
\caption{Semantics of our pattern-based requirement specification language.}
\label{def:semanticSpecification}
\end{figure}

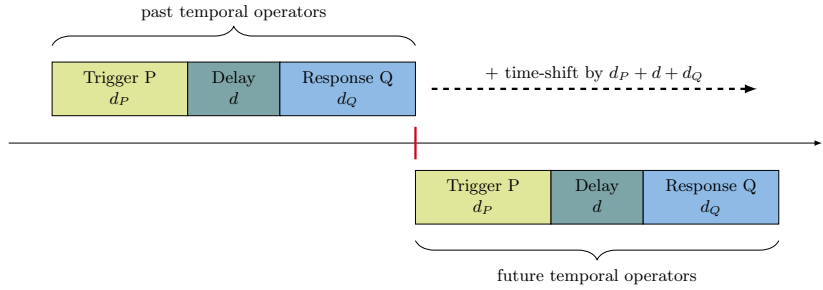
\begin{figure}[t]
	\centering
	\resizebox{0.9\textwidth}{!}{
	\input{images/futureVsPastTimeOperators.tex}%
	}
	\caption{Evaluation of a response pattern with past or future operators. The present is represented by the red tick on the timeline.}
	\label{fig:futureVsPastTime}
\end{figure}

\smallskip\noindent\textbf{Semantics.}\quad 
We define the semantics of requirement specifications using linear temporal logic with quantitative temporal operators to express time durations, that is MTL \cite{mtl}.
The main semantical components are shown in Fig. \ref{def:semanticSpecification} using only future temporal modalities (straightforward and therefore not listed in Fig. \ref{def:semanticSpecification} are the semantics for events and durations, see Appendix \ref{appendix:semantics} for a complete definition). 
We use $D_{\text{Step}}$ to denote the step-size, here in milliseconds. 
We support durations that are multiples of $D_{\text{Step}}$.
An equivalent semantics using past temporal modalities is given in Appendix \ref{appendix:semantics}. 
The difference in terms of a time-shift between the formulations using past (resp. future) operators is illustrated in Fig. \ref{fig:futureVsPastTime}.
The two equivalent semantics support the export of a pattern-based specification into different specification languages. 
For example the specification language of the analysis tool \sldv only supports past temporal operators. 

%% file: images/futureVsPastTimeOperators.tex
\begin{tikzpicture}
\begin{scope}[xshift=-6.7cm, yshift=1cm]
	\node[rectangle,fill=maygreen50, draw, minimum height=0.75cm, minimum width=2.5cm] at(1.25,0) {\begin{tabular}{c}Trigger P\\$d_P$\end{tabular}};
	\node[rectangle, fill=petrol50, draw, minimum height=0.75cm, minimum width=1.7cm] at (3.35, 0) {\begin{tabular}{c}Delay\\$d$\end{tabular}};
	\node[rectangle,fill=blue50, draw, minimum height=0.75cm, minimum width=2.5cm] at(5.45,0) {\begin{tabular}{c}Response Q\\$d_Q$\end{tabular}};
	\draw [decorate,decoration={brace,amplitude=10pt}] (0,0.8) -- (6.7,0.8) node[midway, above=10pt] {past temporal operators};
	
	\draw[-latex, very thick, dashed] (7.0, 0) -- node[midway, above] {$+$ time-shift by $d_P + d + d_Q$} (13, 0); 
\end{scope}

\begin{scope}[yshift=-1cm]
	\node[rectangle,fill=maygreen50, draw, minimum height=0.75cm, minimum width=2.5cm] at(1.25,0) {\begin{tabular}{c}Trigger P\\$d_P$\end{tabular}};
	\node[rectangle, fill=petrol50, draw, minimum height=0.75cm, minimum width=1.7cm] at (3.35, 0) {\begin{tabular}{c}Delay\\$d$\end{tabular}};
	\node[rectangle,fill=blue50, draw, minimum height=0.75cm, minimum width=2.5cm] at(5.45,0) {\begin{tabular}{c}Response Q\\$d_Q$\end{tabular}};
	\draw [decorate, decoration={brace,amplitude=10pt, mirror}] (0,-0.8) -- (6.7,-0.8) node[midway, below=10pt] {\begin{tabular}{c}future temporal operators\end{tabular}};
\end{scope}

\draw[-latex] (-7.5,0) -- (7.5,0);
\draw[very thick, red100] (0,-0.3) -- (0,0.3);
\end{tikzpicture}

%% file: chapters/specificationTool.tex
We have implemented a pattern-based specification tool as a prototype for use inside this research project to support requirement engineers in writing unambiguous and complete textual requirements. 
Our focus was to create a modular tool that is easy to learn and extendible, if in the future a larger set of scopes, patterns or operators needs to be supported.

A user can either import signals, calibratables and constants from a file, or create, change and delete them manually. 
Calibratable parameters remain constant during software execution but can be 
adjusted before the execution for tuning or selecting the possible 
functionalities. 
Captured data includes a name, description, minimum and maximum values, dimensions, a value, the data type and the variable type (signal, calibratable or constant).
With the information of available variables readily available, the tool checks specified events for whether all referenced variables actually exist.

\begin{figure}[t]
	\centering
	\includegraphics[width=0.99\textwidth]{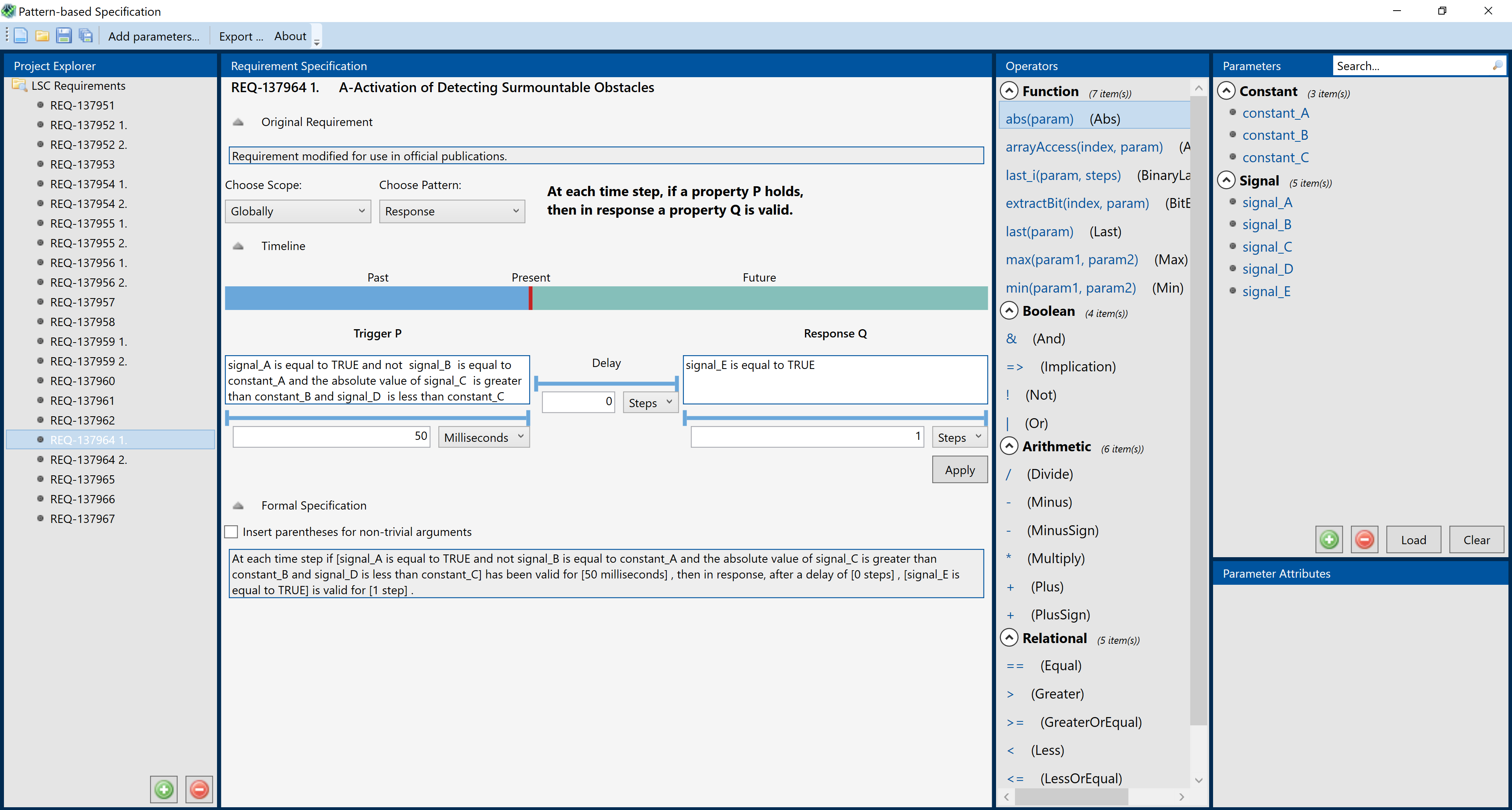}
	\caption{User interface of our pattern-based specification tool.}
	\label{fig:tool}
\end{figure}

The current version of the tool provides export functionality for a selected requirement or for all of them. 
Export formats are  textual (.txt), \sldv\ (.m), \btc (.spec) and C (.c) specifications. 
The last one is compatible with the \href{https://sv-comp.sosy-lab.org/}{SV-COMP} standard \cite{DBLP:conf/tacas/Beyer17} and can be used for formal verification with e.g. the \href{https://monteverdi.informatik.uni-freiburg.de/tomcat/Website/?ui=tool&tool=automizer}{\emph{Ultimate Automizer}} \cite{DBLP:conf/cav/HeizmannHP13}.

The requirement specification panel in Fig. \ref{fig:tool} is the main panel of our tool. 
A scope and a pattern must be selected for the requirement. 
A textual translation of the scope and pattern is given as well as a visualization that shows the time steps where the chosen pattern is evaluated, see Fig. \ref{fig:tool}. 
Events are built using operators, functions, signals, constants and calibration parameters. 
For each event, a duration and a time unit can be specified. 
Additionally, for patterns with more than one event, a time delay between events can be specified, again together with a time unit. 

If the pattern-based specification is incomplete or if it contains specification errors, the lower part of the specification panel provides the list of errors and warnings.
When all issues are resolved, a textual formal specification is generated from the specification.
The modular set-up of the tool allows to add further exporters, e.g. to generate specification in a logic like MTL in a straightforward manner.

Our prototypical implementation supports the functions \texttt{abs(param)},\\ \texttt{min(param1, param2)}, \texttt{max(param1, param2)}, \texttt{last(param)},\\ \texttt{last(param, steps)} and \texttt{extractBit(index, param)}. 
For an explanation see Appendix~\ref{appendix:supportedFunctions}.
Parenthesis expressions can be built using \texttt{(param)} and the basic boolean operators \texttt{not}, \texttt{and}, \texttt{or} and \texttt{implication} are provided. 

%% file: chapters/exportToSldv.tex
\subsection{Export to \sldv}

A formal pattern specification is exported to Simulink in the form of a Matlab script. 
This script generates a specification block inside a model on the currently selected hierarchy level. 
For verification on model-level, the topmost level of a model should be selected, whereas for verification on subsystem-level the topmost level of the subsystem should be selected.
To implement the semantics of $\mathcal{L}$ in Simulink, we use a custom-build, modular and interchangeable block library and existing Simulink logic blocks.

The following requirement is used as a running example to illustrate the various steps:
\begin{example}
	At each time step if [(((\texttt{signal\_A} is equal to \texttt{TRUE}) and ((not \texttt{signal\_B}) is equal to \texttt{constant\_A})) and ((the absolute value of \texttt{signal\_C}) is greater than \texttt{constant\_B})) and (\texttt{signal\_D} is less than \texttt{constant\_C})] has been valid for [$50$ milliseconds], then in response, after a delay of [$0$ steps], [\texttt{signal\_E} is equal to \texttt{TRUE}] is valid for [1 step].
\end{example}





To support the requirement specification for \sldv, we implemented a Simulink library with building blocks for all elements of our requirement specification language. 
The library provides sub-libraries for the specification of scopes, patterns and events.


\begin{figure}[t]
	\centering
	\resizebox{0.8\textwidth}{!}{%
\input{images/figureVerificationSubsystem.tex}
}
	\caption{A sample verification subsystem block.} \label{fig:verificationSubsystem}
\end{figure}
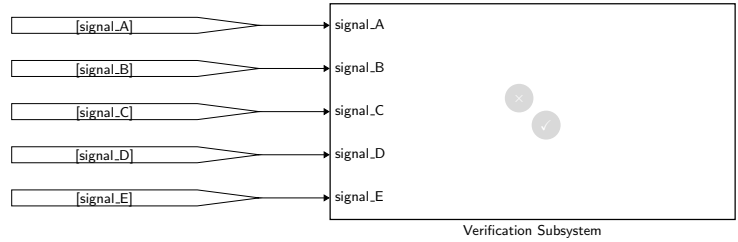

\begin{figure}[b]
	\centering
	\resizebox{0.8\textwidth}{!}{%
\input{images/figureTriggerResponsePattern.tex}
}
	\caption{The \responseTbEb\ pattern of the verification subsystem in Fig. \ref{fig:verificationSubsystem}.} \label{fig:verificationSubsystemContents}
\end{figure}
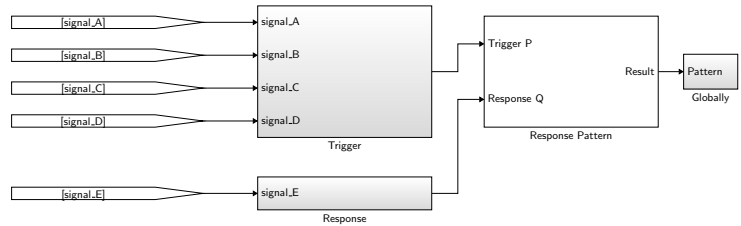

\smallskip\noindent\textbf{Verification subsystem.}\quad
Fig. \ref{fig:verificationSubsystem} shows the topmost generated block, a verification subsystem. 
Its input are all input and output signals of the Simulink model that are used by the generated requirement specification. 
The content of verification subsystems is considered during formal verification but ignored during code generation and is not part of the generated code. 
The top-level verification subsystem contains a separate verification subsystem for each requirement.

The verification subsystem subsumes the implementation of the actual requirements, i.e. encoding the expected functional behavior, by separating it into parts: Transformations on inputs, and implementing timed behavioral aspects.
A requirement specification consists of three parts: a set of events, a pattern and a scope; each is represented by distinct blocks in the library. 
Fig. \ref{fig:verificationSubsystemContents} shows an example requirement specification that consists of a \globally\ scope, a \response\ pattern and two events.

\smallskip\noindent\textbf{Scopes.}\quad
A scope block defines the time steps during which a pattern needs to be evaluated. 
The pattern result is a Boolean input parameter. 
At each simulation step, either the pattern result or \texttt{true} (if the pattern result needs not to be evaluated at the current time step) is the input of a proof objective. 
During formal verification, \sldv analyzes this proof objective. 
A requirement is violated if the input of a proof objective can be \texttt{false} at any simulation step.    

\begin{figure}[t]
	\centering
	\begin{subfigure}{.6\textwidth}%
		\input{images/figureInitially.tex}%
	\end{subfigure}%
	\begin{subfigure}{.35\textwidth}%
	\input{images/figureGlobally.tex}%
\end{subfigure}%
\caption{Proof objectives for the scopes \texttt{initially} and \texttt{globally}.}
\end{figure}

The \initially\ scope evaluates the pattern result only at system start, while the \globally\ scope evaluates the pattern result at each time step.
Fig. \ref{fig:scopeGlobally} and \ref{fig:scopeInitally} show the implementations of scopes \globally\ and \initially, respectively.
The delay block is initialized with the value $1$, while all subsequent output values will be $0$. 
The time shift (see Section \ref{sec:specificationLanguage}) is realized by the {\sffamily Detector} block.

\smallskip\noindent\textbf{Patterns.}\quad
A pattern receives the Boolean signals from the events as inputs along with the time duration and delays between events specified as mask parameters of the pattern.
A pattern block ensures the correct order of events and handles timing aspects like event durations and delays between events. 
Simulink blocks for time durations and delays are provided by our Simulink specification library. 
The output of a pattern block is again a Boolean signal.
In Fig. \ref{fig:verificationSubsystemContents}, the blocks {\sffamily Trigger} and {\sffamily Response} contain the part of signal transformation, whereas the block labeled {\sffamily Response Pattern} represents the details of the duration- and delay checks, as shown in Fig. \ref{fig:triggerResponsePattern}.
Inside this subsystem block, the event order (trigger before response) is established together with the specified time delay between the two events. 
\input{images/figureImplementationTriggerReponsePattern.tex}
In our example, the trigger has to be true for $50$ milliseconds. 
This duration is checked by the {\sffamily Duration Check} block which returns a Boolean true iff its input evaluated to true for a given number of time steps. 
A delay block is then used to account for the response duration and a possible fixed delay between trigger and response.


\smallskip\noindent\textbf{Events.}\quad
Each event is specified in its subsystem.
The event subsystems are connected with the input signals of the verification subsystem using {\sffamily From} blocks. 
An event is built using the blocks provided by our Simulink specification library.
These building blocks must be connected in accordance with the rules of our event grammar. 
The output of an event specification is a Boolean signal.
\input{images/figureTriggerImplementation.tex}
Fig. \ref{fig:triggerImplementation} shows the necessary signal transformations for the trigger of the example requirement.  

\smallskip\noindent\textbf{Connection to the Simulink model.}\quad
After the automated insertion of the verification subsystem at a user-chosen level in the model, the inputs of the verification system need to be connected to the corresponding signals in the model.
Because of possible data dependency issues, we use global data store blocks for accessing the signals. 
For selecting the source signal, we traverse the model in a hierarchical approach and try to find the first match of a named signal matching the one being looked for.
A data store write is then inserted into the model at the matched location, allowing us to generate the corresponding global data store read block next to our verification subsystem.

\subsection{Export to \btcembedded}
We support the export of formalized requirements to \btc's input format, so-called SPEC files. They contain an XML-based structured representation of the requirements and their patterns.
Small transformations are applied during export to match \btc's pattern semantics.
We consider the time step $0$ to be the first time step in the \initially\ scope. 
This means that we start to evaluate the pattern directly after initialization, i.e. before the first computation step. 
In contrast, \btc starts the evaluation after the first computation step. 
It is not possible to check initial variable valuations in BTC, therefore, an error is presented when exporting a requirement with scope initially to BTC.
The generated SPEC files can then be imported into \btcembedded and used for verification.

\subsection{Export to Textual Requirements}
Formally specified requirements can easily be exported to textual form. 
As many engineers and stakeholders without a solid background in formal methods are involved in the design, testing and implementation of the defined software components, it is vital to present the agreed-upon requirements in a textual representation, which is easy to understand, distribute and review. 
Our export feature for textual requirements additionally supports automatically introducing parenthesis around all non-trivial arguments used in the specification to prohibit misinterpretations or misunderstandings of the written specification---a problem we encountered several times in \cite{fm_paper_philipp,fm_paper_johanna}.

\subsection{Export to SV-Comp-style C code}
To enable the use of state-of-the-art academic C code model checkers, we explicitly encode our pattern semantics in C code.
This enables to embed all assumptions and behavior directly in the code, instead of going around it with LTL specifications or similar, as supported by some tools.
We built a boiler-plate framework for initializing parameters and calibratables (enabling verifying with varying calibrations) and updating input variables after every step.
We decided to use the established \texttt{\_\_VERIFIER\_error();} functionality for encoding violations of the behavior allowed by the patterns as supported by many code verifiers such as more than $20$ tools participating in the SV-comp.

\section{Requirement-Based Test Vector Generation}
The automated generation of an \sldv specification can be reused for automated requirement-based test vector generation. 
The Automotive Functional Safety standard ISO26262~\cite{ISO26262} recommends to identify missing test vectors and unintended behavior of the implemented model by:
``For each requirement, a set of test vectors should be generated. 
Afterwards, the structural coverage of the requirement-based test vectors shall be measured according to a suitable coverage metrics. The industry norm recommends different coverage metrics depending on the ASIL-level of the model. In case the coverage metrics reveals uncovered parts of the model, a further analysis is needed: either test vectors are missing or unintended functionality in the model has been detected.''

If requirements are verified using formal verification and the implemented requirement is shown to be valid, additional, manual creation of test vectors should not be necessary. 
Manual creation of test vectors is a tedious work and should be limited to those requirements that are not tested using formal verification. 
We propose to reuse the automated generation of \sldv requirement specification for generating test vectors for these same requirements.
For this purpose, we annotate the generated specification with so-called \emph{test objectives} (see Fig. \ref{fig:orBlockTestObj}) automatically. 
The test objectives specify the signal valuations that must be considered during test-vector generation. 

The set of requirement-based test vectors depends on the chosen coverage metric. 
For \emph{condition coverage}, a set of test vectors is required such that each condition takes every possible value, while for \emph{decision coverage} a set of test vectors must generate every possible outcome for each decision. 
Decision coverage is closely related to \emph{branch coverage}, where conditional and unconditional branches are considered. 
According to ISO26262, branch coverage is suitable for requirement coverage at software unit-level for ASIL A to C. 
However, for ASIL D, \emph{modified condition/decision coverage} (MC/DC) is highly recommended. 
Additionally, it is required that all conditions contributing to a decision must independently affect the outcome of the decision.  

To achieve condition coverage, test objectives must be added to all Boolean input signals. 
For decision coverage, test objectives are needed for all Boolean output signals. 
If test objectives are added to all Boolean output signals and to all Boolean input signals of blocks with more than one input parameter, \emph{condition/decision coverage} is achieved, which guarantees both condition and decision coverage.
For MC/DC coverage, test objectives are hard to generate and currently out of scope of our project. 
One way to at least partly cover MC/DC would be to generate test objectives for all Boolean combinations of possible input signals.
For an {\sffamily OR} block, we currently generate vectors for both outcomes, but "true" could be generated by inputs $01$, $10$ or $11$---by adding additional logic we can enforce all combinations to be generated.

Alternatively, we propose to use the built-in function of \sldv to compute a set of test vectors for MC/DC coverage. 
Unfortunately, this functionality is currently only available on model-level. 
To get requirement-based test vectors for the model, MC/DC must be checked at requirement (i.e. subsystem) level while test vectors must be generated for the complete model. 

To automate the requirement-based test vector generation, we added test objectives for \emph{condition/decision coverage} to all blocks in our Simulink formal specification library.
The relational operators compute Boolean output signals that also must be annotated with test objectives. 
Additional test objectives are necessary for all temporal operators to assure the correct length of generated test cases.
Since we handle Boolean signals only, all test objectives can take the values \texttt{true} and \texttt{false}. 
Fig. \ref{fig:orBlockTestObj} presents the implementation of the annotated Boolean \emph{Or} operator from our specification library.
\begin{figure}[t]
	\input{images/figureTestObjectives.tex}
\end{figure}

Annotating the specification library allows the flexibility of adding/removing test objectives without adapting the source code of the specification tool. 
This enables the user to maintain a set of specification libraries for different coverage metrics or to create a library without any test objective annotations.

%% file: images/figureVerificationSubsystem.tex
	\begin{tikzpicture}[font=\sffamily]%
	\pgfmathsetmacro\widthOfFlag{5.5cm}
	
	\node [rectangle split, rectangle split parts=5, draw, anchor=center, rectangle split draw splits=false,inner ysep=1em,minimum width=9cm, align=left,rectangle split part align={left},thick] at (0,1) (verificationSubsystem)
	{signal\_A\nodepart{two}signal\_B\nodepart{three}signal\_C\nodepart{four}signal\_D\nodepart{five}signal\_E};
	\coordinate (verificationSubsystemCenter) at (verificationSubsystem.center);
	\node [draw=none, below=0.0cm of verificationSubsystem.south] {Verification Subsystem};
	
	\coordinate (verificationSubsystemCenterFail) at ($(verificationSubsystem.center)+(-0.3,+0.3)$);
	\coordinate (verificationSubsystemCenterSuccess) at ($(verificationSubsystem.center)+(+0.3,-0.3)$);
	\node [circle,fill=slGrey,text=white] at (verificationSubsystemCenterFail) {$\mathbf{\times}$};
	\node [circle,fill=slGrey,text=white] at (verificationSubsystemCenterSuccess) {\checkmark};
	
	\coordinate[left=5cm of verificationSubsystem.text west] (invariantWest1);
	\coordinate[left=5cm of verificationSubsystem.two west] (invariantWest2);
	\coordinate[left=5cm of verificationSubsystem.three west] (invariantWest3);
	\coordinate[left=5cm of verificationSubsystem.four west] (invariantWest4);
	\coordinate[left=5cm of verificationSubsystem.five west] (invariantWest5);
	\coordinate[right=2cm of verificationSubsystem.three east] (invariantEast);
	\node[skFlag,draw,flag label=,block width=\widthOfFlag,anchor=center] at (invariantWest1) (flagInvInput1) {[signal\_A]};
	\node[skFlag,draw,flag label=,block width=\widthOfFlag,anchor=center] at (invariantWest2) (flagInvInput2) {[signal\_B]};
	\node[skFlag,draw,flag label=,block width=\widthOfFlag,anchor=center] at (invariantWest3) (flagInvInput3) {[signal\_C]};
	\node[skFlag,draw,flag label=,block width=\widthOfFlag,anchor=center] at (invariantWest4) (flagInvInput4) {[signal\_D]};
	\node[skFlag,draw,flag label=,block width=\widthOfFlag,anchor=center] at (invariantWest5) (flagInvInput5) {[signal\_E]};
	
	\draw [link] (flagInvInput1.portA) -- (verificationSubsystem.text west);
	\draw [link] (flagInvInput2.portA) -- (verificationSubsystem.two west);
	\draw [link] (flagInvInput3.portA) -- (verificationSubsystem.three west);
	\draw [link] (flagInvInput4.portA) -- (verificationSubsystem.four west);
	\draw [link] (flagInvInput5.portA) -- (verificationSubsystem.five west);

	\end{tikzpicture}

%% file: images/figureTriggerResponsePattern.tex
		\begin{tikzpicture}[font=\sffamily]
		\pgfmathsetmacro\widthOfFlag{5.5cm}
		
		\node [rectangle split, rectangle split parts=4, draw, anchor=center, rectangle split draw splits=false,shading = axis, left color=white, right color=slGrey,shading angle=0,inner ysep=1em,minimum width=5cm, align=left,rectangle split part align={left}] at (0,1) (trigger)
		{signal\_A\nodepart{two}signal\_B\nodepart{three}signal\_C\nodepart{four}signal\_D};
		\node [draw=none, below=0.0cm of trigger.south] {Trigger};
		
		\node [rectangle split, rectangle split parts=1, draw, anchor=center, rectangle split draw splits=false,shading = axis, left color=white, right color=slGrey,shading angle=0,inner ysep=1em,minimum width=5cm, align=left,rectangle split part align={left}] at (0,-2.5) (response)
		{signal\_E};
		\node [draw=none, below=0.0cm of response.south] {Response};
		
		\coordinate[left=5cm of trigger.text west] (triggerWest1);
		\coordinate[left=5cm of trigger.two west] (triggerWest2);
		\coordinate[left=5cm of trigger.three west] (triggerWest3);
		\coordinate[left=5cm of trigger.four west] (triggerWest4);
		\coordinate[right=2cm of trigger.east] (triggerEast);
		
		\node[skFlag,draw,flag label=,block width=\widthOfFlag,anchor=center] at (triggerWest1) (flagTrInput1) {[signal\_A]};
		\node[skFlag,draw,flag label=,block width=\widthOfFlag,anchor=center] at (triggerWest2) (flagTrInput2) {[signal\_B]};
		\node[skFlag,draw,flag label=,block width=\widthOfFlag,anchor=center] at (triggerWest3) (flagTrInput3) {[signal\_C]};
		\node[skFlag,draw,flag label=,block width=\widthOfFlag,anchor=center] at (triggerWest4) (flagTrInput4) {[signal\_D]};
		
		\draw [link] (flagTrInput1.portA) -- (trigger.text west);
		\draw [link] (flagTrInput2.portA) -- (trigger.two west);
		\draw [link] (flagTrInput3.portA) -- (trigger.three west);
		\draw [link] (flagTrInput4.portA) -- (trigger.four west);
		
		\coordinate[left=5cm of response.text west] (responseWest1);
		\node[skFlag,draw,flag label=,block width=\widthOfFlag,anchor=center] at (responseWest1) (flagRespInput1) {[signal\_E]};
		\draw [link] (flagRespInput1.portA) -- (response.text west);

		\coordinate[right=4cm of trigger.east] (patternLoc);
		\node [rectangle split, rectangle split parts=2, draw, anchor=center, rectangle split draw splits=false,inner ysep=2em,minimum width=5cm, align=left,rectangle split part align={left}] at (patternLoc) (responsePattern)
		{Trigger P\nodepart{two}Response Q};
		\coordinate[left=0.75cm of responsePattern.text west] (responsePatternInput1Coord);
		\coordinate[left=0.75cm of responsePattern.two west] (responsePatternInput2Coord);
		\coordinate[left=0cm of responsePattern.east] (responsePatternResultTextCoord);
		\node[draw=none,anchor=east] at (responsePatternResultTextCoord) {Result};
		
		\node [draw=none, below=0.0cm of responsePattern.south] {Response Pattern};
		\coordinate[right=1.5cm of responsePattern.east] (responsePatternEast);
		
		\draw [link] (trigger.east) -| (responsePatternInput1Coord) -- (responsePattern.text west);
		\draw [link] (response.east) -| (responsePatternInput2Coord) -- (responsePattern.two west);
		
		\node [rectangle split, rectangle split horizontal, rectangle split parts=1, draw, anchor=center, rectangle split draw splits=false,shading = axis, left color=white, right color=slGrey,shading angle=0,minimum height=1cm,rectangle split part align={left}] at (responsePatternEast) (scope)
		{Pattern\quad\quad};
		\node [draw=none, below=0.0cm of scope.south] {Globally};

		\draw [link] (responsePattern.east) -- (scope.west);
		\end{tikzpicture}

%% file: images/figureInitially.tex
	\centering
	\resizebox{0.99\textwidth}{!}{%
	\begin{tikzpicture}[font=\sffamily]
	\pgfmathsetmacro\widthOfFlag{5.5cm}
	
	\node [rectangle split, rectangle split parts=2, draw, anchor=center, rectangle split draw splits=false,inner ysep=2em,minimum width=2cm, align=left,rectangle split part align={left}] at (0,0) (implies)
	{A\nodepart{two}B};
	\node [draw=none, below=0.0cm of implies.south] {Implies};
	\node [draw=none, left=0.0cm of implies.east,anchor=east] {A $\Rightarrow$ B};
	
	\node[draw, left=3.25cm of implies.text west] (sigDelay) {$\text{Z}^{-1}$};
	\node[draw=none,below=0cm of sigDelay.south,anchor=north] {Signal Delay};
	
	\node[draw,rectangle, left=0.5cm of implies.text west, minimum width=2cm, minimum height=1cm] (detectorBlock) {};
	\node[draw=none,below=0cm of detectorBlock.south,anchor=north] {Detector};
	\node[draw=none,right=0cm of detectorBlock.west,anchor=west] (detectorBlockIn) {\footnotesize In};
	\node[draw=none,left=0cm of detectorBlock.east,anchor=east] (detectorBlockOut) {\footnotesize Out};
	
	\coordinate (a) at ($ (detectorBlockIn.north east)!0.1!(detectorBlockIn.south east) $);
	\coordinate (aLow) at ($ (detectorBlockIn.north east)!0.9!(detectorBlockIn.south east) $);
	\coordinate (b) at ($ (detectorBlockOut.north west)!0.1!(detectorBlockOut.south west) $);
	\coordinate (xDiff) at ($ (a)!0.0833!(b) - (a) $);
	\coordinate (yDiff) at ($ (detectorBlockIn.north east) - (detectorBlockIn.south east) $);
	
	\draw[dash pattern=on 3pt off 1pt] ($(a) + (xDiff)$) -- ($(a) + 2*(xDiff)$) -- ($(a) + 2*(xDiff) + 0.3*(yDiff) $)  -- ($(a) + 8*(xDiff) + 0.3*(yDiff) $)  -- ($(a) + 8*(xDiff) $) -- ($(a) + 11*(xDiff) $);
	\draw ($(aLow) + (xDiff)$) -- ($(aLow) + 4*(xDiff)$) -- ($(aLow) + 4*(xDiff) + 0.3*(yDiff) $)  -- ($(aLow) + 10*(xDiff) + 0.3*(yDiff) $)  -- ($(aLow) + 10*(xDiff) $) -- ($(aLow) + 11*(xDiff) $);
	
	\coordinate[left=5.5cm of implies.text west] (coordConstant);
	\node[skConstant,draw,cstval=0] at (coordConstant) (constant) {};
	\node[draw=none,below=0.4cm of constant.center,anchor=north] {Constant};
	
	\coordinate[left=5.5cm of implies.two west] (coordInput);
	\node [skInport,draw] at (coordInput) (input) {};
	\node[draw=none,below=0.2cm of input.center,anchor=north] {Pattern};
	
	\node [verificationObj, right=1cm of implies.east] (verificationObj) {\textbf{P}};
	
	\draw[link] (constant.outA) -- (sigDelay.west);
	\draw[link] (sigDelay.east) -- (detectorBlock.west);
	\draw[link] (detectorBlock.east) -- (implies.text west);
	\draw[link] (input.portA) -- (implies.two west);
	
	\draw[toVerificationObj] (implies.east) -- (verificationObj.west);
	\end{tikzpicture}}
	\caption{Scope \texttt{initially}} \label{fig:scopeInitally}

%% file: images/figureGlobally.tex
	\centering
	\begin{tikzpicture}[font=\sffamily]
	\pgfmathsetmacro\widthOfFlag{5.5cm}
	
	\node [skInport,draw] at (0,0) (input1) {};
	\node [verificationObj, right=1cm of input1.portA] (verificationObj1) {\textbf{P}};
	
	\draw[toVerificationObj] (input1.portA) -- (verificationObj1.west);
	\end{tikzpicture}
	\caption{Scope \texttt{globally}} \label{fig:scopeGlobally}

%% file: images/figureImplementationTriggerReponsePattern.tex
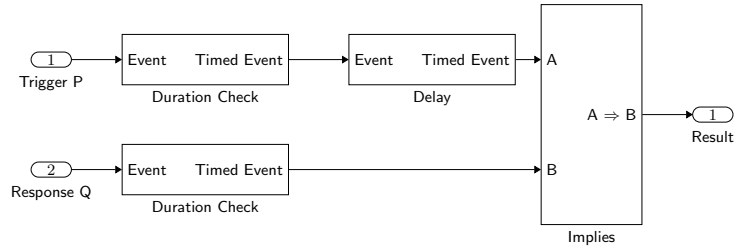
\begin{figure}
	\centering
	\resizebox{0.8\textwidth}{!}{%
		\begin{tikzpicture}[font=\sffamily]
		\pgfmathsetmacro\widthOfFlag{5.5cm}
		
		\node [rectangle split, rectangle split parts=2, draw, anchor=center, rectangle split draw splits=false,inner ysep=3em,minimum width=2cm, align=left,rectangle split part align={left}] at (0,0) (implies)
		{A\nodepart{two}B};
		\node [draw=none, below=0.0cm of implies.south] {Implies};
		\node [draw=none, left=0.0cm of implies.east,anchor=east] {A $\Rightarrow$ B};
		
		\node [rectangle split, rectangle split horizontal, rectangle split parts=2, draw, anchor=center, rectangle split draw splits=false,minimum height=1cm,rectangle split part align={left, right}, left=0.5cm of implies.text west] (delay)
		{Event\nodepart{two}\quad Timed Event};
		\node [draw=none, below=0.0cm of delay.south] {Delay};
		
		\node [rectangle split, rectangle split horizontal, rectangle split parts=2, draw, anchor=center, rectangle split draw splits=false,minimum height=1cm,rectangle split part align={left, right}, left=5cm of implies.text west] (durationT)
		{Event\nodepart{two}\quad Timed Event};
		\node [draw=none, below=0.0cm of durationT.south] {Duration Check};
		
		\node [rectangle split, rectangle split horizontal, rectangle split parts=2, draw, anchor=center, rectangle split draw splits=false,minimum height=1cm,rectangle split part align={left, right}, left=5cm of implies.two west] (durationQ)
		{Event\nodepart{two}\quad Timed Event};
		\node [draw=none, below=0.0cm of durationQ.south] {Duration Check};
		
		\node[skInport,draw, port number=1, left=1cm of durationT,anchor=portA] (inputT) {};
		\node [draw=none, below=0.2cm of inputT.center] {Trigger P};
		
		\node[skInport,draw, port number=2, left=1cm of durationQ,anchor=portA] (inputR) {};
		\node [draw=none, below=0.2cm of inputR.center] {Response Q};
		
		\node[skOutport,draw, port number=1, right=1cm of implies.east,anchor=portA] (output) {};
		\node [draw=none, below=0.2cm of output.center] {Result};
		
		\draw[link] (inputT.portA) -- (durationT.west);
		\draw[link] (inputR.portA) -- (durationQ.west);
		
		\draw[link] (durationT.east) -- (delay.west);
		
		\draw[link] (delay.east) -- (implies.text west);
		\draw[link] (durationQ.east) -- (implies.two west);
		
		\draw[link] (implies.east) -- (output.portA);
		\end{tikzpicture}}
	\caption{The implementation of the Trigger/Response pattern.} \label{fig:triggerResponsePattern}
\end{figure}

%% file: images/figureTriggerImplementation.tex
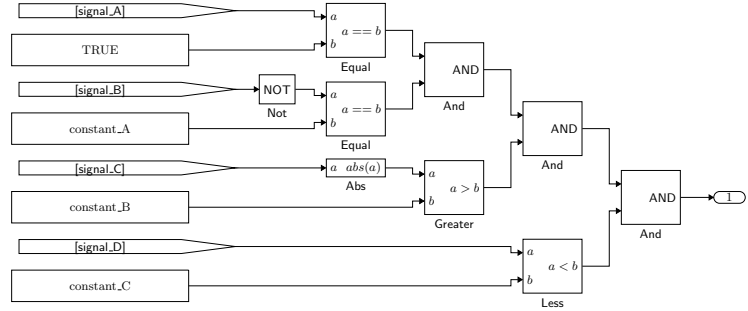
\begin{figure}
	\centering
	\resizebox{0.8\textwidth}{!}{%
	\begin{tikzpicture}[font=\sffamily]%
	\pgfmathsetmacro\widthOfFlag{5.5cm}
	\pgfmathsetmacro\widthOfConstant{4.5cm}
	\pgfmathsetmacro\widthOfBlock{1.5cm}
	\pgfmathsetmacro\ysepBlock{0.8em}
	
	\node[skFlag,draw,flag label=,block width=\widthOfFlag,anchor=center] at (0,0) (flagInput1) {[signal\_A]};
	\node[skConstant,draw,cstval=TRUE,block width=\widthOfConstant] at (0,-1) (constantInput1) {};
	\node[skFlag,draw,flag label=,block width=\widthOfFlag,anchor=center] at (0,-2) (flagInput2) {[signal\_B]};
	\node[skConstant,draw,cstval=constant\_A,block width=\widthOfConstant] at (0,-3) (constantInput2) {};
	\node[skFlag,draw,flag label=,block width=\widthOfFlag,anchor=center] at (0,-4) (flagInput3) {[signal\_C]};
	\node[skConstant,draw,cstval=constant\_B,block width=\widthOfConstant] at (0,-5) (constantInput3) {};
	\node[skFlag,draw,flag label=,block width=\widthOfFlag,anchor=center] at (0,-6) (flagInput4) {[signal\_D]};
	\node[skConstant,draw,cstval=constant\_C,block width=\widthOfConstant] at (0,-7) (constantInput4) {};
	
	\coordinate (lvl0Blk0Center) at ($ (flagInput1)!0.5!(constantInput1) +(6.5,0) $);
	\node [rectangle split, rectangle split parts=2, draw, anchor=center, rectangle split draw splits=false,inner ysep=\ysepBlock,minimum width=\widthOfBlock, align=left,rectangle split part align={left}] at (lvl0Blk0Center) (lvl0Blk0)
	{\nodepart{two}};
	\node [draw=none, below=0.0cm of lvl0Blk0.south] {Equal};
	\node [draw=none,anchor=east] at (lvl0Blk0.east) {$a==b$};
	\node [draw=none,anchor=west] at (lvl0Blk0.text west) {$a$};
	\node [draw=none,anchor=west] at (lvl0Blk0.two west) {$b$};
	
	\coordinate (lvl0BlkNotCenter) at ($ (flagInput2) + (4.5,0) $);
	\node [draw, anchor=center,inner ysep=\ysepBlock, align=left] at (lvl0BlkNotCenter) (lvl0BlkNot) {NOT};
	\node [draw=none, below=0.0cm of lvl0BlkNot.south] {Not};
	
	\coordinate (lvl0Blk1Center) at ($ (flagInput2)!0.5!(constantInput2) +(6.5,0) $);
	\node [rectangle split, rectangle split parts=2, draw, anchor=center, rectangle split draw splits=false,inner ysep=\ysepBlock,minimum width=\widthOfBlock, align=left,rectangle split part align={left}] at (lvl0Blk1Center) (lvl0Blk1)
	{\nodepart{two}};
	\node [draw=none, below=0.0cm of lvl0Blk1.south] {Equal};
	\node [draw=none,anchor=east] at (lvl0Blk1.east) {$a==b$};
	\node [draw=none,anchor=west] at (lvl0Blk1.text west) {$a$};
	\node [draw=none,anchor=west] at (lvl0Blk1.two west) {$b$};
	
	\coordinate (lvl0Blk2Center) at ($ (flagInput3) + (6.5,0) $);
	\node [draw, anchor=center,inner ysep=0.7em,minimum width=\widthOfBlock, align=left] at (lvl0Blk2Center) (lvl0Blk2)
	{\nodepart{two}};
	\node [draw=none, below=0.0cm of lvl0Blk2.south] {Abs};
	\node [draw=none,anchor=east] at (lvl0Blk2.east) {$abs(a)$};
	\node [draw=none,anchor=west] at (lvl0Blk2.west) {$a$};
	
	\coordinate (lvl1Blk0Center) at ($ (lvl0Blk0Center)!0.5!(lvl0Blk1Center) +(2.5,0) $);
	\node [rectangle split, rectangle split parts=2, draw, anchor=center, rectangle split draw splits=false,inner ysep=\ysepBlock,minimum width=\widthOfBlock, align=left,rectangle split part align={left}] at (lvl1Blk0Center) (lvl1Blk0)
	{\nodepart{two}};
	\node [draw=none, below=0.0cm of lvl1Blk0.south] {And};
	\node [draw=none,anchor=east] at (lvl1Blk0.east) {AND};
	
	\coordinate (lvl1Blk1Center) at ($ (flagInput3)!0.5!(constantInput3) +(9.0,0) $);
	\node [rectangle split, rectangle split parts=2, draw, anchor=center, rectangle split draw splits=false,inner ysep=\ysepBlock,minimum width=\widthOfBlock, align=left,rectangle split part align={left}] at (lvl1Blk1Center) (lvl1Blk1)
	{\nodepart{two}};
	\node [draw=none, below=0.0cm of lvl1Blk1.south] {Greater};
	\node [draw=none,anchor=east] at (lvl1Blk1.east) {$a>b$};
	\node [draw=none,anchor=west] at (lvl1Blk1.text west) {$a$};
	\node [draw=none,anchor=west] at (lvl1Blk1.two west) {$b$};
	
	\coordinate (lvl2Blk0Center) at ($ (lvl1Blk0Center)!0.5!(lvl1Blk1Center) + (2.5,0) $);
	\node [rectangle split, rectangle split parts=2, draw, anchor=center, rectangle split draw splits=false,inner ysep=\ysepBlock,minimum width=\widthOfBlock, align=left,rectangle split part align={left}] at (lvl2Blk0Center) (lvl2Blk0)
	{\nodepart{two}};
	\node [draw=none, below=0.0cm of lvl2Blk0.south] {And};
	\node [draw=none,anchor=east] at (lvl2Blk0.east) {AND};
	
	\coordinate (lvl2Blk1Center) at ($ (flagInput4)!0.5!(constantInput4) + (11.5,0) $);
	\node [rectangle split, rectangle split parts=2, draw, anchor=center, rectangle split draw splits=false,inner ysep=\ysepBlock,minimum width=\widthOfBlock, align=left,rectangle split part align={left}] at (lvl2Blk1Center) (lvl2Blk1)
	{\nodepart{two}};
	\node [draw=none, below=0.0cm of lvl2Blk1.south] {Less};
	\node [draw=none,anchor=east] at (lvl2Blk1.east) {$a<b$};
	\node [draw=none,anchor=west] at (lvl2Blk1.text west) {$a$};
	\node [draw=none,anchor=west] at (lvl2Blk1.two west) {$b$};
	
	\coordinate (lvl3Blk0Center) at ($ (lvl2Blk0Center)!0.5!(lvl2Blk1Center) + (2.5,0) $);
	\node [rectangle split, rectangle split parts=2, draw, anchor=center, rectangle split draw splits=false,inner ysep=\ysepBlock,minimum width=\widthOfBlock, align=left,rectangle split part align={left}] at (lvl3Blk0Center) (lvl3Blk0)
	{\nodepart{two}};
	\node [draw=none, below=0.0cm of lvl3Blk0.south] {And};
	\node [draw=none,anchor=east] at (lvl3Blk0.east) {AND};
	
	\coordinate (lvl0Blk0In1) at ($ (lvl0Blk0.text west) + (-0.3,0) $);
	\coordinate (lvl0Blk0In2) at ($ (lvl0Blk0.two west) + (-0.3,0) $);
	\draw [link] (flagInput1.portA) -| (lvl0Blk0In1) -- (lvl0Blk0.text west);
	\draw [link] (constantInput1.portA) -| (lvl0Blk0In2) -- (lvl0Blk0.two west);
	
	\draw [link] (flagInput2.portA) -- (lvl0BlkNot.west);
	
	\coordinate (lvl0Blk1In1) at ($ (lvl0Blk1.text west) + (-0.3,0) $);
	\coordinate (lvl0Blk1In2) at ($ (lvl0Blk1.two west) + (-0.3,0) $);
	\draw [link] (lvl0BlkNot.east) -| (lvl0Blk1In1) -- (lvl0Blk1.text west);
	\draw [link] (constantInput2.portA) -| (lvl0Blk1In2) -- (lvl0Blk1.two west);
	
	\draw [link] (flagInput3.portA) -- (lvl0Blk2.west);
	
	\coordinate (lvl1Blk0In1) at ($ (lvl1Blk0.text west) + (-0.3,0) $);
	\coordinate (lvl1Blk0In2) at ($ (lvl1Blk0.two west) + (-0.3,0) $);
	\draw [link] (lvl0Blk0.east) -| (lvl1Blk0In1) -- (lvl1Blk0.text west);
	\draw [link] (lvl0Blk1.east) -| (lvl1Blk0In2) -- (lvl1Blk0.two west);
	
	\coordinate (lvl1Blk1In1) at ($ (lvl1Blk1.text west) + (-0.3,0) $);
	\coordinate (lvl1Blk1In2) at ($ (lvl1Blk1.two west) + (-0.3,0) $);
	\draw [link] (lvl0Blk2.east) -| (lvl1Blk1In1) -- (lvl1Blk1.text west);
	\draw [link] (constantInput3.portA) -| (lvl1Blk1In2) -- (lvl1Blk1.two west);
	
	\coordinate (lvl2Blk0In1) at ($ (lvl2Blk0.text west) + (-0.3,0) $);
	\coordinate (lvl2Blk0In2) at ($ (lvl2Blk0.two west) + (-0.3,0) $);
	\draw [link] (lvl1Blk0.east) -| (lvl2Blk0In1) -- (lvl2Blk0.text west);
	\draw [link] (lvl1Blk1.east) -| (lvl2Blk0In2) -- (lvl2Blk0.two west);
	
	\coordinate (lvl2Blk1In1) at ($ (lvl2Blk1.text west) + (-0.3,0) $);
	\coordinate (lvl2Blk1In2) at ($ (lvl2Blk1.two west) + (-0.3,0) $);
	\draw [link] (flagInput4.portA) -| (lvl2Blk1In1) -- (lvl2Blk1.text west);
	\draw [link] (constantInput4.portA) -| (lvl2Blk1In2) -- (lvl2Blk1.two west);
	
	\coordinate (lvl3Blk0In1) at ($ (lvl3Blk0.text west) + (-0.3,0) $);
	\coordinate (lvl3Blk0In2) at ($ (lvl3Blk0.two west) + (-0.3,0) $);
	\draw [link] (lvl2Blk0.east) -| (lvl3Blk0In1) -- (lvl3Blk0.text west);
	\draw [link] (lvl2Blk1.east) -| (lvl3Blk0In2) -- (lvl3Blk0.two west);
	
	\node [skOutport,draw,right of=lvl3Blk0Center,xshift=1cm,port number=1] (output) {};
	\draw [link] (lvl3Blk0.east) -- (output.portA);
	
	\end{tikzpicture}}
	\caption{The logic of the trigger condition of the example requirement.} \label{fig:triggerImplementation}
\end{figure}

%% file: images/figureTestObjectives.tex
\centering
\begin{tikzpicture}[font=\sffamily]
\pgfmathsetmacro\widthOfFlag{5.5cm}
\node [rectangle split, rectangle split parts=2, draw, anchor=center, rectangle split draw splits=false,inner ysep=2em,minimum width=1.5cm, align=left,rectangle split part align={left}] at (0,0) (orBlock)
{\nodepart{two}};
\coordinate (orCenter) at (orBlock.center);
\node [draw=none, below=0.0cm of orBlock.south] {or};
\node [draw=none] at (orCenter) {OR};

\coordinate[left=1cm of orBlock.text west] (orWest1);
\coordinate[left=1cm of orBlock.two west] (orWest2);
\coordinate[right=1cm of orBlock.east] (orEast);

\node [testObj] at (orWest1) (testObj1) {\textbf{O}};
\node [testObjLabel, above=0.0cm of testObj1.north] {\{true, false\}};
\node [skInport,draw,left of=orWest1,xshift=-0.5cm] (inputOr1) {};

\draw [toTestObj] (inputOr1.portA) -- (testObj1.west);
\draw [fromTestObj] (testObj1.east) -- (orBlock.text west);

\node [testObj] at (orWest2) (testObj2) {\textbf{O}};
\node [testObjLabel, above=0.0cm of testObj2.north] {\{true, false\}};
\node [skInport,draw,left of=orWest2,xshift=-0.5cm,port number=2] (inputOr2) {};

\draw [toTestObj] (inputOr2.portA) -- (testObj2.west);
\draw [fromTestObj] (testObj2.east) -- (orBlock.two west);

\node [testObj] at (orEast) (testObj3) {\textbf{O}};
\node [testObjLabel, above=0.0cm of testObj3.north] {\{true, false\}};
\node [skOutport,draw,right of=orEast,xshift=0.5cm,port number=1] (outputOr) {};

\draw [toTestObj] (orBlock.east) -- (testObj3.west);
\draw [fromTestObj] (testObj3.east) -- (outputOr.portA);

\end{tikzpicture}
\caption{A logic {\sffamily OR} block with test objectives attached.} \label{fig:orBlockTestObj}

%% file: chapters/conclusion.tex
In this paper we presented a prototypical pattern-based specification tool together with automated translations to \sldv and \btcembedded together with an adaption of the \sldv input for automated test-case generation.
This corresponds to the vision of enabling engineers to specify requirements with formal semantics \emph{once} and then applying the requirements in \emph{multiple analyses}.
The tool was designed as a prototype for use inside this research project as a proof of concept.

Although a big step towards a highly automated automotive verification process has been made within this project and 
investigations by Ford have been producing encouraging results,
this is only a proof-of-concept and many open problems still need to be resolved.

As future work we plan the extension of our pattern set with a few
further relevant elements like time- and event-bounded response
patterns. 
We plan to tackle the automated translation of textual legacy requirements into formal notation.
Scripts are needed to further automate verification at different
development levels with suitable configuration parameters, and to
trigger the verification process if changes are applied to the model
or the requirements.  Another module should monitor the verification
results and automatically report conspicuous behavior if the
comparison with previous results reveals deviations.
In case of invalid verification results, counterexamples
should be analyzed.  

We plan to use the export of formalized requirements to SV-COMP like C-code patterns in order to benchmark academic C-code model checkers on industrial examples against commercial tools.

%% file: chapters/appendix/eventGrammar.tex
\section{Appendix}
\label{sec:eventGrammar}

\subsubsection{Syntax}
In the following we present the formal grammar for the specification of events.

	\begin{verbatim}
	grammar EventGrammar;

	////////////////////////////// 
	//GRAMMAR RULES (Parser)    //
	////////////////////////////// 
	////////////////////////////// 
	// EVENT (Start rule)       //
	////////////////////////////// 
	event : expr=booleanExpression EOF;
	
	////////////////////////////// 
	// BOOLEAN EXPRESSIONS		//
	////////////////////////////// 
	booleanExpression : booleanTerm (booleanImplication booleanTerm)*;
	booleanTerm : booleanFactor (booleanAddition booleanFactor)*;	
	booleanFactor : booleanEqualityTerm (booleanMultiplication booleanEqualityTerm)*;
	booleanEqualityTerm : booleanBaseTerm (booleanEquality booleanBaseTerm)*;
	booleanBaseTerm
	  : booleanAtom	
	  | booleanNotExpression
	  | booleanParExpression	
	  | relationalExpression
	  | booleanFunctionExpression
	booleanNotExpression : booleanNot booleanBaseTerm;
	booleanParExpression : leftPar booleanExpression rightPar;
	relationalExpression : arithmeticExpression relationalOperator arithmeticExpression;
	booleanFunctionExpression : booleanBinaryFunctionExpression;
	booleanBinaryFunctionExpression
	  : BIT_EXTRACTION_TEXTUAL arithmeticExpression OF arithmeticExpression
	  | booleanBinaryFunction leftPar arithmeticExpression 
	     comma arithmeticExpression rightPar;
	
	////////////////////////////// 
	// ARITHMETIC EXPRESSIONS  	//
	////////////////////////////// 
	arithmeticExpression : arithmeticFactor (arithmeticAddition arithmeticFactor)*;
	arithmeticFactor : arithmeticBaseTerm (arithmeticMultiplication arithmeticBaseTerm)*;
	arithmeticBaseTerm 
	  : arithmeticAtom 				
	  | arithmeticParExpression 		
	  | arithmeticSignExpression
	  | arithmeticFunction			
	arithmeticParExpression  : leftPar arithmeticExpression rightPar;
	arithmeticSignExpression : arithmeticSign arithmeticBaseTerm;
	//Functions
	arithmeticFunction 
	  : arithmeticUnaryFunctionExpression
	  | arithmeticBinaryFunctionExpression
	  | temporalUnaryFunctionExpression
	  | temporalBinaryFunctionExpression;
	arithmeticUnaryFunctionExpression
	  : arithmeticUnaryFunction leftPar arithmeticExpression rightPar
	  | arithmeticUnaryFunctionTextual arithmeticExpression;
	arithmeticBinaryFunctionExpression
	  : arithmeticBinaryFunction leftPar arithmeticExpression
	     comma arithmeticExpression rightPar
	  | arithemticBinaryFunctionTextual arithmeticExpression
	     AND arithmeticExpression;
	temporalUnaryFunctionExpression
	  : temporalUnaryFunction leftPar arithmeticExpression rightPar
	  | temporalUnaryFunctionTextual arithmeticExpression;
	temporalBinaryFunctionExpression
	  : temporalBinaryFunction leftPar arithmeticExpression 
	     comma arithmeticBaseTerm rightPar
	  | LAST_BINARY_TEXTUAL_0 arithmeticExpression 
	     arithmeticBaseTerm LAST_BINARY_TEXTUAL_1;
	
	///////////////
	//Operators  //
	///////////////
	booleanNot : NOT;   
	booleanAddition : OR;
	booleanMultiplication : AND;
	booleanImplication : IMPLICATION;
	booleanEquality : EQ;	
	relationalOperator
	  : EQ 
	  | GT   
	  | GTEQ    
	  | LT   
	  | LTEQ;	
	arithmeticAddition
	  : PLUS 
	  | MINUS;
	arithmeticMultiplication
	  : TIMES
	  | DIV;
	arithmeticSign
	  : PLUS
	  | MINUS;
	
	///////////////
	//Functions  //
	///////////////
	booleanBinaryFunction : BIT_EXTRACTION_FUNCTION;	
	arithmeticUnaryFunction : ABS_FUNCTION;
	arithmeticUnaryFunctionTextual : ABS_TEXTUAL;
	arithmeticBinaryFunction
	  : MIN_FUNCTION
	  | MAX_FUNCTION;
	arithemticBinaryFunctionTextual
	  : MIN_TEXTUAL
	  | MAX_TEXTUAL;
	
	//temporal
	temporalUnaryFunction : LAST_UNARY_FUNCTION;
	temporalUnaryFunctionTextual : LAST_UNARY_TEXTUAL;
	temporalBinaryFunction : LAST_BINARY_FUNCTION;
	
	/////////////////
	//Parentheses  //
	/////////////////
	leftPar : LPAR;
	rightPar : RPAR;
	
	/////////////////
	//Other        //
	/////////////////
	comma : COMMA;
	
	/////////////////////////////////
	//Identifiers and Constants    //
	/////////////////////////////////
	booleanAtom
	  : booleanConstant
	  | booleanIdentifier;
	arithmeticAtom 
	  : arithmeticConstant
	  | arithmeticIdentifier;	
	booleanConstant
	  : TRUE
	  | FALSE; 
	arithmeticConstant
	  : INTEGER 
	  | FLOATINGPOINT;
	booleanIdentifier : IDENTIFIER;
	arithmeticIdentifier : IDENTIFIER;
		
	////////////////////////////// 
	//TOKENS (Lexer rules)      //
	////////////////////////////// 
	/////////////////
	//Constants    //
	/////////////////
	TRUE
	  : 'TRUE'
	  | 'True'
	  | 'true';
	FALSE
	  : 'FALSE'
	  | 'False'
	  | 'false';
	  
	/////////////////
	//Operators    //
	/////////////////
	//boolean
	OR
	  : '|'
	  | 'or';
	AND
	  : '&'
	  | 'and';
	NOT
	  : '!'
	  | 'not';   
	IMPLICATION
	  : '=>'
	  | 'implies';   
	//arithmetic
	PLUS
	  : '+'
	  | 'plus';
	MINUS
	  : '-'
	  | 'minus';
	TIMES
	  : '*'
	  | 'multiplied with';
	DIV
	  : '/'
	  | 'divided by';
	//relational	
	GT
	  : '>'
	  | 'is greater than';
	GTEQ
	  : '>='
	  | 'is greater or equal to';
	LT
	  : '<'
	  | 'is less than';
	LTEQ
	  : '<='
	  | 'is less or equal to';
	EQ
	  : '='
	  | 'is equal to';
	
	/////////////////
	//Parentheses  //
	/////////////////
	LPAR
	  : '('
	  | 'left parenthesis';
	RPAR
	  : ')'
	  | 'right parenthesis';
	
	/////////////////
	//Other        //
	/////////////////
	OF : 'of';
	COMMA : ',';
	
	/////////////////
	//Functions    //
	/////////////////
	BIT_EXTRACTION_TEXTUAL : 'bit';
	BIT_EXTRACTION_FUNCTION : 'extractBit';
	ABS_FUNCTION : 'abs';
	ABS_TEXTUAL : 'the absolute value of';
	MIN_FUNCTION : 'min';
	MIN_TEXTUAL : 'the minimum of';
	MAX_FUNCTION : 'max';
	MAX_TEXTUAL : 'the maximum of';
	LAST_UNARY_FUNCTION : 'last';
	LAST_UNARY_TEXTUAL : 'the previous value of' ;
	LAST_BINARY_FUNCTION : 'last';
	LAST_BINARY_TEXTUAL_0 : 'the value of';
	LAST_BINARY_TEXTUAL_1 : 'steps ago' ;
	
	/////////////////
	//Identifiers  //
	/////////////////
	IDENTIFIER : CHAR (CHAR | SPECIAL_CHAR | DIGIT)*;
	CHAR
	 : ('a'..'z')
	 | ('A'..'Z');
	SPECIAL_CHAR : '_';
	
	/////////////////
	//Numbers      //
	/////////////////
	INTEGER : DIGIT+;
	FLOATINGPOINT : DIGIT+ '.' DIGIT+;
	DIGIT :('0'..'9');   
	
	/////////////////
	//White spaces //
	/////////////////
	WS : [ \r\t\n] + -> skip // skip spaces, tabs, newlines;
	\end{verbatim}


\subsubsection{Semantics}\label{appendix:semantics}

\noindent The semantics of $w$ for temporal operators that reach to the past is given by the following MTL formulas:\\[0.5em]
\begin{enumerate}
	\item $P$,\\[0.5em]
	if $s=\initially,\ p=\invariant \text{ and } E=(P)$,\\[1em]
	\item $\Box_{[0,\infty]}\ P$,\\[0.5em]
	if $s=\globally,\ p=\invariant \text{ and } E=(P)$,\\[1em]
	\item $\lozenge_{[t_p + t_d + t_q, t_p + t_d + t_q]} \left(\left(\blacklozenge_{[-t_d - t_q, t_d - t_q]} (P\ \ \mathcal{S}_{[t_p, t_p]}\ \texttt{true})\right) \rightarrow (Q\ \ \mathcal{S}_{[t_q, t_q]}\ \texttt{true})\right)$,\\[0.5em]
	if $s=\initially,\ p=\responseTbEb \text{ and } E=(P,Q)$\\
	(Note, that a time shift of $t_p + t_d + t_q$ is needed for this scope/pattern combination.),\\[1em]
	\item $\Box_{[t_p + t_d + t_q,\infty]}\ \left[\left(\blacklozenge_{[-t_d - t_q, -t_d - t_q]} (P\ \ \mathcal{S}_{[t_p, t_p]}\ \texttt{true})\right) \rightarrow (Q\ \ \mathcal{S}_{[t_q, t_q]}\ \texttt{true})\right]$,\\[0.5em]
	if $s=\globally,\ p=\responseTbEb \text{ and } E=(P,Q)$\\
	(Note, that a time shift of $t_p + t_d + t_q$ is needed for this scope/pattern combination.).\\[1em]
\end{enumerate}

\subsubsection{Supported Functions}\label{appendix:supportedFunctions}
\ \\\ \\
\begin{tabular}{ll}
	Operator & Meaning\\
	\hline
	\texttt{abs(param)} & Absolute value of \texttt{param}.\\
	\texttt{min(param1, param2)} & The minimum of \texttt{param1} and \texttt{param2}.\\
	\texttt{max(param1, param2)} & The maximum of \texttt{param1} and \texttt{param2}.\\
	\texttt{last(param)} & The value of \texttt{param} one step ago.\\
	\texttt{last(param, steps)} & The value of \texttt{param} \texttt{steps} step ago.\\
	\texttt{extractBit(index, param)} & The bit at position \texttt{index} of \texttt{param}.
\end{tabular}

%% file: chapters/appendix/specificationLibrary.tex
\section{Appendix}
\label{app:specificationLib}
\begin{figure}[h]
	\centering
	\includegraphics[width=0.4\textwidth]{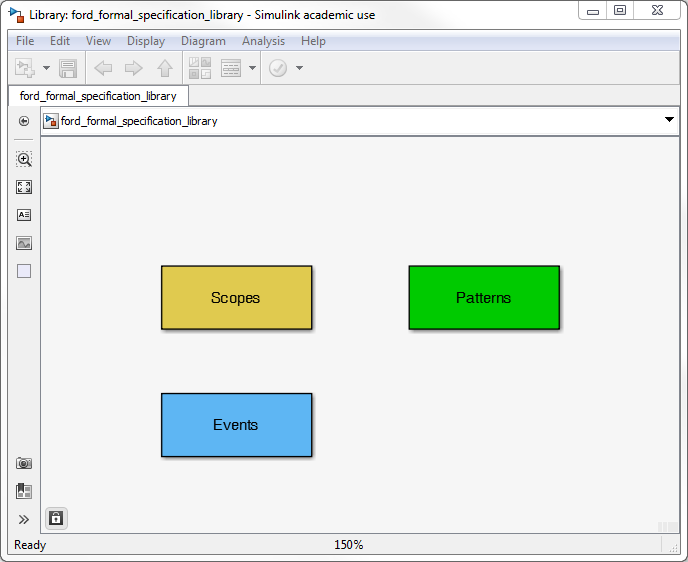}
	\caption{Simulink library with sub-libraries for the specification of scopes, patterns and events.}
	\label{fig:library}
\end{figure}

In this appendix, we present the Simulink library for the formal specification of properties for \sldv.
We defined sub-libraries for the specification of scopes, patterns, and events (see Figure \ref{fig:library}).
Additionally, the library contains a set of auxiliary scripts that can be called to create a specification. 

\subsection{Scopes}
Our specification library supports the set $\mathcal{S} = \{\initially, \globally\}$ of scopes. A scope defines the time steps in which a pattern is evaluated.
\simulinkBlock
{\initially} 
{Scope} 
{At system start, [Pattern].} 
{\includegraphics[width=0.35\textwidth]{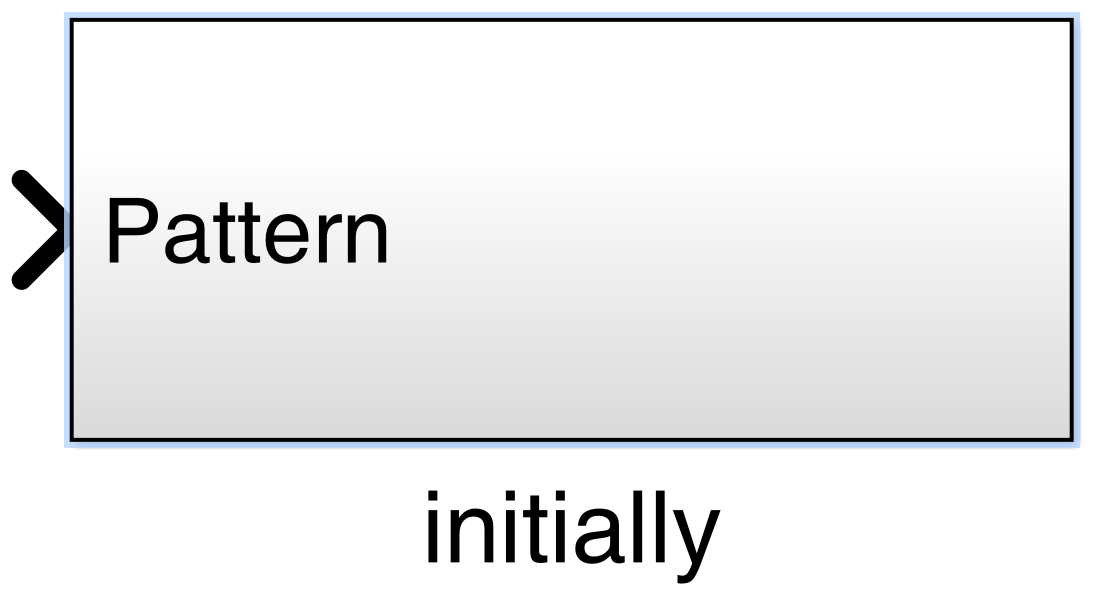}} 
{\item Input parameter \customHochkommata{Pattern}: The pattern that is evaluated at system start.} 
{\includegraphics[width=0.5\textwidth]{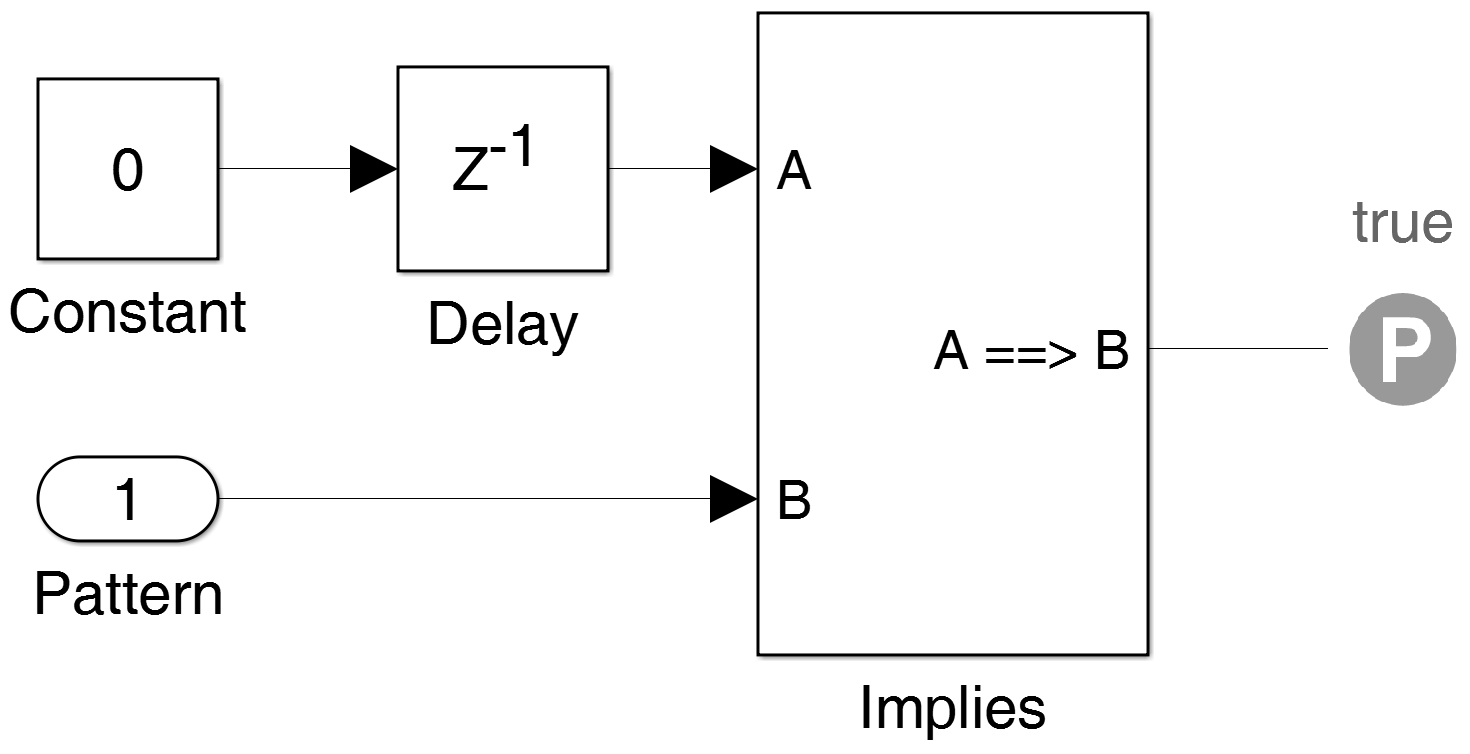}} 
{\item No scope shift is performed. For more information on the time shift, we refer to Section \ref{sec:specificationLanguage}. \item Delay mask parameter \customHochkommata{Delay length}: 1 \item Delay mask parameter \customHochkommata{Initial condition}: 1} 

\simulinkBlock
{\initially} 
{Scope} 
{At system start, [Pattern].} 
{\includegraphics[width=0.35\textwidth]{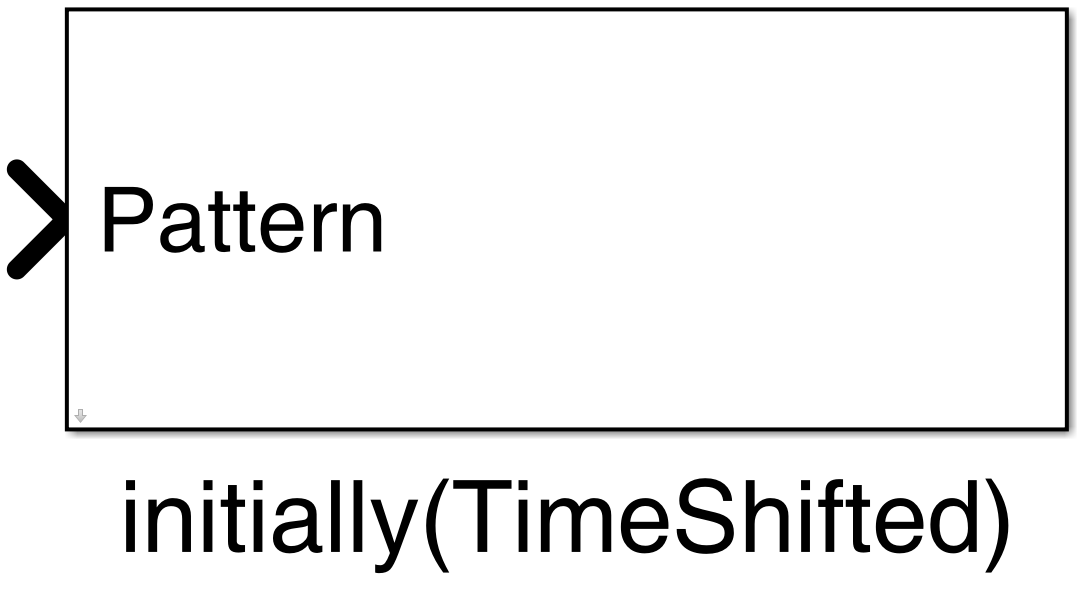}} 
{\item Input parameter \customHochkommata{Pattern}: The pattern that is evaluated at system start.} 
{\includegraphics[width=0.7\textwidth]{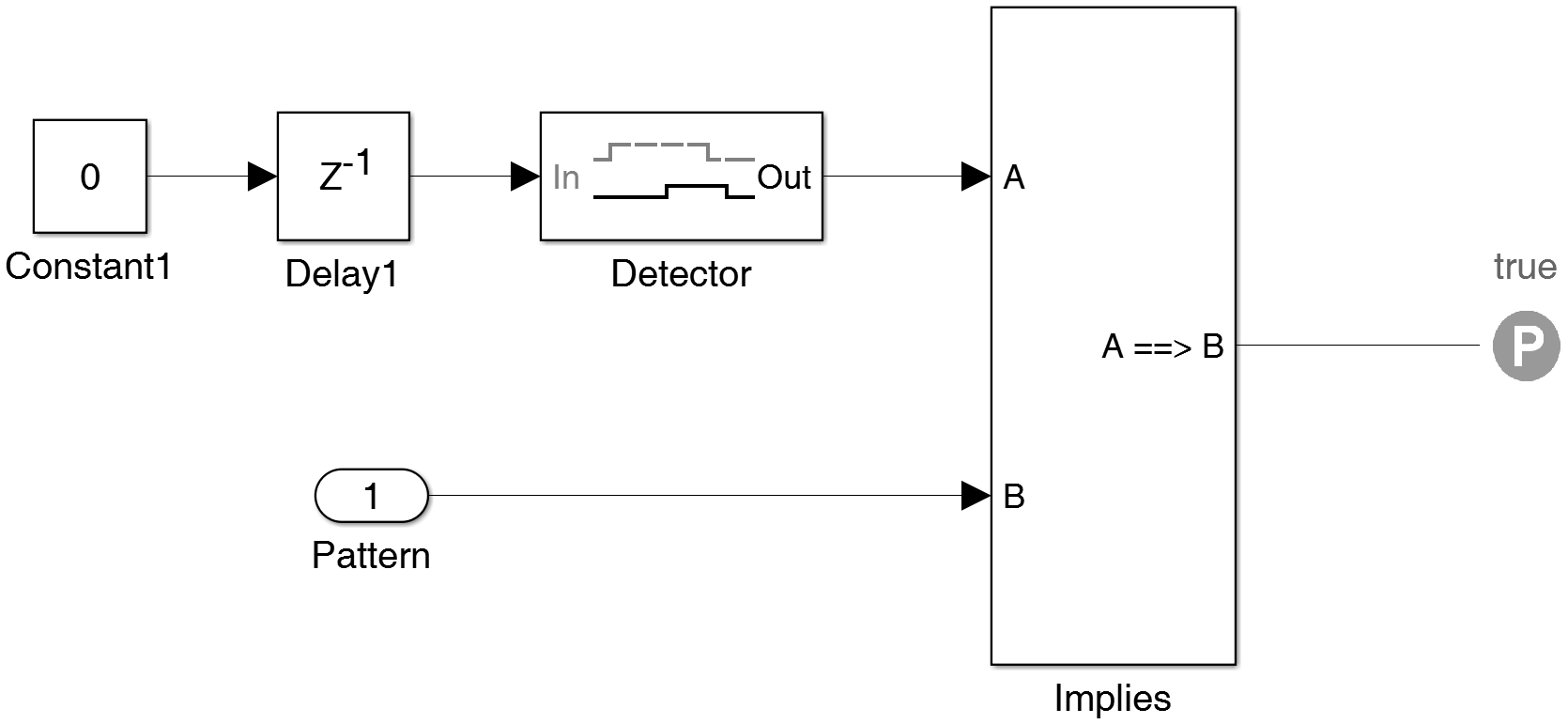}} 
{\item A scope shift is performed. For more information on the time shift, we refer to Section \ref{sec:specificationLanguage}. \item Mask parameter \customHochkommata{Time shift}: The time step to which the scope must be shifted. \item Delay mask parameter \customHochkommata{Delay length}: 1 \item Delay mask parameter \customHochkommata{Initial condition}: 1 \item Detector mask parameter \customHochkommata{Step for input detection}: 1 \item Detector mask parameter \customHochkommata{Time steps for delay}: $d-1$ \item Detector mask parameter \customHochkommata{Time steps for output duration}: 1} 

\simulinkBlock
{\globally} 
{Scope} 
{At each time step, [Pattern].} 
{\includegraphics[width=0.35\textwidth]{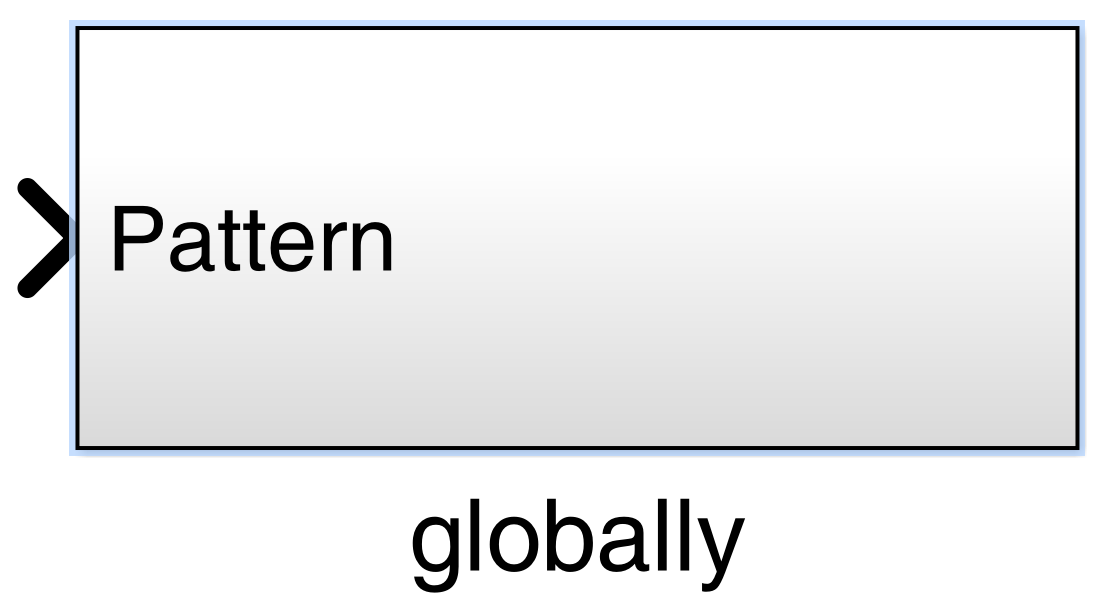}} 
{\item Input parameter \customHochkommata{Pattern}: The pattern that is evaluated at each time step.} 
{\includegraphics[width=0.5\textwidth]{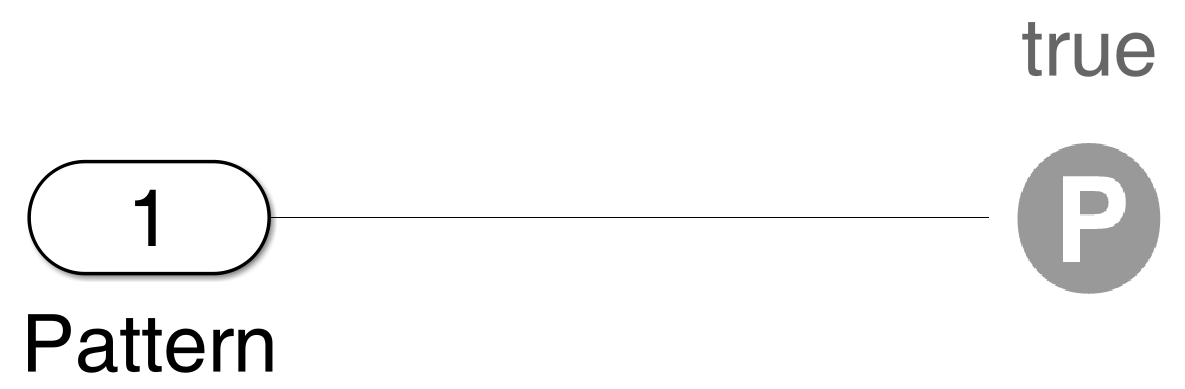}} 
{\item No scope shift is performed. For more information on the time shift, we refer to Section \ref{sec:specificationLanguage}. } 


\newpage
\subsection{Patterns}
The Simulink library for formal specification supports the set $\mathcal{P} = \{\invariant, \responseTbEb\}$ of patterns and offers Simulink blocks for durations and delays. A pattern is a template for the definition of properties. A set of events with time durations and delays between events must be specified.  
\simulinkBlock
	{\invariant} 
	{Pattern} 
	{[P] holds.} 
	{\includegraphics[width=0.45\textwidth]{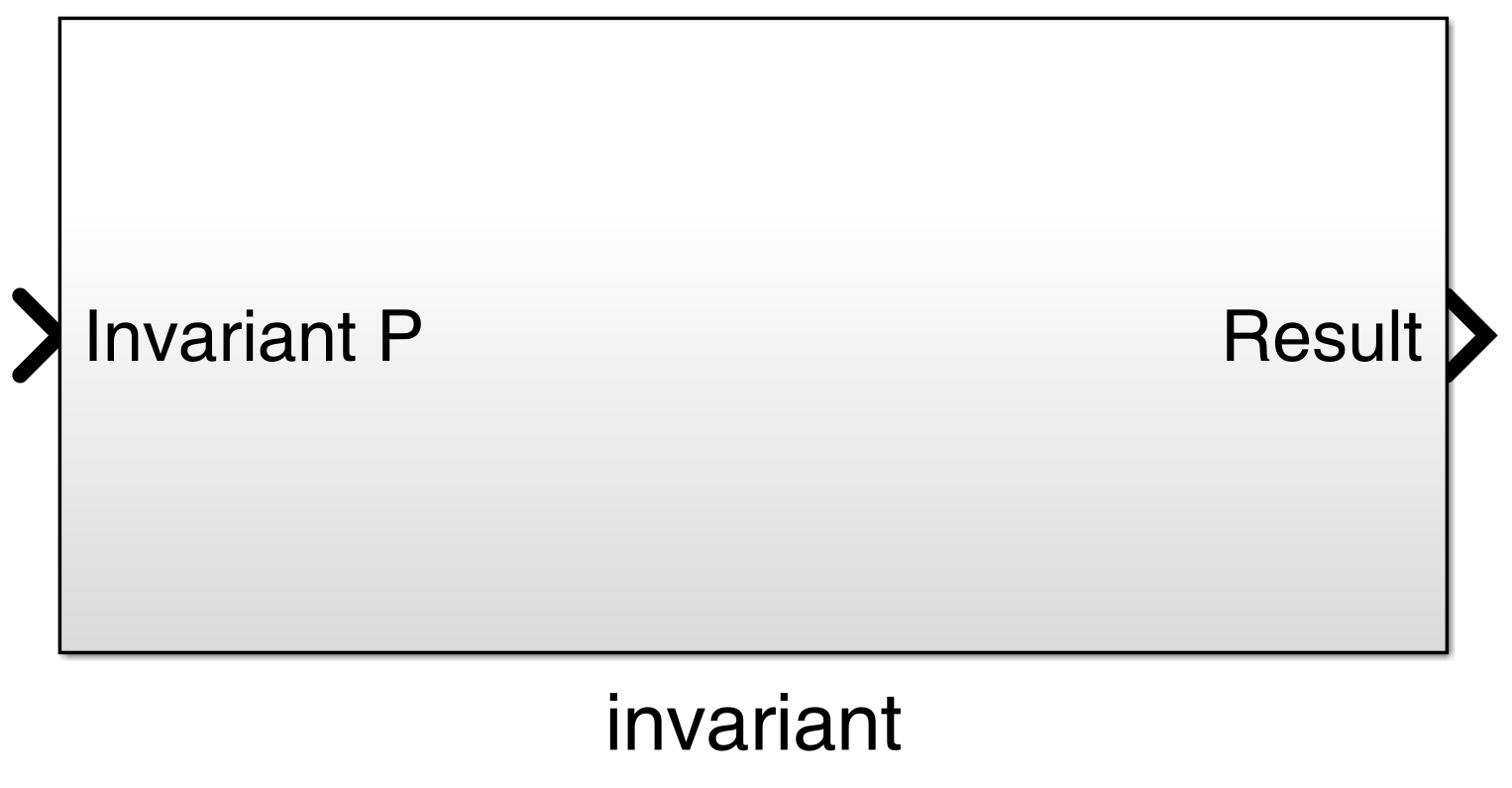}} 
	{\item Input parameter \customHochkommata{P}: The invariant event. \item Output parameter \customHochkommata{Result}: The invariant pattern result (Boolean).} 
	{\includegraphics[width=0.75\textwidth]{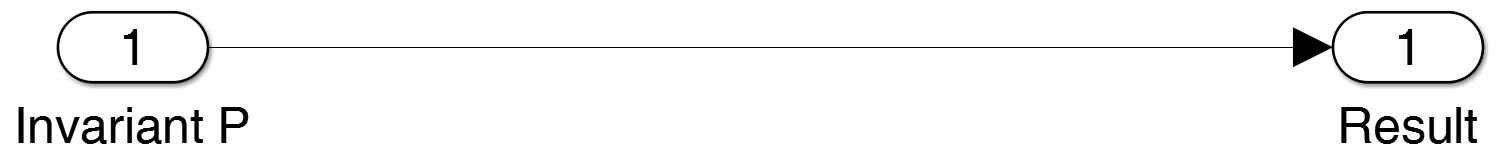}} 
	{\item --} 
	
\simulinkBlock
	{\responseTbEb} 
	{Pattern} 
	{If [P] has been valid for [Trigger duration], then in response, after a delay of [Delay] , [Q] is valid for [Response duration].} 
	{\includegraphics[width=0.35\textwidth]{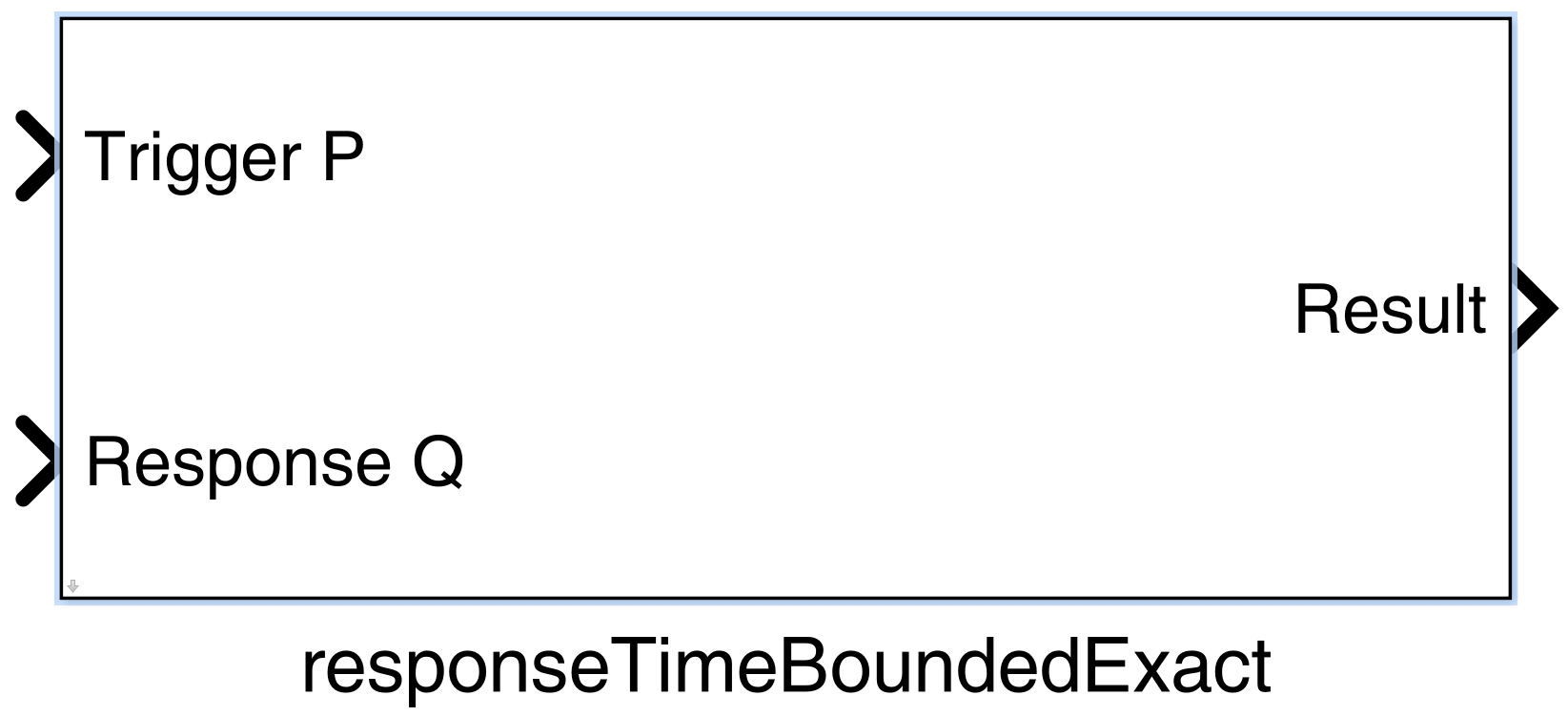}} 
	{\item Input parameter \customHochkommata{Trigger P}/\customHochkommata{Response Q}: The trigger/response event. \item Output parameter \customHochkommata{Result}: The response pattern result (Boolean). \item Mask parameter \customHochkommata{Delay}: The delay between the trigger and the response event. \item Mask parameter: \customHochkommata{Trigger duration}/\customHochkommata{Response duration}: The duration of the trigger/response event.} 
	{\includegraphics[width=0.75\textwidth]{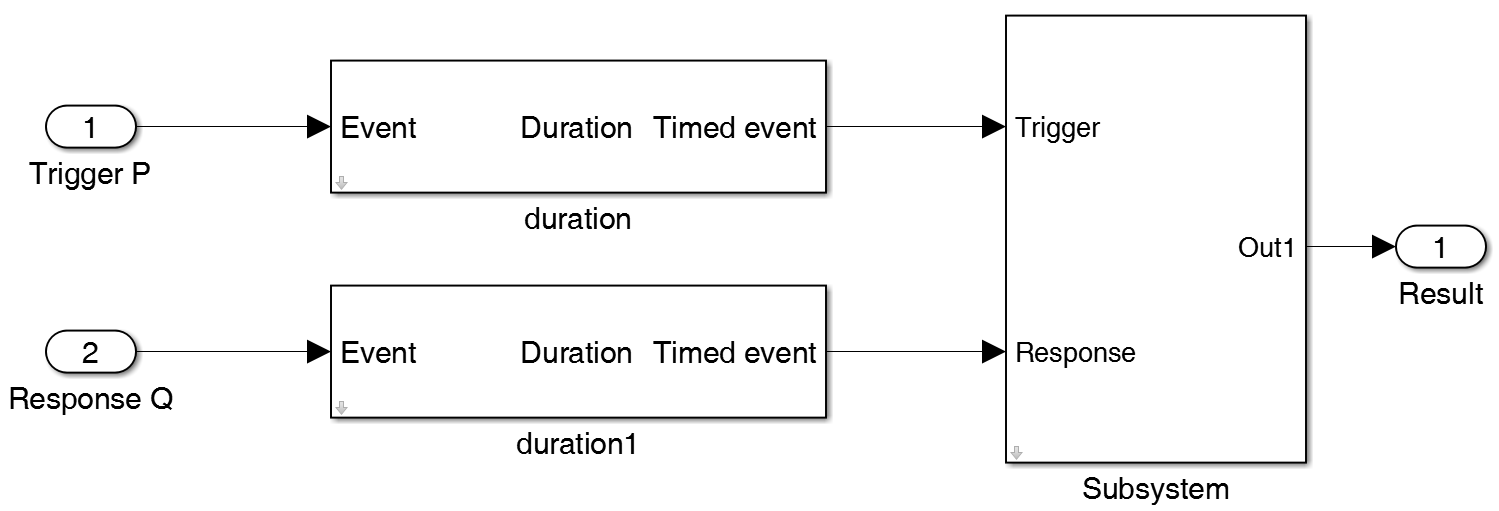}\\[2em]Subsystem implementation:\\\includegraphics[width=0.75\textwidth]{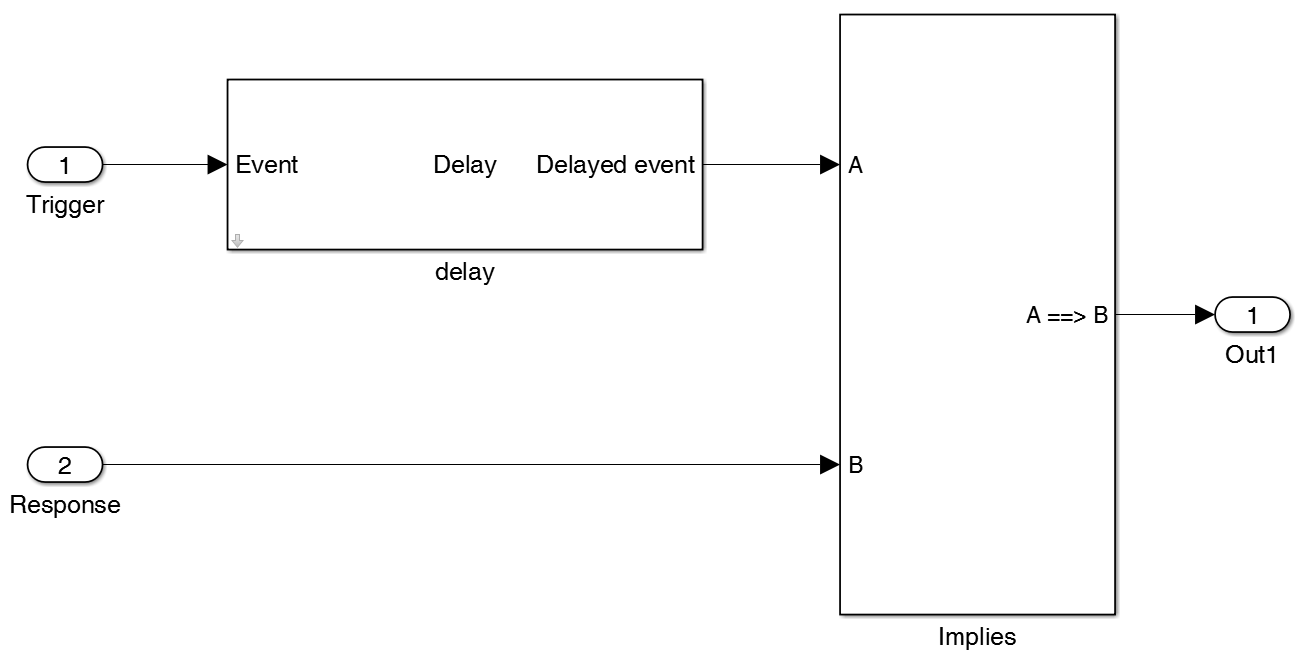}} 
	{\item Duration/Duration1 mask parameter: Trigger/Response duration \item Delay mask parameter: Delay + Response duration} 

\simulinkBlock
	{\texttt{duration}} 
	{Pattern} 
	{[Event is valid for] Time duration simulation steps.} 
	{\includegraphics[width=0.4\textwidth]{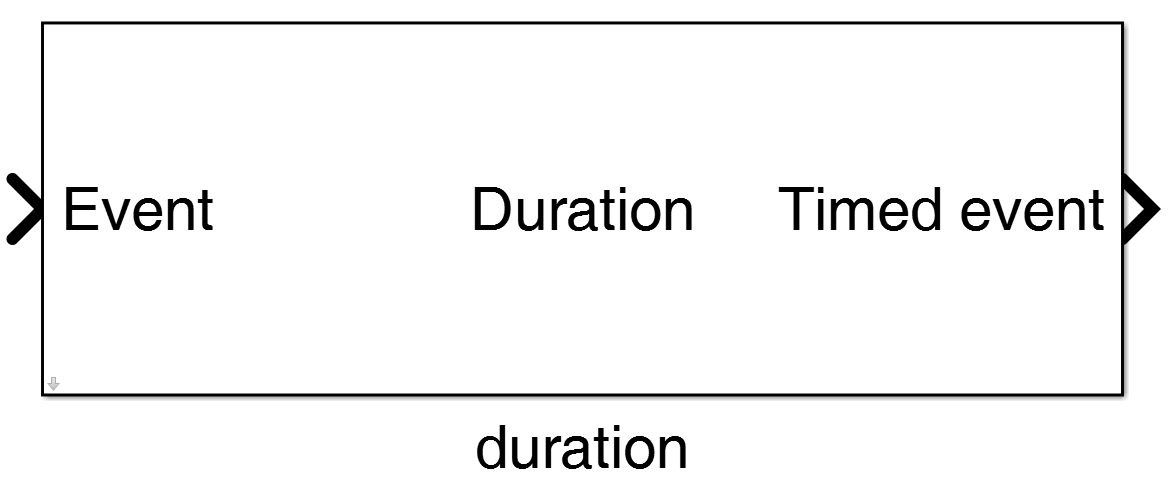}} 
	{\item Input parameter \customHochkommata{Event}: The event for which a duration is specified. \item Output parameter \customHochkommata{Timed event}: The timed event result (Boolean). \item Mask parameter \customHochkommata{Time duration}: The duration of the event in simulation steps.} 
	{\includegraphics[width=0.75\textwidth]{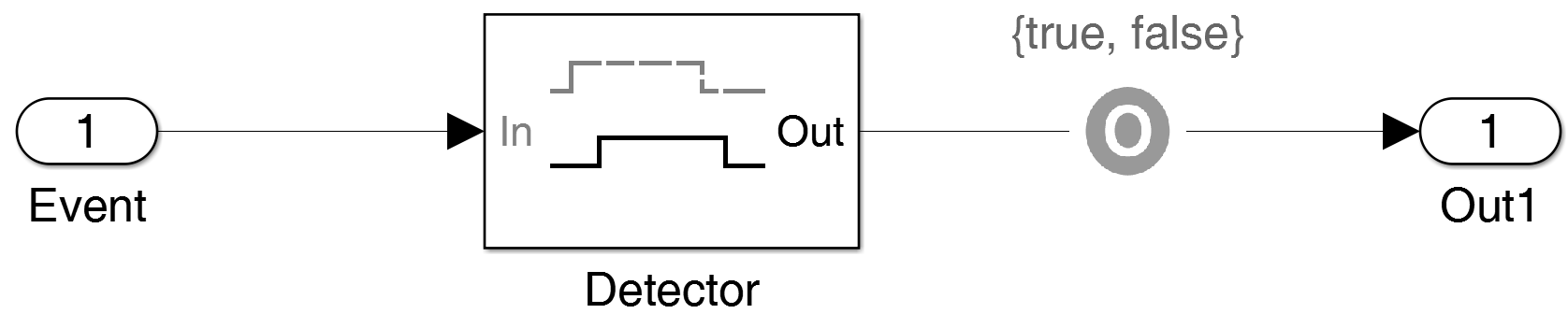}} 
	{\item Detector mask parameter \customHochkommata{External reset}: No \item Detector mask parameter \customHochkommata{Time steps for input detection}: time Duration \item Detector mask parameter \customHochkommata{Time steps for delay (optional)}: 0 \item Detector mask parameter \customHochkommata{Time steps for output duration}: 1 \item A test objective is added to ensure generated test vectors of the correct length.} 

\simulinkBlock
	{\texttt{delay}} 
	{Pattern} 
	{after a delay of Time duration simulation steps, [Event is valid].} 
	{\includegraphics[width=0.4\textwidth]{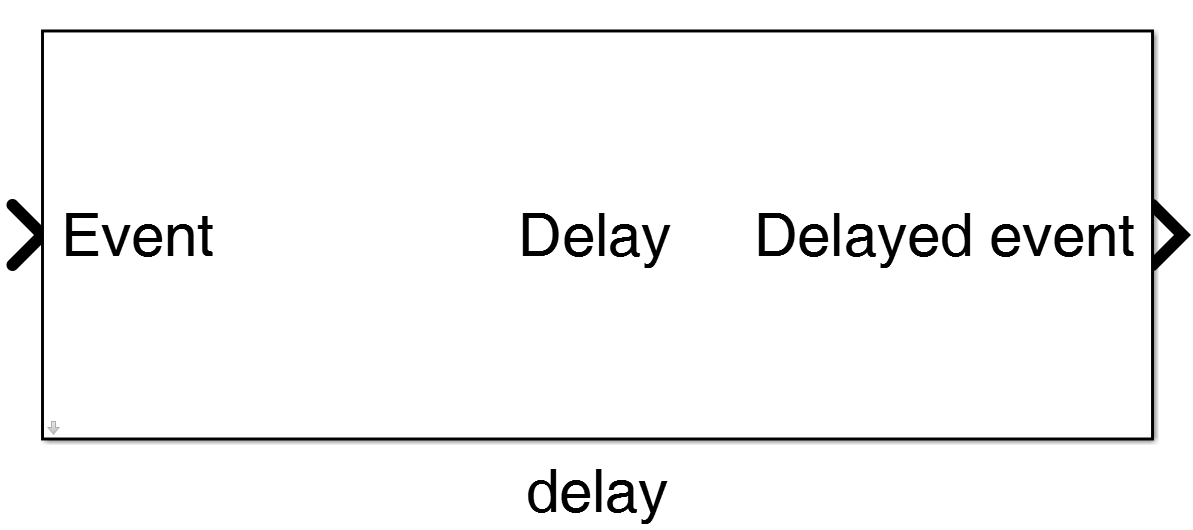}} 
	{\item Input parameter \customHochkommata{Event}: The event which is delayed. \item Output parameter \customHochkommata{Delayed event}: The delayed event result (Boolean). \item Mask parameter \customHochkommata{Time Delay}: The delay of the event in simulation steps.} 
	{\includegraphics[width=0.75\textwidth]{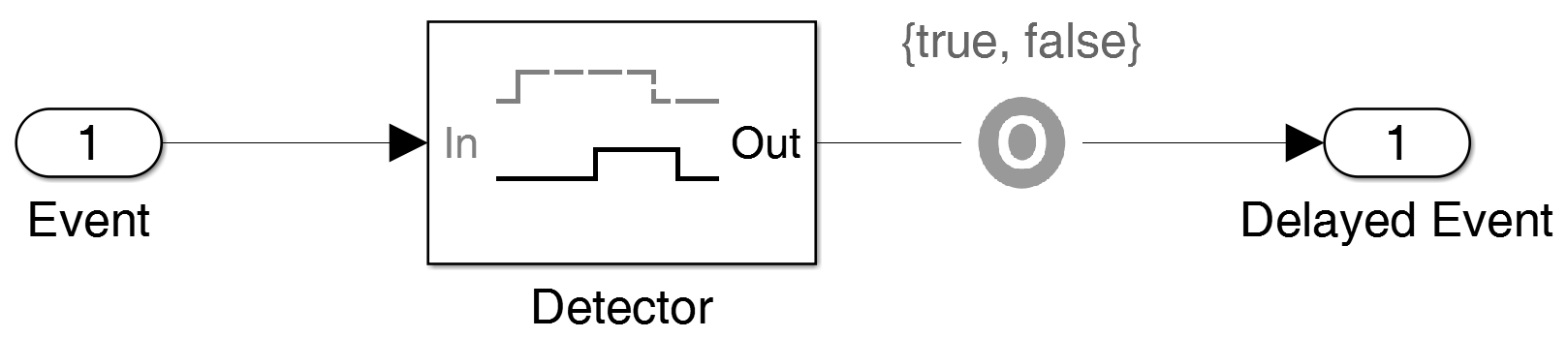}} 
	{\item Detector mask parameter \customHochkommata{External reset}: No \item Detector mask parameter \customHochkommata{Time steps for input detection}: 1 \item Detector mask parameter \customHochkommata{Time steps for delay (optional)}: time Duration \item Detector mask parameter \customHochkommata{Time steps for output duration}: 1 \item A test objective is added to ensure generated test vectors of the correct length.} 

\newpage
\subsection{Events}
\begin{figure}[h]
	\centering
	\includegraphics[width=0.4\textwidth]{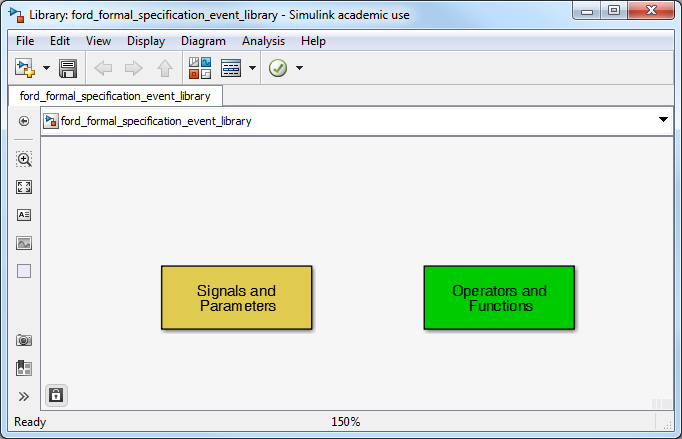}
	\caption{Simulink library for events with sub-libraries for the specification of signals and parameters and for operators and functions.}
	\label{fig:eventLib}
\end{figure}

Events are specified in a subsystem using building blocks from the specification library for signals, constants, calibration parameters, operators and functions.
For each input signal of the event, a \customHochkommata{signal in} block is used and its output is connected to the input of a \customHochkommata{signal goto} block. This provides the interface and allows to access the signal using a \customHochkommata{signal from} block during event specification. Our event library contains sub-libraries for signals and parameters and for operators and funcations (see Figure \ref{fig:eventLib}).

\subsubsection{Signals, Constants and Calibration Parameters}
\ \\
\simulinkBlock
	{\texttt{input/output signal}} 
	{Event -- Signals and Parameters} 
	{--} 
	{\includegraphics[width=0.2\textwidth]{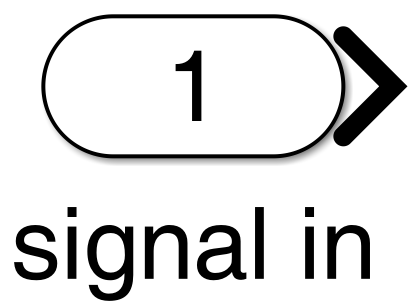} \hspace*{2cm}\includegraphics[width=0.2\textwidth]{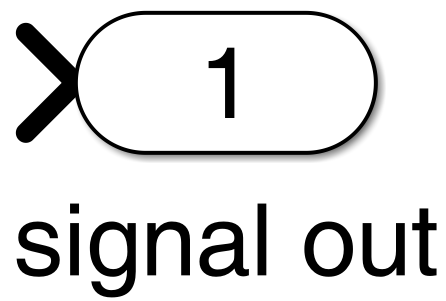}} 
	{\item Mask parameter \customHochkommata{Port number}: The number of the corresponding input/output port.} 
	{--} 
	{\item Used to define the interface for (verification) subsystems.} 

\simulinkBlock
	{\texttt{signal}} 
	{Event -- Signals and Parameters} 
	{signal name, e.g. signal\_A } 
	{\includegraphics[width=0.2\textwidth]{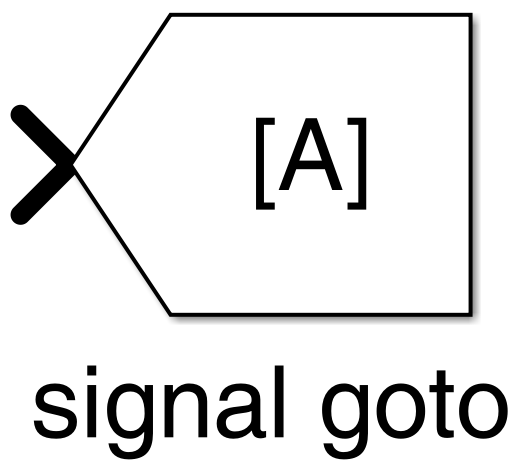} \hspace*{2cm} \includegraphics[width=0.2\textwidth]{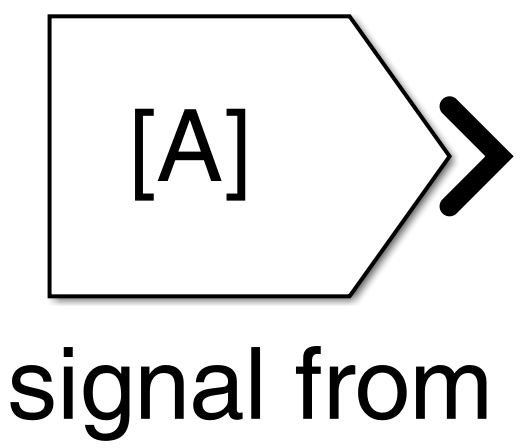}} 
	{\item Mask parameter \customHochkommata{Goto Tag}: The tag of the corresponding signal goto block.} 
	{--} 
	{\item Each signal in block is connected with a signal goto block to make the signal accessible during event specification. \item Signal from blocks are used for the specification of signals inside the event specification.} 

\simulinkBlock
	{\texttt{constant}} 
	{Event -- Signals and Parameters} 
	{constant name, e.g. constant\_A } 
	{\includegraphics[width=0.3\textwidth]{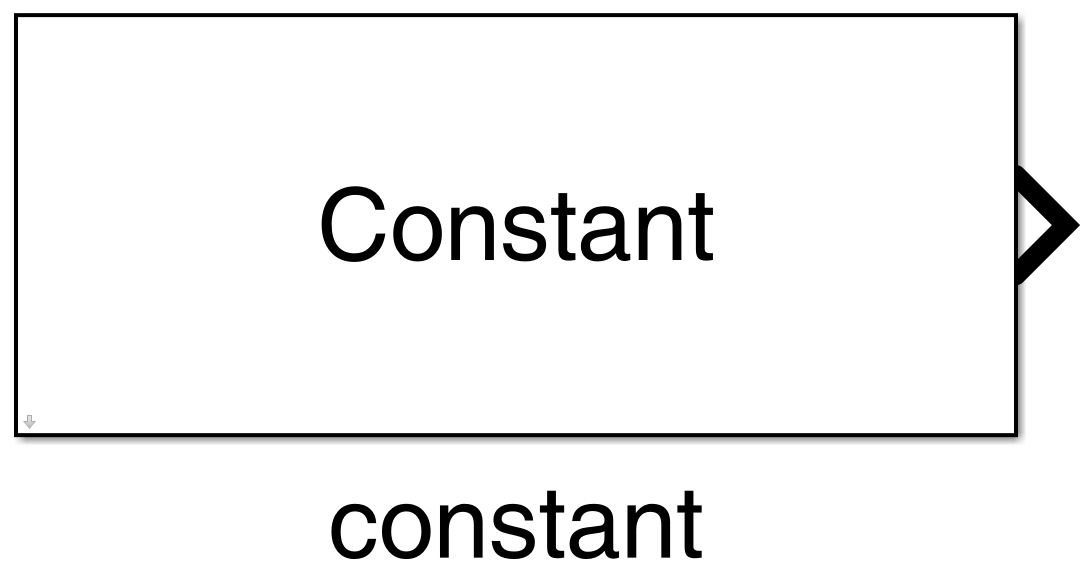}} 
	{\item Mask parameter \customHochkommata{Constant value}: The numerical value of the constant.} 
	{\includegraphics[width=0.45\textwidth]{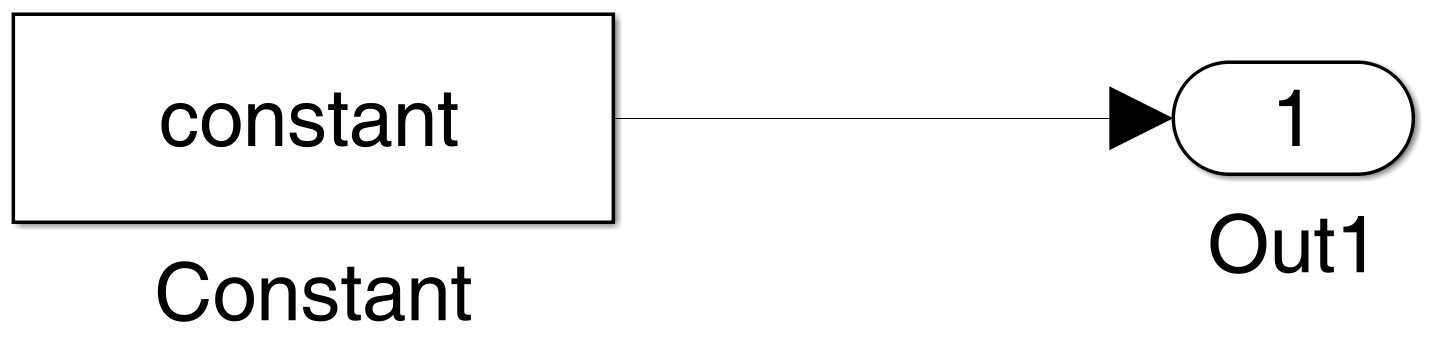}} 
	{\item --} 
	
\simulinkBlock
	{\texttt{calibration parameter}} 
	{Event -- Signals and Parameters} 
	{signal name, e.g. calibration\_A } 
	{\includegraphics[width=0.3\textwidth]{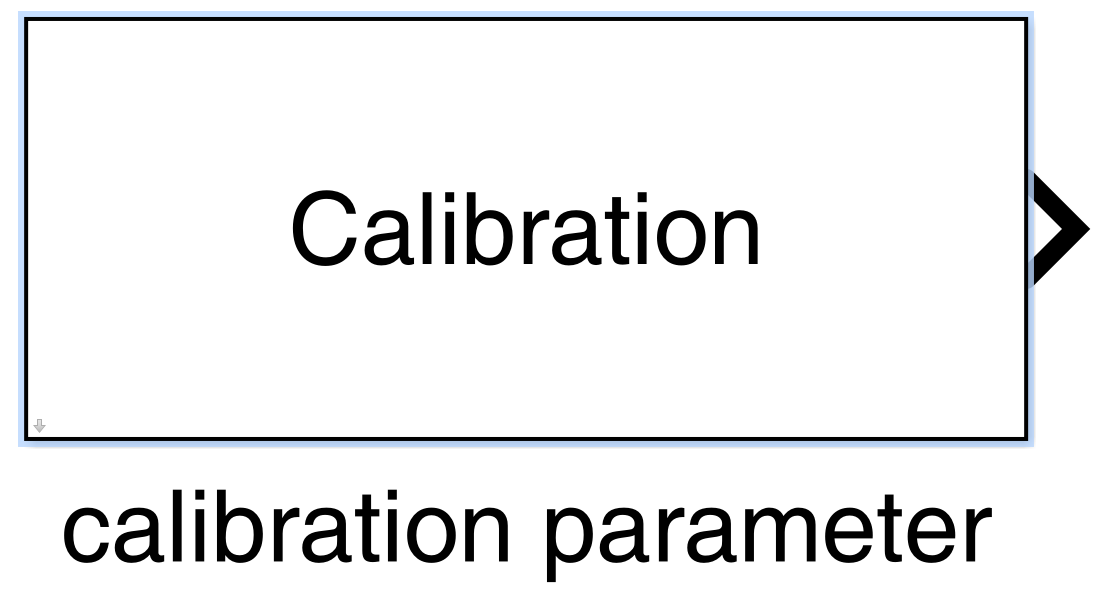}} 
	{\item Mask parameter \customHochkommata{Calibration parameter}: The name of the corresponding calibration parameter.} 
	{\includegraphics[width=0.45\textwidth]{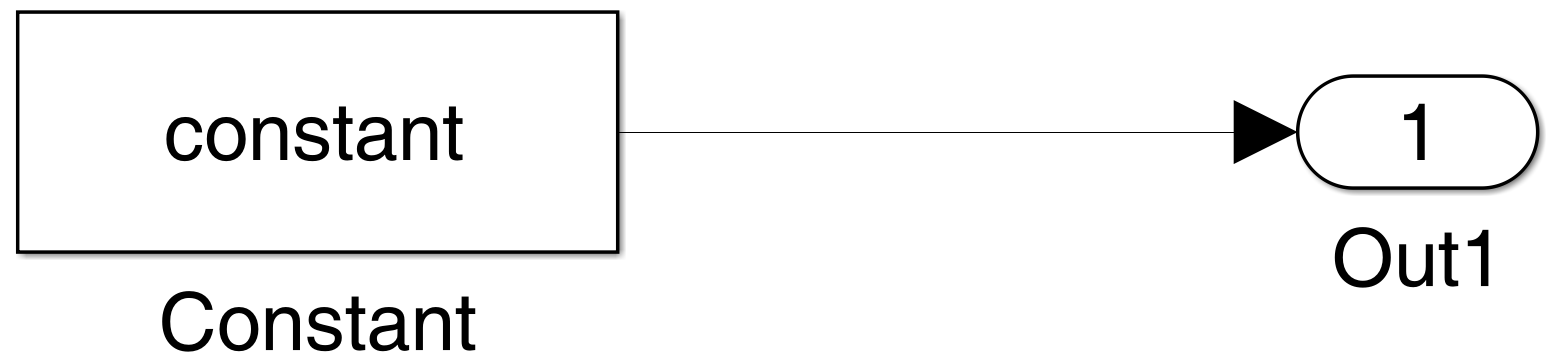}} 
	{\item --} 
	
\subsubsection{Operators and Functions}
\begin{figure}[h]
	\centering
	\includegraphics[width=0.45\textwidth]{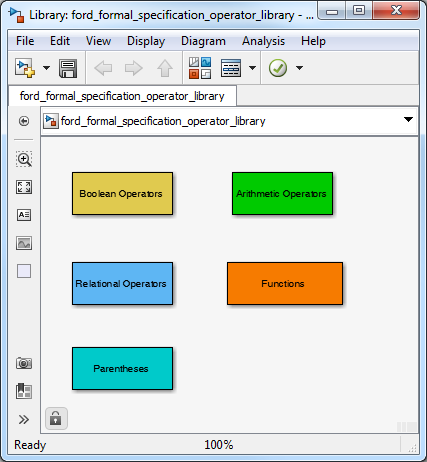}
	\caption{Simulink library for events with sub-libraries for the specification of signals and parameters and for operators and functions.}
	\label{fig:operatorsLib}
\end{figure}

The sub-library for operators and functions is further divided into sub-libraries for Boolean operators, arithmetic operators, relational operators, functions and parentheses (see Figure \ref{fig:operatorsLib}).

\simulinkBlock
	{\texttt{not}} 
	{Event -- Operators and Functions -- Boolean Operators} 
	{not [In1]} 
	{\includegraphics[width=0.15\textwidth]{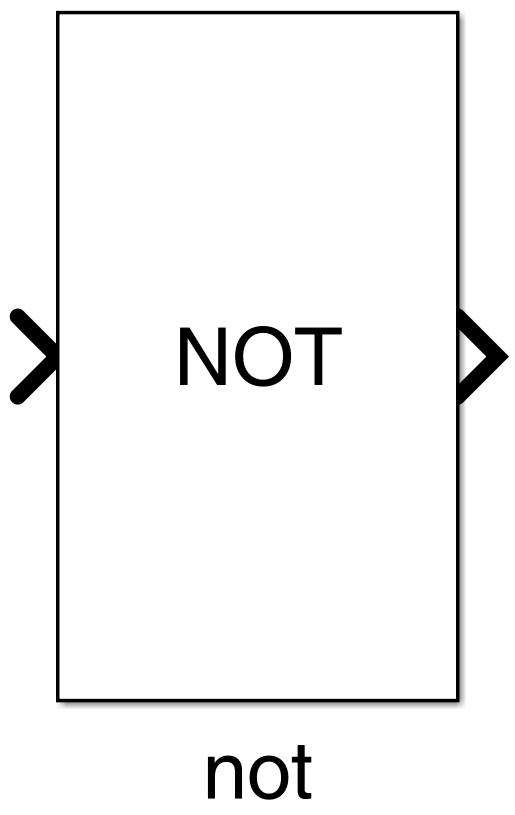}} 
	{\item Input parameter \customHochkommata{In1}: input that is negated. \item Output parameter \customHochkommata{Out1}: negation of the input.} 
	{\includegraphics[width=0.85\textwidth]{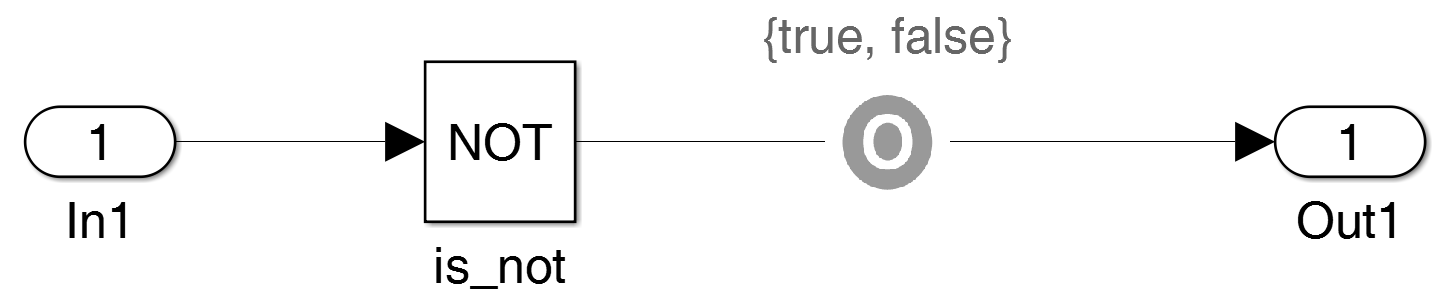}} 
	{\item A test objective is added for the generation of test vectors for true/false result.} 

\simulinkBlock
	{\texttt{and}} 
	{Event -- Operators and Functions -- Boolean Operators} 
	{[In1] and [In2]} 
	{\includegraphics[width=0.15\textwidth]{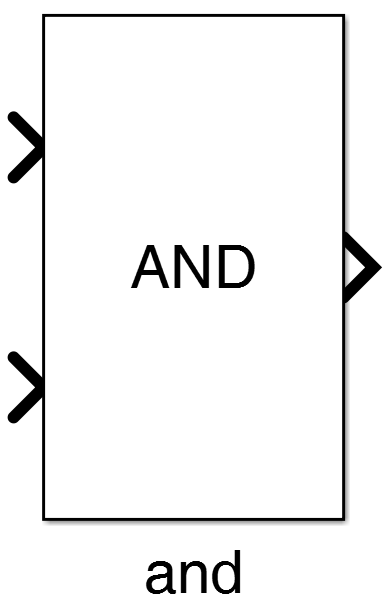}} 
	{\item Input parameter \customHochkommata{In1}/\customHochkommata{In2}: The inputs to which the and operator is applied. \item Output parameter \customHochkommata{Out1}: The logical and of the two input parameters.} 
	{\includegraphics[width=0.85\textwidth]{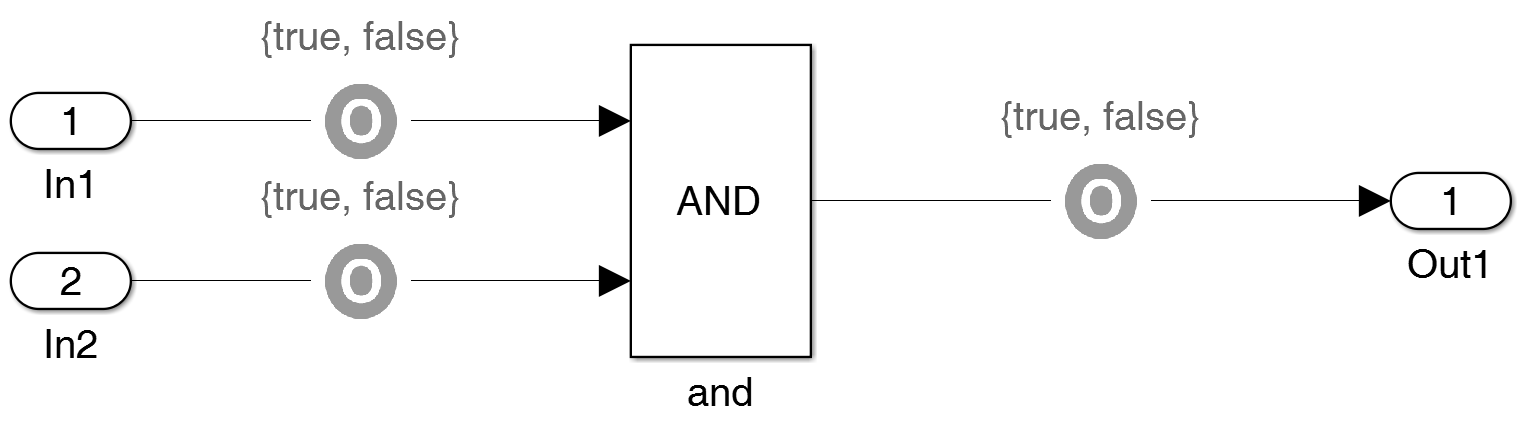}} 
	{\item Test objectives are added for the generation of test vectors for all possible input combinations. \item A test objective is added for the generation of test vectors for true/false result.} 

\simulinkBlock
	{\texttt{or}} 
	{Event -- Operators and Functions -- Boolean Operators} 
	{[In1] or [In2]} 
	{\includegraphics[width=0.15\textwidth]{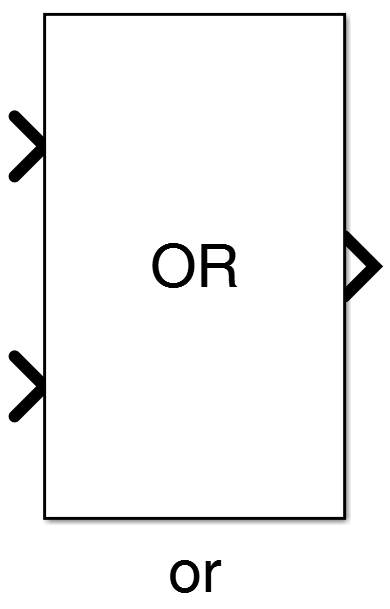}} 
	{\item Input parameter \customHochkommata{In1}/\customHochkommata{In2}: The inputs to which the or operator is applied. \item Output parameter \customHochkommata{Out1}: The logical or of the two input parameters.} 
	{\includegraphics[width=0.85\textwidth]{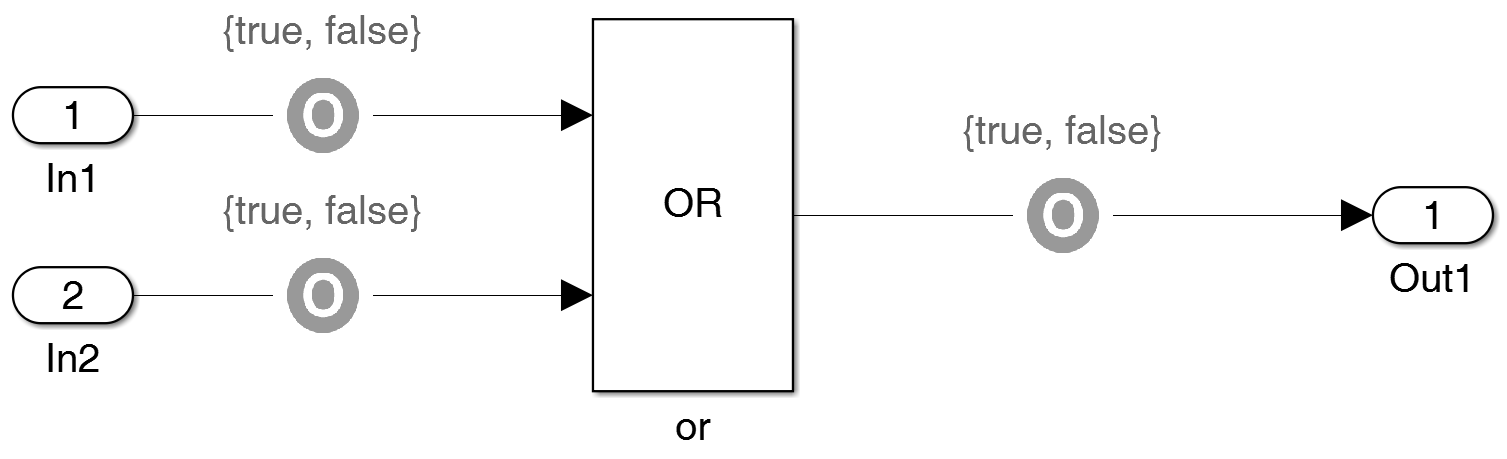}} 
	{\item Test objectives are added for the generation of test vectors for all possible input combinations. \item A test objective is added for the generation of test vectors for true/false result.} 

\simulinkBlock
	{\texttt{implies}} 
	{Event -- Operators and Functions -- Boolean Operators} 
	{[A] implies [B]} 
	{\includegraphics[width=0.15\textwidth]{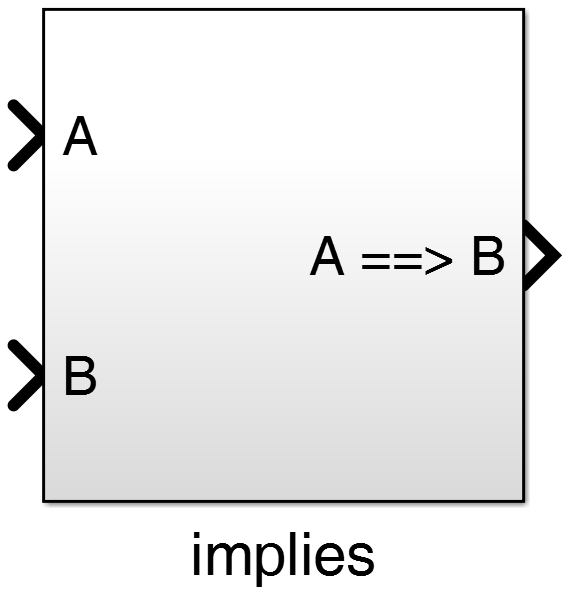}} 
	{\item Input parameter \customHochkommata{A}/\customHochkommata{B}: The inputs to which the implication operator is applied. \item Output parameter \customHochkommata{A $\Rightarrow$ B}: The logical implication of the two input parameters.} 
	{\includegraphics[width=0.85\textwidth]{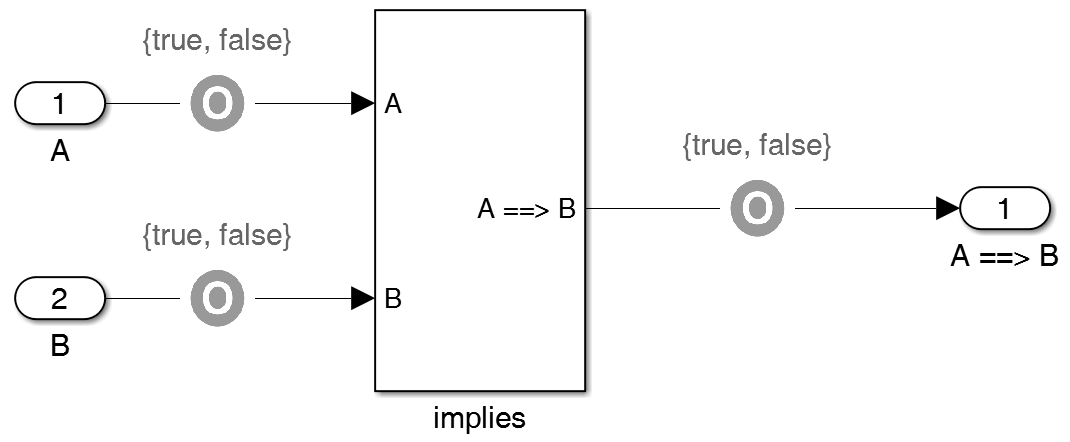}} 
	{\item Test objectives are added for the generation of test vectors for all possible input combinations. \item A test objective is added for the generation of test vectors for true/false result.} 

\simulinkBlock
	{\texttt{plus}} 
	{Event -- Operators and Functions -- Arithmetic Operators} 
	{[a] plus [b]} 
	{\includegraphics[width=0.15\textwidth]{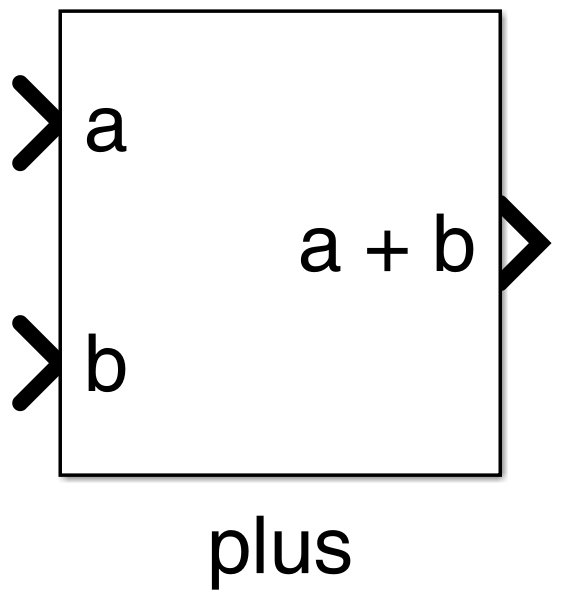}} 
	{\item Input parameter \customHochkommata{a}/\customHochkommata{b}: The inputs to which the addition operator is applied. \item Output parameter \customHochkommata{a+b}: The addition of the two input parameters.} 
	{\includegraphics[width=0.85\textwidth]{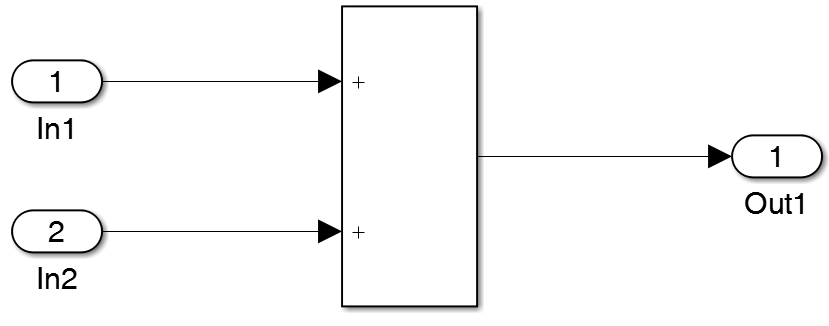}} 
	{\item --} 

\simulinkBlock
	{\texttt{minus}} 
	{Event -- Operators and Functions -- Arithmetic Operators} 
	{[a] minus [b]} 
	{\includegraphics[width=0.15\textwidth]{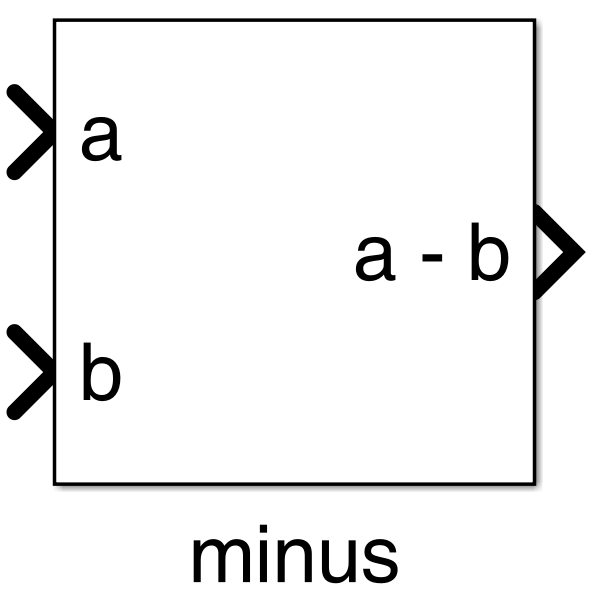}} 
	{\item Input parameter \customHochkommata{In1}/\customHochkommata{In2}: The inputs to which the subtraction operator is applied. \item Output parameter \customHochkommata{a-b}: The subtraction of the two input parameters.} 
	{\includegraphics[width=0.85\textwidth]{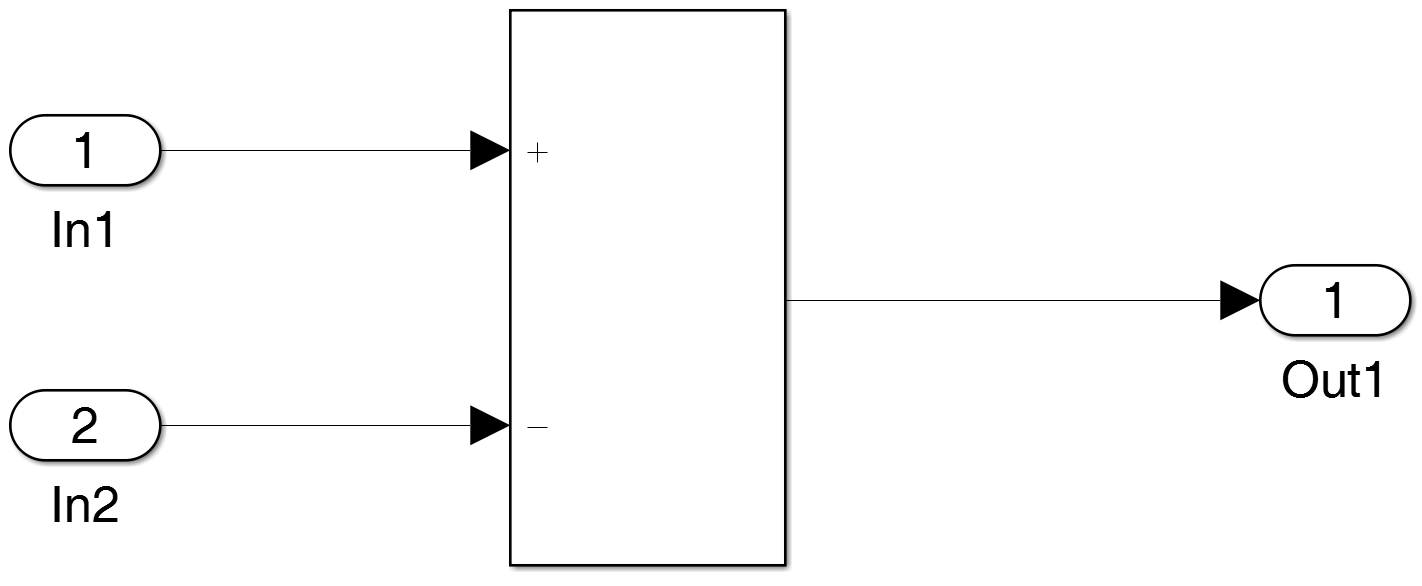}} 
	{\item --} 
	
\simulinkBlock
	{\texttt{multiplication}} 
	{Event -- Operators and Functions -- Arithmetic Operators} 
	{[a] multiplied with [b]} 
	{\includegraphics[width=0.15\textwidth]{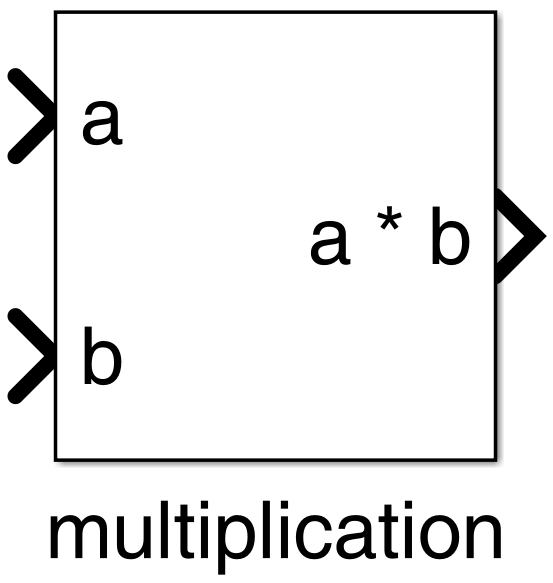}} 
	{\item Input parameter \customHochkommata{a}/\customHochkommata{b}: The inputs to which the multiplication operator is applied. \item Output parameter \customHochkommata{a*b}: The multiplication of the two input parameters. \item Mask parameter \customHochkommata{Number of inputs}: \customHochkommata{2}. \item Mask parameter \customHochkommata{Multiplication}: Element-wise(.*).} 
	{} 
	{\item --} 
	
\simulinkBlock
	{\texttt{division}} 
	{Event -- Operators and Functions -- Arithmetic Operators} 
	{[a] divided by [b]} 
	{\includegraphics[width=0.15\textwidth]{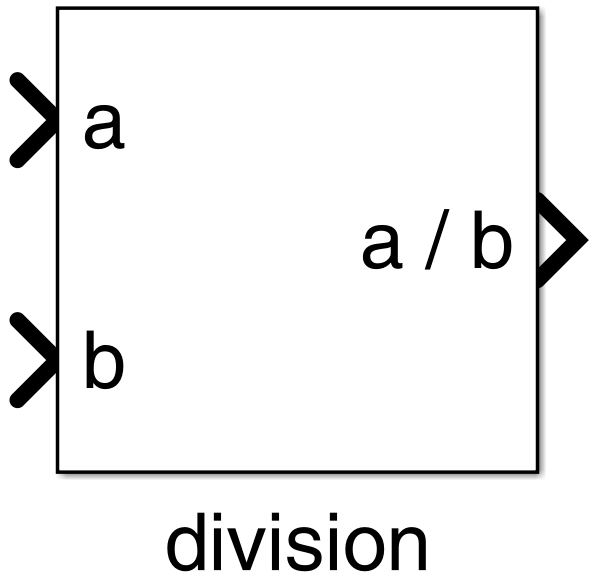}} 
	{\item Input parameter \customHochkommata{a}/\customHochkommata{b}: The inputs to which the division operator is applied. \item Output parameter \customHochkommata{a/b}: The division of the two input parameters. \item Mask parameter \customHochkommata{Number of inputs}: \customHochkommata{*/}. \item Mask parameter \customHochkommata{Multiplication}: Element-wise(.*).} 
	{} 
	{\item --} 
	
\simulinkBlock
	{\texttt{plus sign}} 
	{Event -- Operators and Functions -- Arithmetic Operators} 
	{plus [a]} 
	{\includegraphics[width=0.15\textwidth]{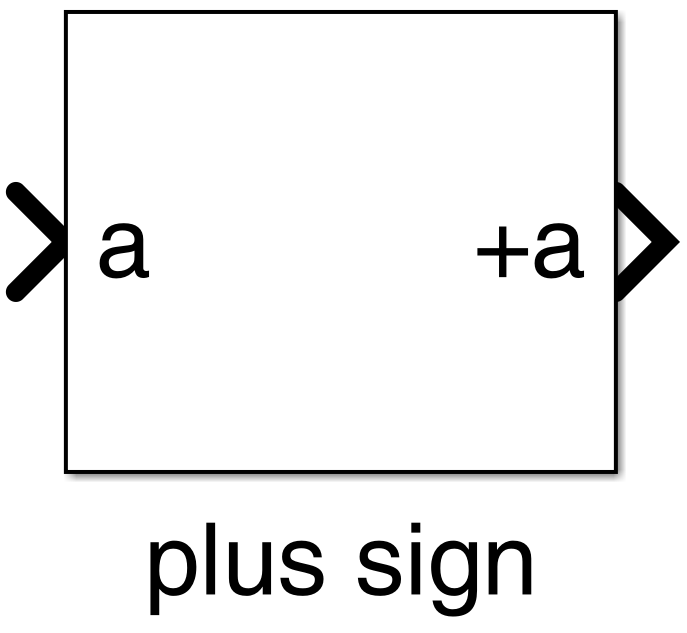}} 
	{\item Input parameter \customHochkommata{a}: The inputs to which the positive sign is applied. \item Output parameter \customHochkommata{+a}: The computation result.} 
	{\includegraphics[width=0.65\textwidth]{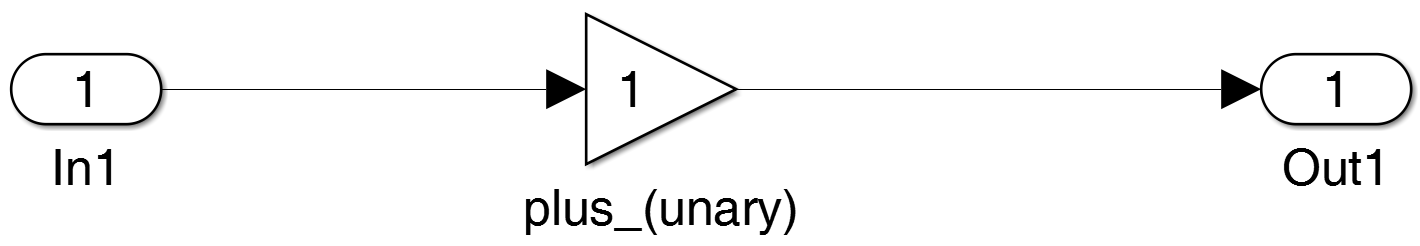}} 
	{\item --} 
	
\simulinkBlock
	{\texttt{minus sign}} 
	{Event -- Operators and Functions -- Arithmetic Operators} 
	{minus [a]} 
	{\includegraphics[width=0.15\textwidth]{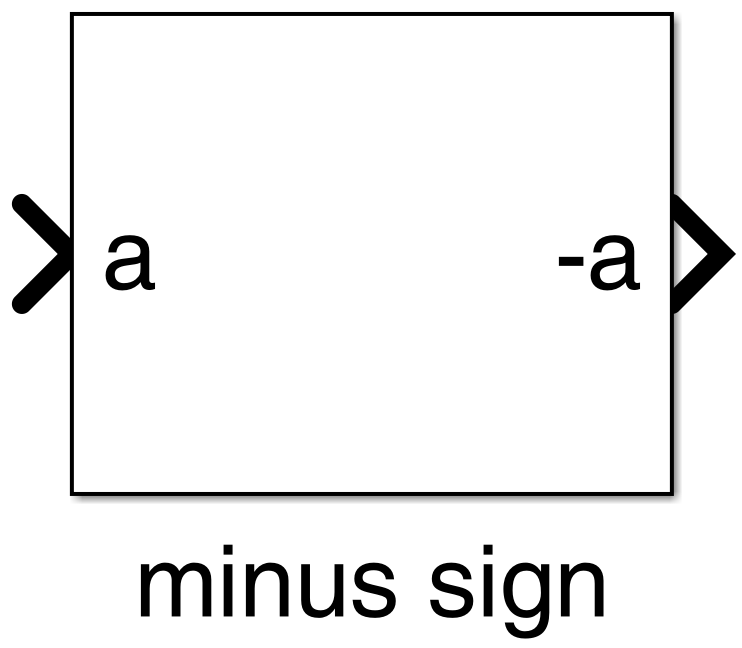}} 
	{\item Input parameter \customHochkommata{a}: The inputs to which the negative sign is applied. \item Output parameter \customHochkommata{-a}: The computation result.} 
	{\includegraphics[width=0.65\textwidth]{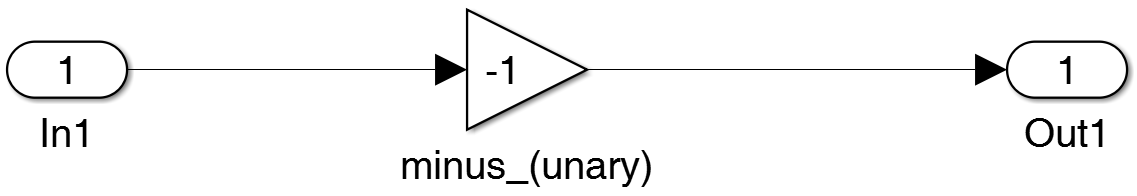}} 
	{\item --} 

\simulinkBlock
	{\texttt{less}} 
	{Event -- Operators and Functions -- Relational Operators} 
	{[a] is less than [b]} 
	{\includegraphics[width=0.15\textwidth]{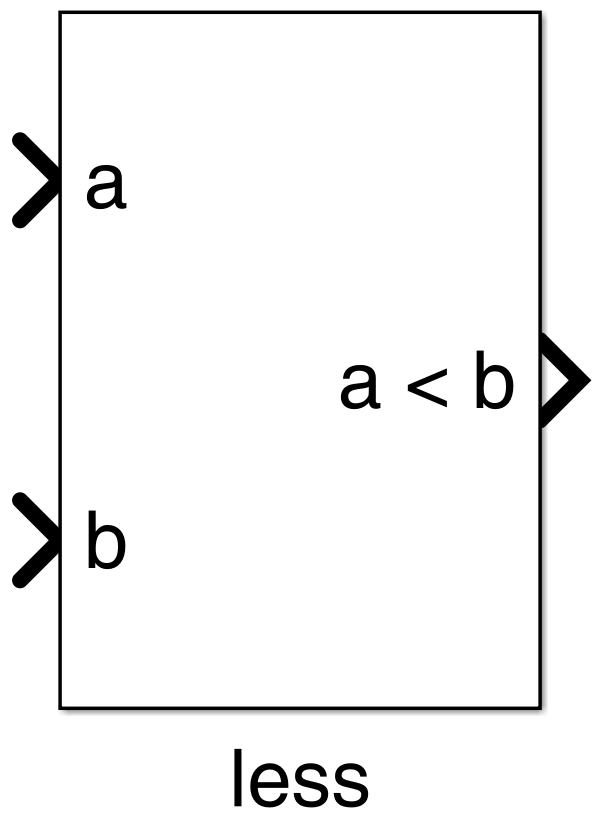}} 
	{\item Input parameter \customHochkommata{a}/\customHochkommata{b}: The inputs which are compared using the less operator. \item Output parameter \customHochkommata{a$<$b}: The comparison result.} 
	{\includegraphics[width=0.65\textwidth]{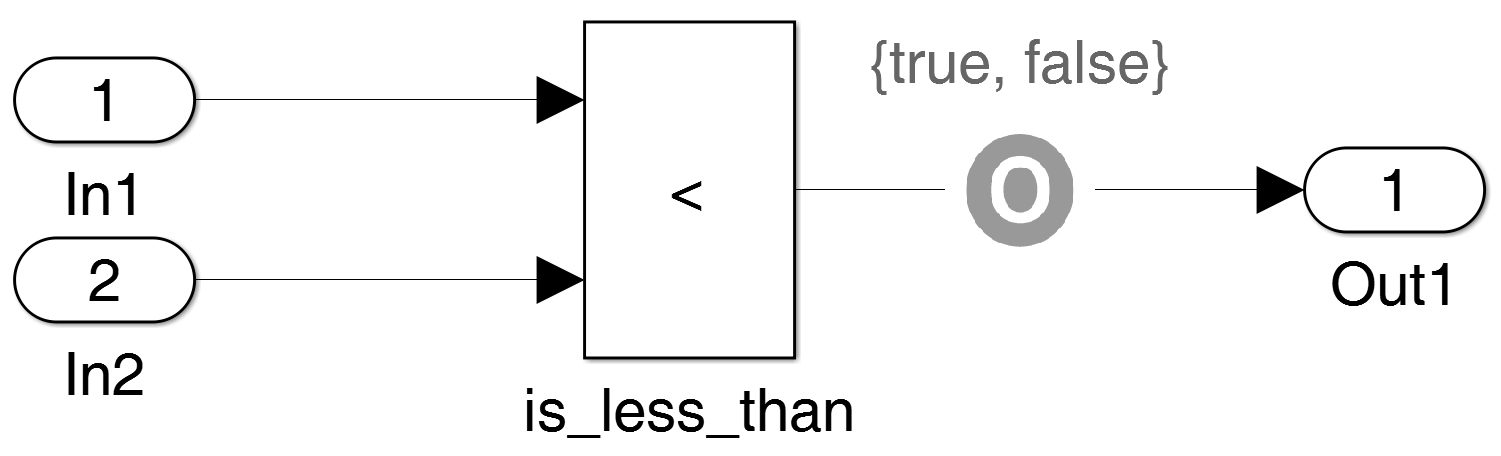}} 
	{\item A test objective is added for the generation of test vectors for true/false result.} 
	
\simulinkBlock
	{\texttt{less or equal}} 
	{Event -- Operators and Functions -- Relational Operators} 
	{[a] is less or equal to [b]} 
	{\includegraphics[width=0.15\textwidth]{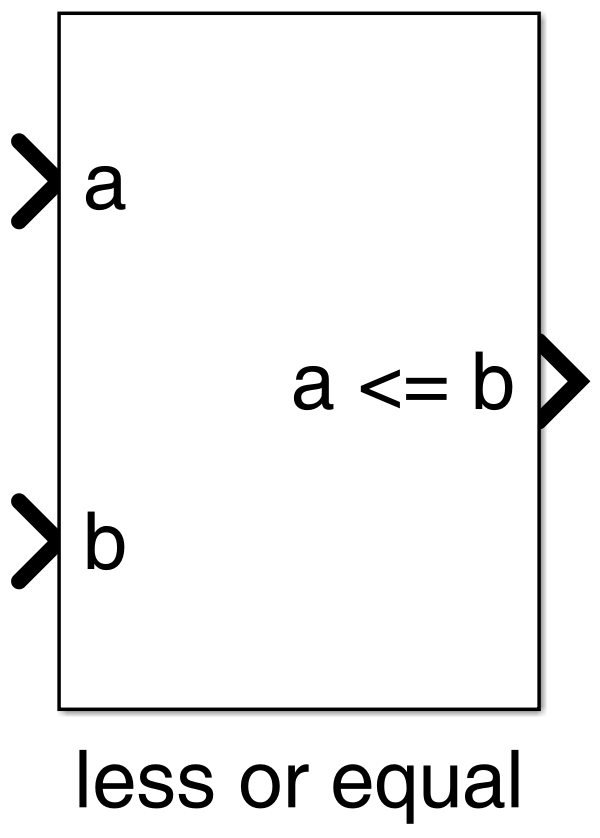}} 
	{\item Input parameter \customHochkommata{a}/\customHochkommata{b}: The inputs which are compared using the less or equal operator. \item Output parameter \customHochkommata{a$\leq$b}: The comparison result.} 
	{\includegraphics[width=0.65\textwidth]{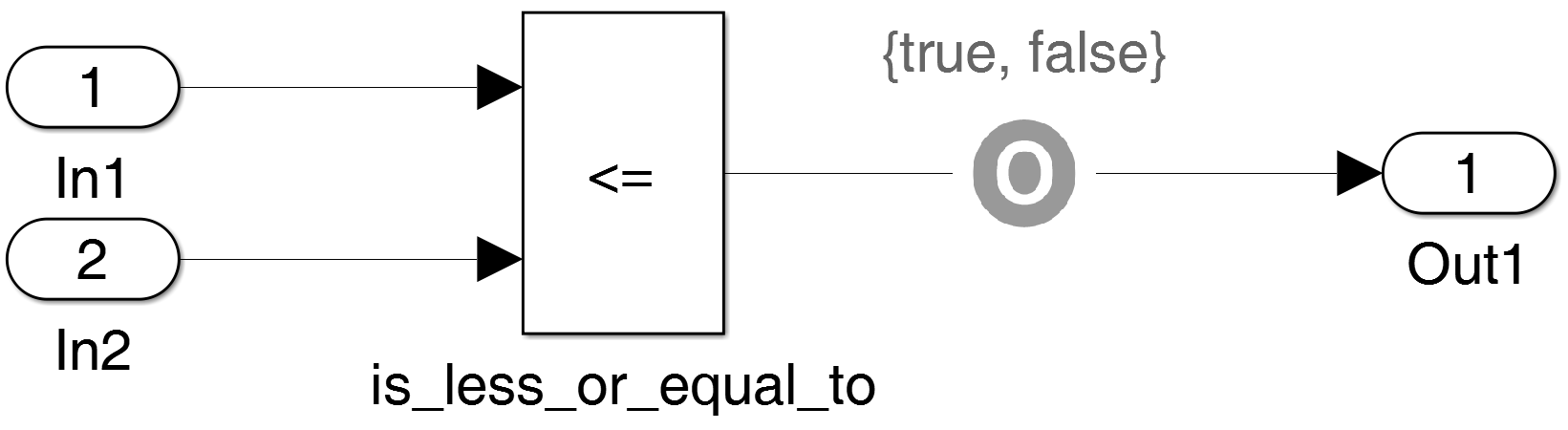}} 
	{\item A test objective is added for the generation of test vectors for true/false result.} 
	
\simulinkBlock
	{\texttt{greater}} 
	{Event -- Operators and Functions -- Relational Operators} 
	{[a] is greater than [b]} 
	{\includegraphics[width=0.15\textwidth]{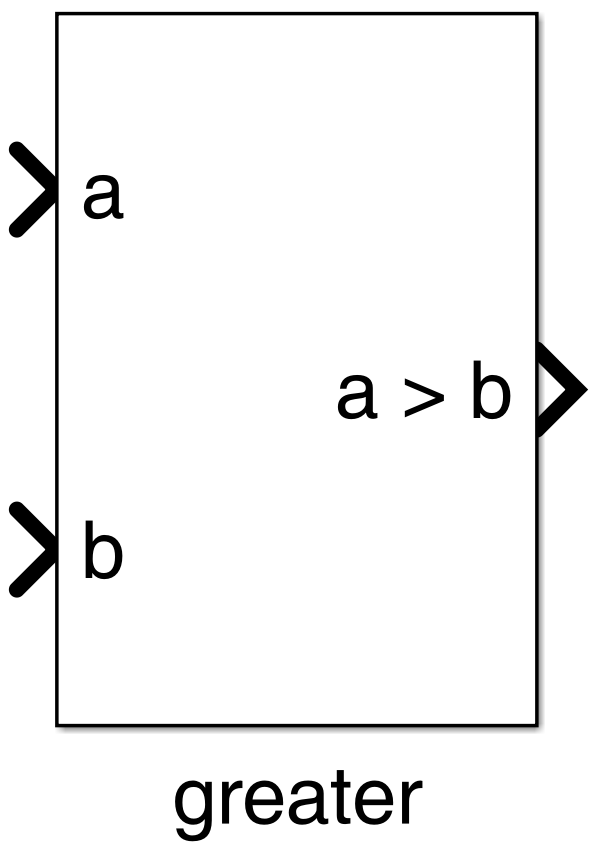}} 
	{\item Input parameter \customHochkommata{a}/\customHochkommata{b}: The inputs which are compared using the greater operator. \item Output parameter \customHochkommata{a$>$b}: The comparison result.} 
	{\includegraphics[width=0.65\textwidth]{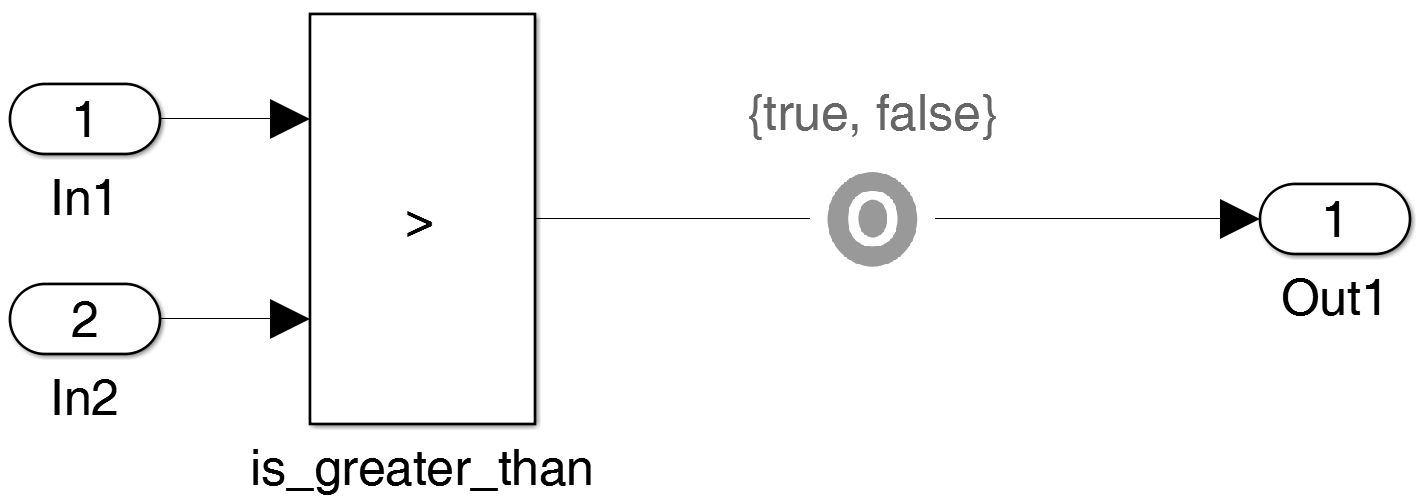}} 
	{\item A test objective is added for the generation of test vectors for true/false result.} 
	
\simulinkBlock
	{\texttt{greater or equal}} 
	{Event -- Operators and Functions -- Relational Operators} 
	{[a] is greater or equal to [b]} 
	{\includegraphics[width=0.15\textwidth]{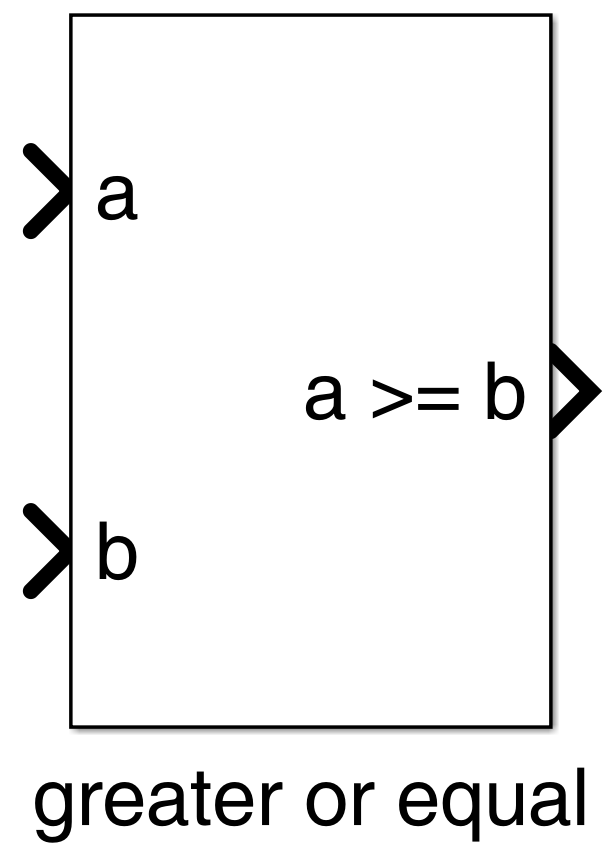}} 
	{\item Input parameter \customHochkommata{a}/\customHochkommata{b}: The inputs which are compared using the greater or equal operator. \item Output parameter \customHochkommata{a$\geq$b}: The comparison result.} 
	{\includegraphics[width=0.65\textwidth]{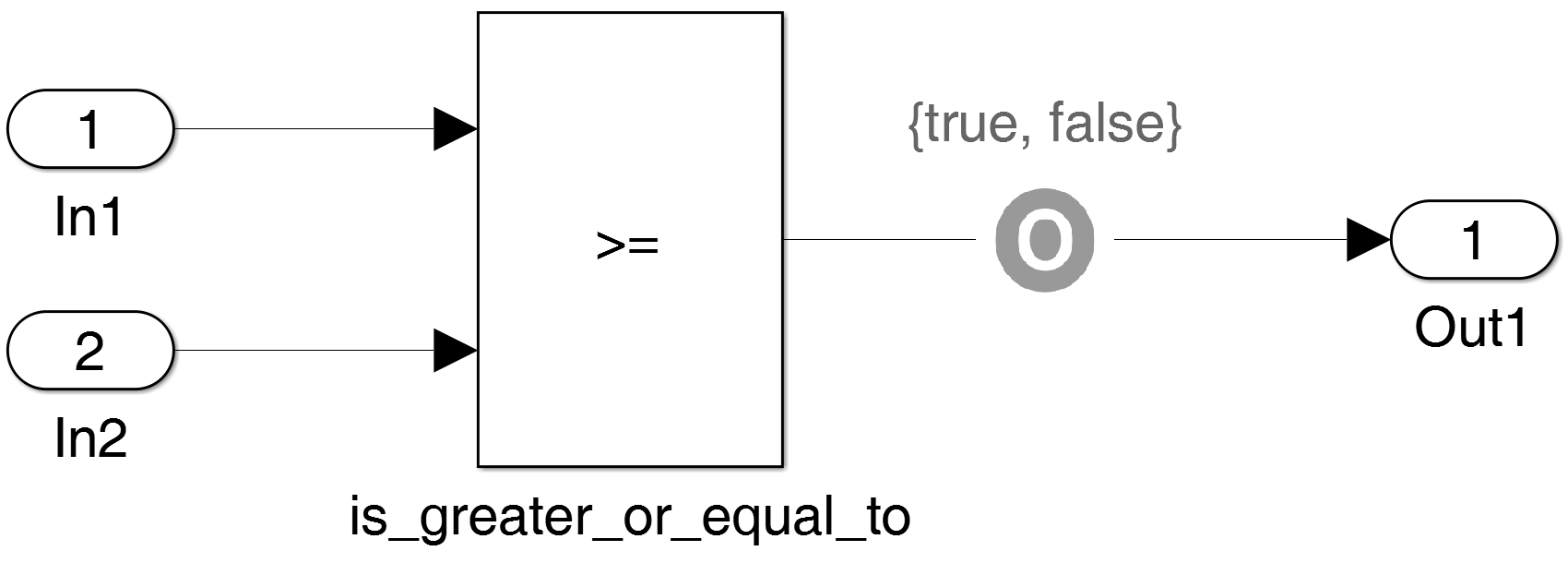}} 
	{\item A test objective is added for the generation of test vectors for true/false result.} 
	
\simulinkBlock
	{\texttt{equal}} 
	{Event -- Operators and Functions -- Relational Operators} 
	{[a] is equal to [b]} 
	{\includegraphics[width=0.15\textwidth]{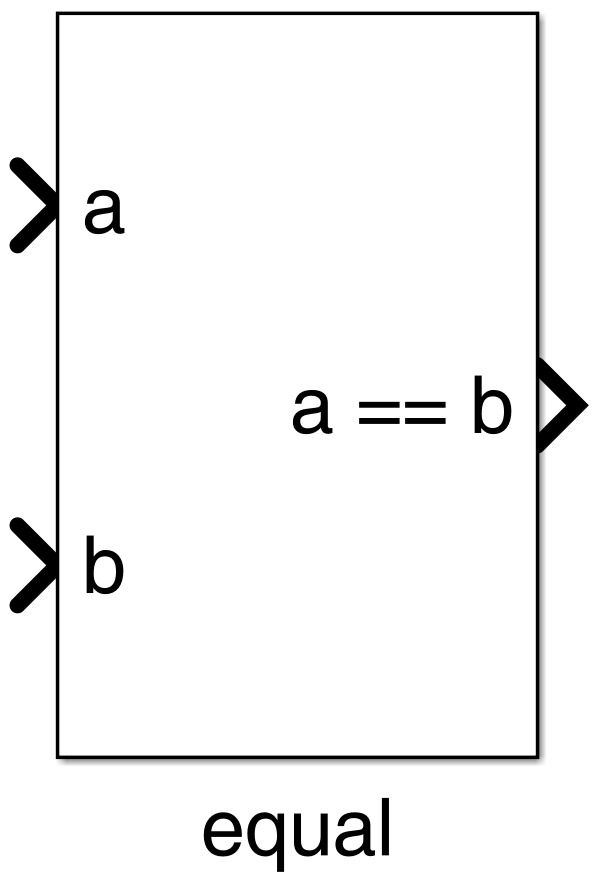}} 
	{\item Input parameter \customHochkommata{a}/\customHochkommata{b}: The inputs which are compared using the equal operator. \item Output parameter \customHochkommata{a$==$b}: The comparison result.} 
	{\includegraphics[width=0.65\textwidth]{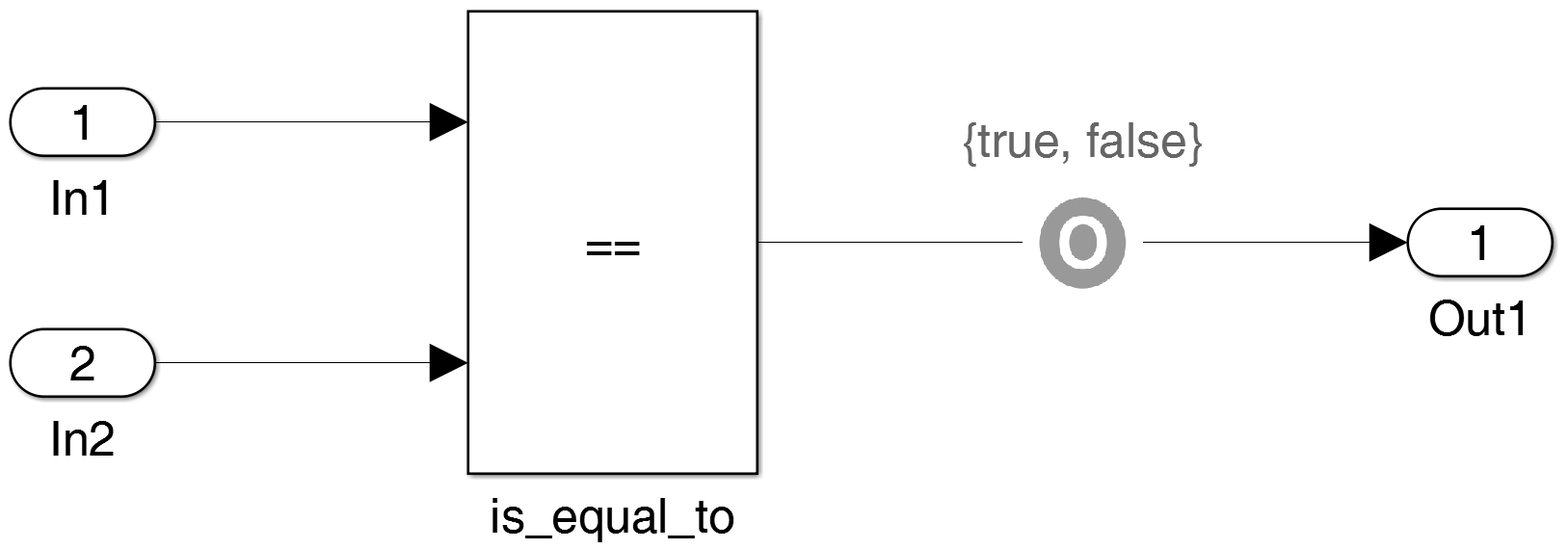}} 
	{\item A test objective is added for the generation of test vectors for true/false result.} 
	
\simulinkBlock
	{\texttt{min}} 
	{Event -- Operators and Functions -- Functions} 
	{the minimum of [a] and [b]} 
	{\includegraphics[width=0.25\textwidth]{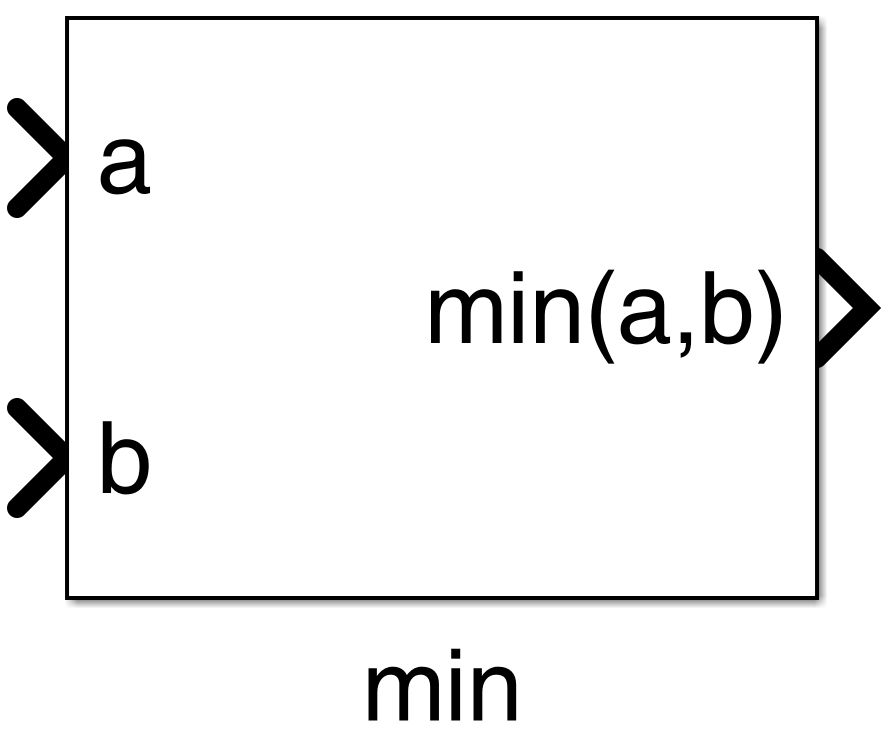}} 
	{\item Input parameter \customHochkommata{a}/\customHochkommata{b}: The inputs from which the minimum is selected. \item Output parameter \customHochkommata{min(a,b)}: The minimum of the two inputs. \item Mask parameter \customHochkommata{Function}: min. \item Mask parameter \customHochkommata{Number of input ports}: 2.} 
	{} 
	{\item --} 
	
\simulinkBlock
	{\texttt{max}} 
	{Event -- Operators and Functions -- Functions} 
	{the maximum of [a] and [b]} 
	{\includegraphics[width=0.25\textwidth]{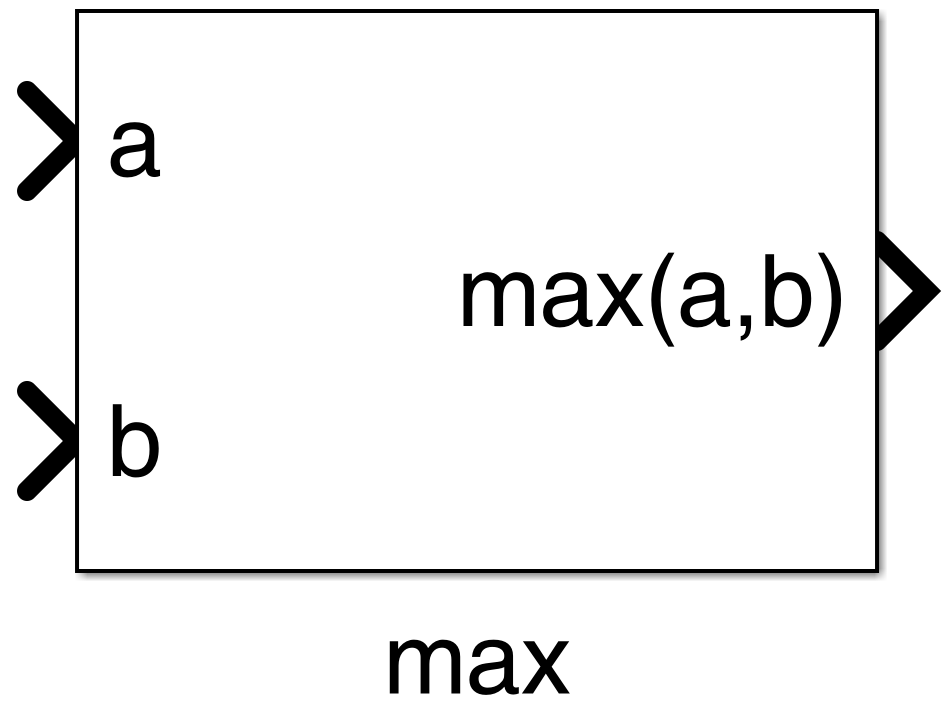}} 
	{\item Input parameter \customHochkommata{a}/\customHochkommata{b}: The inputs from which the maximum is selected. \item Output parameter \customHochkommata{max(a,b)}: The maximum of the two inputs. \item Mask parameter \customHochkommata{Function}: max. \item Mask parameter \customHochkommata{Number of input ports}: 2.} 
	{} 
	{\item --} 
	
\simulinkBlock
	{\texttt{abs}} 
	{Event -- Operators and Functions -- Functions} 
	{the absolute value of [a]} 
	{\includegraphics[width=0.25\textwidth]{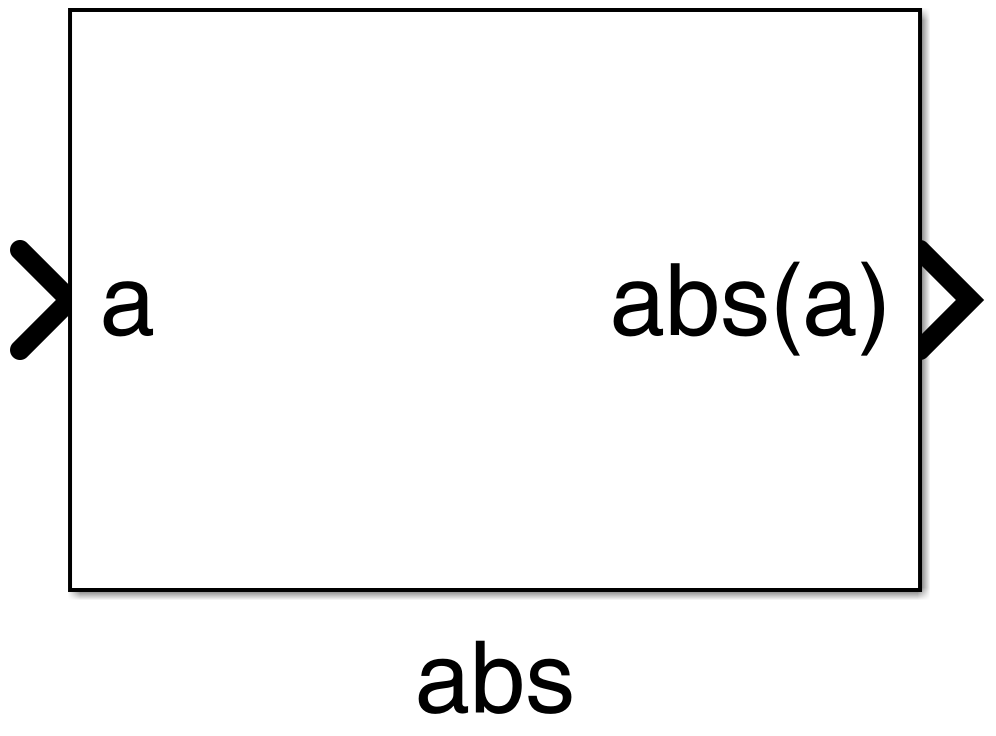}} 
	{\item Input parameter \customHochkommata{a}: The inputs for which the absolute value is computed. \item Output parameter \customHochkommata{abs(a)}: The absolute value of the input.} 
	{} 
	{\item --} 
	
\simulinkBlock
	{\texttt{last}} 
	{Event -- Operators and Functions -- Functions} 
	{the value of [a] [delay] simulation steps ago} 
	{\includegraphics[width=0.25\textwidth]{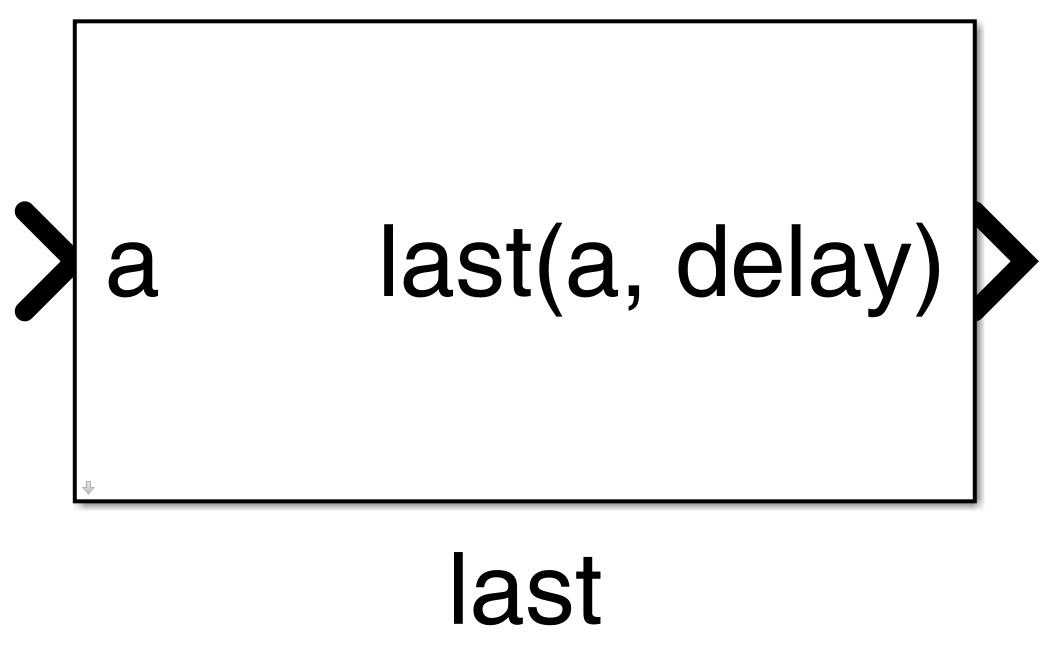}} 
	{\item Input parameter \customHochkommata{a}: The input for which the past value is computed. \item Output parameter \customHochkommata{last(a,delay)}: The value of the input a delay simulation steps ago. \item Mask parameter \customHochkommata{Time delay}: The number of steps the last operator reaches.} 
	{\includegraphics[width=0.65\textwidth]{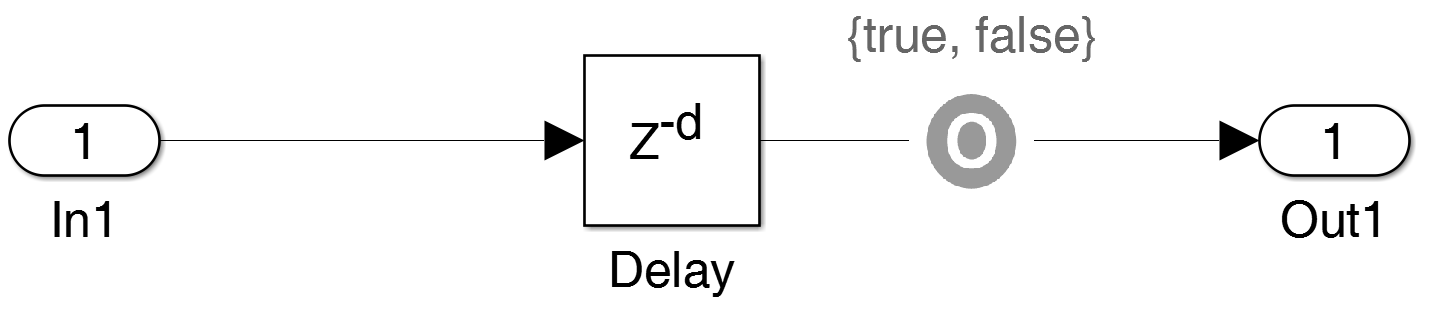}} 
	{\item Mask parameter \customHochkommata{Delay length} of Delay block: Source= Dialog, Value = delay. \item A test objective is added to ensure generated test vectors of the correct length.} 
		
\simulinkBlock
	{\texttt{extractBit}} 
	{Event -- Operators and Functions -- Functions} 
	{bit [bit] of [a]} 
	{\includegraphics[width=0.25\textwidth]{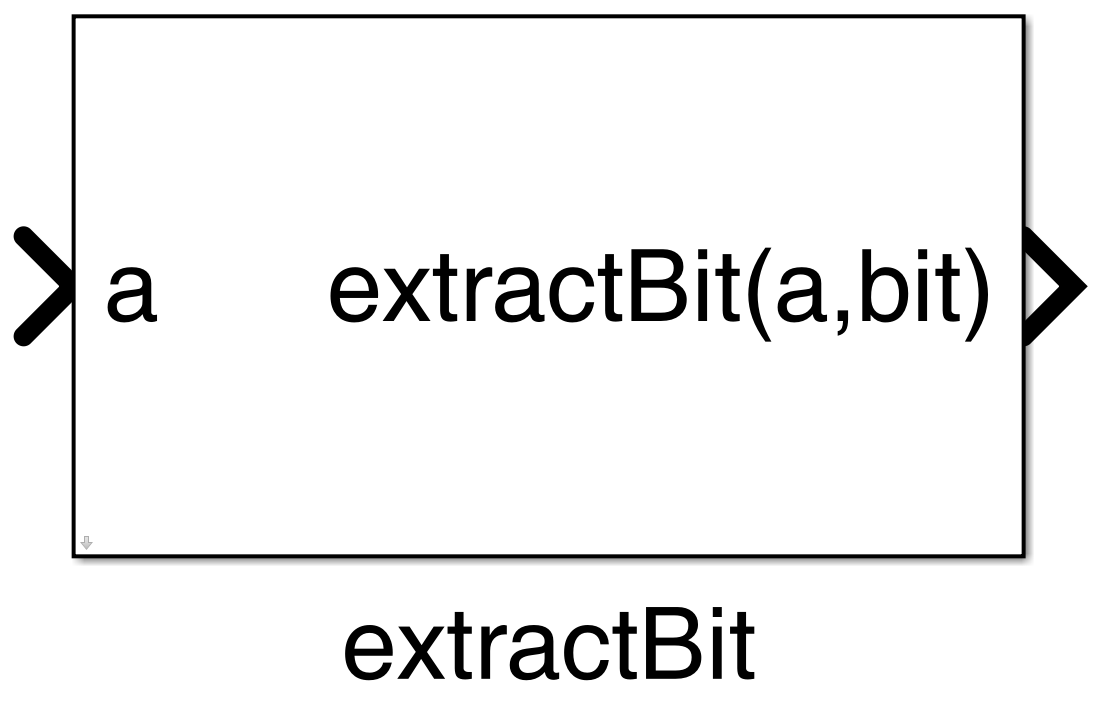}} 
	{\item Input parameter \customHochkommata{a}: The input from which a bit is extracted. \item Output parameter \customHochkommata{extractBit(a,bit)}: The value of bit \customHochkommata{bit} of the input a. \item Mask parameter \customHochkommata{Bit}: The number of the bit that is to be extracted.} 
	{\includegraphics[width=0.85\textwidth]{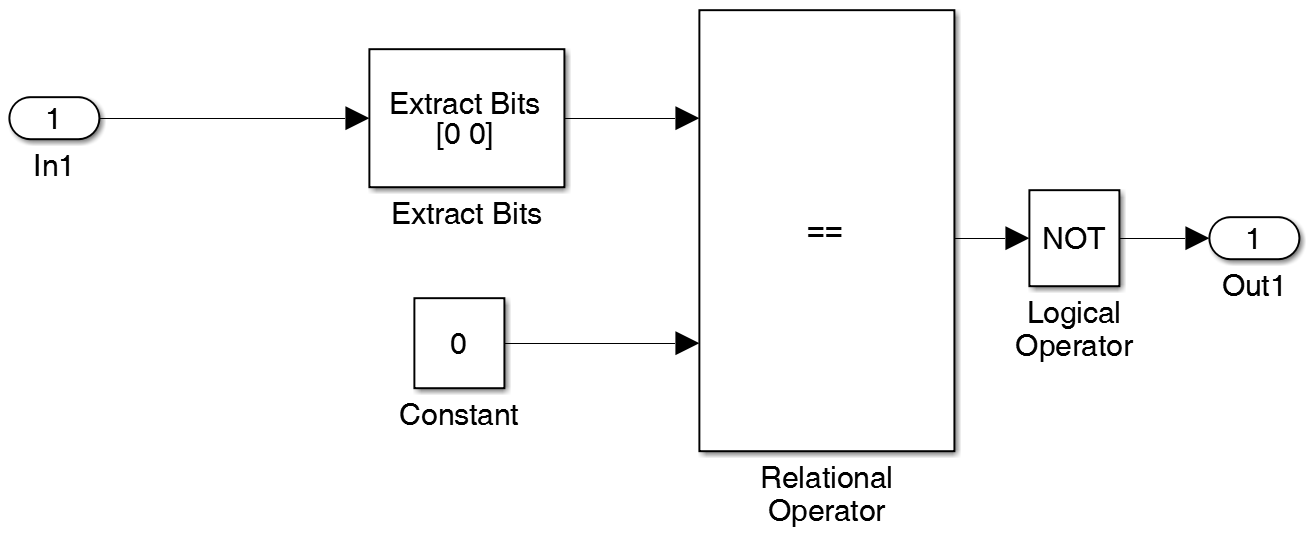}} 
	{\item Mask parameter \customHochkommata{Bits to extract} of Extract Bits block: Range of bits. \item Mask parameter \customHochkommata{Bit indices} of Extract Bits block: [bit, bit].} 

\simulinkBlock
	{\texttt{parenthesis}} 
	{Event -- Operators and Functions -- Functions} 
	{([In1])} 
	{\includegraphics[width=0.25\textwidth]{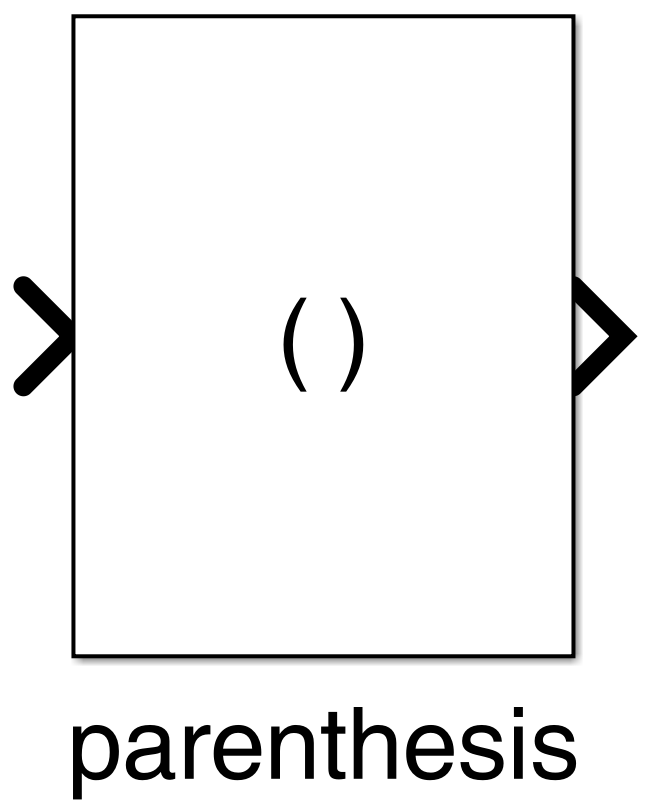}} 
	{\item Input parameter \customHochkommata{In1}: The input that should be surrounded by a parenthesis. \item Output parameter \customHochkommata{Out1}: The computed parenthesis expression.} 
	{\includegraphics[width=0.65\textwidth]{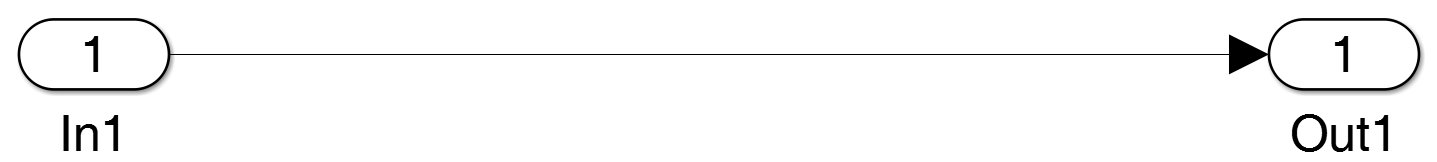}} 
	{\item --} 
	
\subsection{Scripts}
Currently, the library is shipped with a single script that defines a matlab function to connect the inputs of a specification block with matching signals in the Simulink model. 
 
 \simulinkFunction{addSubsystemConnection}{\item Input parameter \customHochkommata{systempath}: The path to the top-level verification subsystem of the specification. \item Input parameter \customHochkommata{blockName}: The name of the top-level verification subsystem of the specification. \item Input parameter \customHochkommata{signalNames}: A list of strings that contains the names of the input signals of the top-level specification block.}{This function connects the inports of a verification subsystem with the matching signals in the selected (sub)model. For each input signal in the list of signal names, a search for a matching signal is started. In a first step, the input and output signals of the selected (sub)system are searched. If no matching signal name is found, all nested subsystems are searched, starting with the innermost subsystem. If the search is successful, a Data Store Write block is attached to the signal source. Additionally, a corresponding Date Store block is added to the selected (sub)system and the corresponding inport is connected with the output of a Data Store Read block. If no matching signal is found, an error message is printed.}